\documentclass[twocolumn]{aastex63}

\newcommand{\logg}{$\log g$\,}
\newcommand{\teff}{$T_\mathrm{eff}$}
\newcommand{\feh}{[Fe/H]}
\newcommand{\fehphot}{[Fe/H]$_{\mathrm{phot}}$}
\newcommand{\alphafe}{[$\alpha$/Fe]}
\newcommand{\rproj}{$R_{\mathrm{proj}}$}
\newcommand{\kms}{km s$^{-1}$}
\usepackage{mathrsfs}
\usepackage{csvsimple}
\usepackage[normalem]{ulem}
\usepackage{gensymb}

\newcommand{\psub}{$p_{\mathrm{sub}}$}
\newcommand{\meanfeh}{$\langle$\feh $\rangle$}
\newcommand{\meanalphafe}{$\langle$\alphafe $\rangle$}
\newcommand{\sigmaalphafe}{$\sigma$\alphafe}
\newcommand{\sigmafeh}{$\sigma$\feh}

\newcommand{\gradientfehicsm}{$-0.0075\pm0.0003$}

\newcommand{\gradientfehicss}{$-0.0050\pm0.0003$}
\newcommand{\gradientfehicsms}{$-0.0050\pm0.0003$}
\newcommand{\gradientfehicsmslimited}{$-0.0094\pm0.0003$}

\newcommand{\gradientalphaism}{$+0.0040\pm0.0006$}
\newcommand{\gradientalphaicsm}{$+0.0020\pm0.0003$}
\newcommand{\gradientalphaiss}{$+0.0001\pm0.0008$}
\newcommand{\gradientalphaicss}{$+0.0002\pm0.0006$}
\newcommand{\gradientalphaicsms}{$+0.0026\pm0.0004$}
\newcommand{\gradientalphaicsmlimited}{$+0.0055\pm0.0020$}
\newcommand{\gradientalphaicsmslimited}{$+0.0030\pm0.0017$}

\newcommand{\meanfehGSS}{$-0.859\pm0.004$}
\newcommand{\meanalphafeGSS}{$0.107\pm0.017$}
\newcommand{\meanfehGSSsmooth}{$-1.080\pm0.005$}
\newcommand{\meanalphafeGSSsmooth}{$0.340\pm0.012$}

\newcommand{\meanfehGSSenv}{$-1.459\pm0.011$}
\newcommand{\meanalphafeGSSenv}{$0.066\pm0.026$}
\newcommand{\meanfehGSSenvsmooth}{$-1.391\pm0.013$}
\newcommand{\meanalphafeGSSenvsmooth}{$0.670\pm0.032$}

\newcommand{\meanfehStC}{$-1.313\pm0.007$}
\newcommand{\meanalphafeStC}{$0.199\pm0.020$}
\newcommand{\meanfehStCsmooth}{$-1.599\pm0.007$}
\newcommand{\meanalphafeStCsmooth}{$0.394\pm0.018$}

\shorttitle{\feh\ and \alphafe\ in the M31 Halo}
\shortauthors{Wojno et al.}

\begin{document}

\title[Coadded abundances in the M31 halo]{Elemental abundances in M31: Individual and Coadded Spectroscopic [Fe/H] and [$\alpha$/Fe] throughout the M31 Halo with SPLASH}
\correspondingauthor{J. Leigh Wojno}
\email{wojno@mpia-hd.mpg.de}

\author[0000-0002-3233-3032]{J. Leigh Wojno}
\affiliation{The William H. Miller III Department of Physics \& Astronomy, Bloomberg Center for Physics and Astronomy, Johns Hopkins University, 3400 N. Charles Street, Baltimore, MD 21218}
\affiliation{Max-Planck Institute for Astronomy, Königstuhl 17, D-69117 Heidelberg, Germany}

\author[0000-0003-0394-8377]{Karoline M. Gilbert}
\affiliation{The William H. Miller III Department of Physics \& Astronomy, Bloomberg Center for Physics and Astronomy, Johns Hopkins University, 3400 N. Charles Street, Baltimore, MD 21218}
\affiliation{Space Telescope Science Institute,
3700 San Martin Dr.,
Baltimore, MD 21218, USA}

\author[0000-0001-6196-5162]{Evan N. Kirby}
\affiliation{Department of Physics and Astronomy, University of Notre Dame, Notre Dame, IN 46556, USA}

\author[0000-0002-9933-9551]{Ivanna Escala}
\altaffiliation{Carnegie-Princeton Fellow}
\affiliation{Department of Astrophysical Sciences, Princeton University, 4 Ivy Lane, Princeton, NJ~08544}
\affiliation{The Observatories of the Carnegie Institution for Science, 813 Santa Barbara St, Pasadena, CA 91101}

\author[0000-0001-8867-4234]{Puragra Guhathakurta}
\affiliation{Department of Astronomy \& Astrophysics, University of California, Santa Cruz, 1156 High Street, Santa Cruz, CA 95064, USA}

\author[0000-0002-1691-8217]{Rachael L. Beaton}
\altaffiliation{Hubble Fellow}
\altaffiliation{Carnegie-Princeton Fellow}
\affiliation{Department of Astrophysical Sciences, Princeton University, 4 Ivy Lane, Princeton, NJ~08544}
\affiliation{The Observatories of the Carnegie Institution for Science, 813 Santa Barbara St., Pasadena, CA~91101}

\author[0000-0001-9690-4159]{Jason Kalirai}
\affiliation{John Hopkins Applied Physics Laboratory, 11101 Johns Hopkins Road, Laurel, MD 20723}

\author[0000-0002-9053-860X]{Masashi Chiba}
\affiliation{Astronomical Institute, Tohoku University, Aoba-ku, Sendai 980-8578, Japan}

\author[0000-0003-2025-3147]{Steven R. Majewski}
\affiliation{Department of Astronomy, University of Virginia, Charlottesville, VA 22904-4325, USA}


\begin{abstract}
We present spectroscopic chemical abundances of red giant branch (RGB) stars in Andromeda (M31), using medium resolution ($R\sim6000$) spectra obtained via the Spectroscopic and Photometric Landscape of Andromeda's Stellar Halo (SPLASH) survey. In addition to individual chemical abundances, we coadd low signal-to-noise ratio (S/N) spectra of stars to obtain a high enough to measure average \feh\ and \alphafe\ abundances. We obtain individual and coadded measurements for \feh\ and \alphafe\ for M31 halo stars, covering a range of 9--180 kpc in projected radius from the center of M31. With these measurements, we greatly increase the number of outer halo (\rproj $> 50$ kpc) M31 stars with spectroscopic \feh\ and \alphafe, adding abundance measurements for 45 individual stars and 33 coadds from a pool of an additional 174 stars. We measure the spectroscopic metallicity (\feh) gradient, finding a negative radial gradient of $-0.0050\pm0.0003$ for all stars in the halo, consistent with gradient measurements obtained using photometric metallicities. Using the first measurements of \alphafe\ for M31 halo stars covering a large range of projected radii, we find a positive gradient ($+0.0026\pm0.0004$) in \alphafe\ as a function of projected radius. We also explore the distribution in \feh--\alphafe\ space as a function of projected radius for both individual and coadded measurements in the smooth halo, and compare these measurements to those stars potentially associated with substructure. These spectroscopic abundance distributions highlight the substantial evidence that M31 has had an appreciably different formation and merger history compared to our own Galaxy.
\end{abstract}


\section{Introduction} \label{sec:intro}
The stellar halos of late-type galaxies contain a wealth of information detailing the formation and merger history of the host galaxy.
The earliest models of halo formation followed two distinct pathways: rapid radial collapse \citep{Eggen62}, and the accretion model \citep{Searle78}. Modern halo formation scenarios follow cosmologically-motivated prescriptions for hierarchical formation \citep[e.g.][]{Bullock05}, where galactic halos are built up through the accumulation of smaller satellites \citep{cole94,johnston96,bullock01,helmi08}. These satellites are typically assimilated into the host galaxy, making it difficult to determine which stars were formed in-situ, and which were formed in a galaxy that has since been accreted. 

Our nearest spiral neighbor, the Andromeda galaxy (M31), offers a particularly illuminating case for understanding the stellar halo of a massive galaxy. 
Within its halo, we observe a kinematically hot classical halo component \citep[e.g.][]{guhathakurta05,kalirai06b}, many distinct debris streams
\citep[e.g.][]{ibata01,zucker04,ibata07,McConnachie09}, ancient globular clusters \citep[e.g.][]{mackey19}, 
and intact dwarf spheroidal galaxies (\citealt{mcconnachie12}; see \citealt{mcconnachie18} for a recent review of these components). 
In M31, accretion events where tidal debris has not yet  been phased mixed can be identified as overdensities in star count maps \citep[e.g.][]{ibata01,ferguson02}, and as kinematic peaks in the velocity distribution of the halo \citep[e.g.][]{kalirai06b,gilbert09}. The largest and most massive of these features is the Giant Stellar Stream \citep[GSS, ][]{ibata01}, which likely is the result of a recent ($<4$ Gyr ago) accretion of a $M_* \sim 10^{9-10}$M$_{\odot}$ satellite galaxy \citep[e.g.,][]{fardal13,hammer18,dsouza18}. In addition, it has been shown that a number of additional debris features in the halo of M31 may be related to this interaction \citep{fardal06,fardal07,fardal08,fardal12,gilbert07,escala22,dey22}. These substructure features cover a large radial extent, and can be found out to a projected radius of $\sim 90$ kpc from M31's center.

In our own Galaxy, 6D phase space information is now available for hundreds of thousands of individual stars \citep[e.g.][]{gaiadr3,apogee_dr17}. In contrast, we rely on a significantly less complete kinematical data set for stars in M31, where kinematical information for individual stars is typically limited to line-of-sight velocity measurements. 
However, abundance measurements are attainable for individual stars in M31, and can be obtained using either photometry or spectroscopy.
Therefore, because it is directly measurable and not dependent on modeling missing dimensions for interpretation, at present it is most straightforward to utilize the chemical abundance information available from these M31 stars to reconstruct the evolutionary history of M31's halo. 

Metallicity and elemental abundance measurements are extremely informative with respect to exploring the star formation and assembly history of a galaxy. Hierarchical formation scenarios predict a spread in metallicity as a function of distance from the center of the galaxy, as accreted populations are likely to have
a wide range of star formation histories that also differ from that of the host galaxy itself. Star formation and chemical enrichment timescales are typically driven by the mass of a system, and therefore the mass distribution, as well as time of accretion of accreted galaxies, imprint on the observed abundances in the host. Generally, cosmological models \citep[e.g.][]{samland03,kobayashi11,grand17,tissera17,monachesi19} predict   metallicity gradients as a function of radial distance for disk and halo populations, as the inner regions are more dense and therefore have a higher star formation efficiency (reaching higher metallicities), compared to the outer regions, which are sparser and have lower rates of star formation. 

In a similar vein, the distribution of the ratio of $\alpha$-elements (e.g., Mg, Si, Ca, Ti) to Fe (\alphafe\footnote{Throughout the paper, we use the standard definition for elemental abundance ratios: 
[X/H] $= \mathrm{log}_{10}\left(\frac{N_{\mathrm{X}}}{N_{\mathrm{H}}} \right)_{\mathrm{star}} -\mathrm{log}_{10}\left( \frac{N_{\mathrm{X}}}{N_{\mathrm{H}}}\right)_{\odot}$.}) as a function of \feh\ directly correlates to the enrichment history of a system: stars that explode as Type II supernova produce  $\alpha$-elements and Fe. When Type Ia supernova begin to contribute their ejecta, which are much richer
in Fe than $\alpha$-elements, to the interstellar medium, the \alphafe\ ratio decreases as a function of increasing \feh\@. The position of the decrease in \alphafe\ is often referred to as a ``knee'' of the distribution, and where this knee falls as a function of \feh\ provides vital information on the star formation and enrichment history of the environment where those stars were formed. In low-mass galaxies, this ``knee'' is observed at lower values of \feh\ compared to higher mass systems \citep{mcwilliam97,Tolstoy09}.

Compared to the MW, the halo of M31 is overall significantly more metal-rich for comparable radial ranges \citep{mould86,durrell94,rich96}. While the metal-rich nature of the M31 halo is well documented, measuring the presence (or lack thereof) of stellar metallicity gradients across the halo of M31 has required surveys over a substantial radial range, which has not been possible until relatively recently. Early work, restricted to limited ranges in projected distance from M31, found little evidence for a metallicity gradient. Using CFHT photometry of individual stars, \citet{durrell01, durrell04} found no evidence for a radial metallicity gradient out to 30 kpc. \citet{ferguson02} carried out a panoramic imaging study of M31 using the Wide Field Camera on the Isaac Newton Telescope, and while they found some variation in the color (and therefore metallicity) of RGB stars as a function of position across the M31 halo, they did not find evidence for a gradient. In addition, \citet{bellazzini03} found ``remarkably uniform" metallicity distributions of 16 fields observed with HST photometry out to a projected radius of 35 kpc. 

These early results conflict with more recent photometric and spectroscopic studies that found evidence for a metallicity gradient in the M31 halo. The Pan-Andromeda Archaeological Survey \citep[PAndAS;][]{McConnachie09} obtained comprehensive photometric coverage of the M31 halo out to $\sim150$ kpc in projected radii from the center of M31. 
Studies using these data found evidence for low-metallicity stellar populations at large projected radii, based on comparing the position of stars in the color-magnitude diagram to stellar isochrones 
\citep{irwin05,chapman06,ibata14}. A number of other photometric surveys were conducted to explore evidence for a metallicity gradient in the halo of M31, with varying results (\citealt{richardson09, tanaka10}; see \citealt{gilbert14} and references therein for a comprehensive history).

Using the DEIMOS instrument on the 10m Keck II telescope, the Spectroscopic and Photometric Landscape of Andromeda's Stellar Halo (SPLASH) survey obtained spectra for thousands of red giant branch stars throughout the disk, halo, and satellites of M31
\citep[e.g.,][]{tollerud12,gilbert12,dorman12}.
With the first SPLASH fields, \cite{kalirai06b} 
found a significant metallicity gradient (with both CMD-based and Ca II triplet-based \feh\ estimates) using over 250 secure M31 member stars, observed with Keck/DEIMOS spectroscopy, from 12 fields spanning $12-165$ kpc in projected radius from the center of M31.
\citet{gilbert14} analyzed a much larger SPLASH halo sample (over 1500 individual RGB stars). 
They corroborated the presence of a large scale metallicity gradient, using both photometric metallicity measurements and Ca II triplet based metallicity measurements, for individual stars over a similar range ($10-90$ kpc) in projected radii. However, these prior metallicity measurements 
are all indirect measurements of \feh\ that rely on a significant number of assumptions, and do not provide estimates for \alphafe. Therefore, a substantial effort has been made to obtain spectral abundance measurements via spectral synthesis for both individual and coadded stars observed as part of SPLASH. 

\citet{vargas14} presented the first spectroscopic abundance measurements in M31's halo using SPLASH data, obtaining  spectroscopic \feh\ and \alphafe\ abundances for four stars in the outer halo ($70 < $\rproj$< 140$ kpc) of M31. They found these outer halo stars to be metal poor (\feh$\sim -2.2$ to $-1.4$), with an average \alphafe\ of $0.2\pm0.2$. \citet{gilbert20} added to this sample 21 outer halo stars with spectroscopic \feh\ measurements, and seven outer halo stars with both \feh\ and \alphafe\ abundance measurements. These outer halo stars were also measured to be relatively \alphafe-enhanced and metal-poor, and were found to be more similar to the outer halo of the MW than the inner halo of M31. Most recently, for a sample of stars in the inner regions ($8 < $ \rproj\ $ < 34$ kpc) of the M31 halo with spectra observed using long exposures with the 600ZD grating, \citet{escala20b} found a similar ($-0.025\pm0.002$ dex kpc$^{-1}$) spectroscopic \feh\ gradient, and a flat gradient in \alphafe. \citet{escala21} also found the GSS to be \alphafe-enhanced compared to the smooth halo, with a mean of $\sim0.4 $ dex and a large spread in \alphafe. 

In this work, we greatly expand on previous spectroscopic abundance measurements (\feh\ and \alphafe) in M31's halo using the SPLASH halo sample of \citet{gilbert18}, which includes fields covering a large radial extent (9\,--\,180 kpc) in projected radius from the center of M31. 
We measure both individual and coadded abundances:
for M31 spectra without sufficient S/N to measure individual abundances, we coadd spectra to obtain a S/N ratio high enough to measure mean \feh\ and \alphafe\ abundances \citep{yang13,wojno20}. In particular, we greatly improve the number of M31 stars with abundance measurements at large values of \rproj. For stars with \rproj\ $> 50$kpc, we measure abundances for 45 individual stars, with another 174 stars represented in 33 coadds. Previously, only six stars in M31's halo at \rproj\ $> 50$kpc had measured \alphafe\ and \feh\ abundances \citep{vargas14,gilbert20}. 

\begin{figure*}
 	\includegraphics[width=\textwidth]{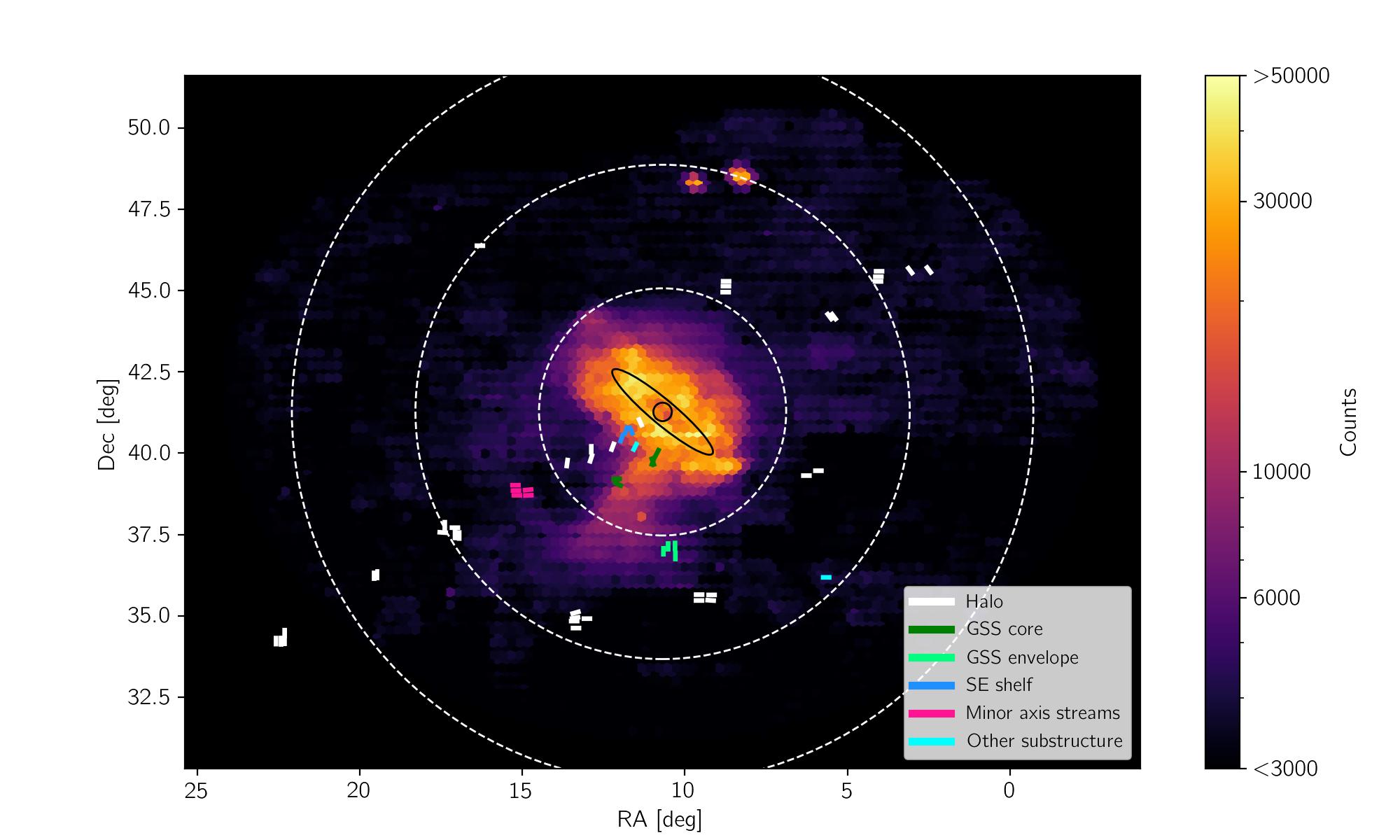}
    \caption{The halo of M31 and the spectroscopic slitmask positions for our sample of SPLASH stars. The colored 2D histogram indicates the density of stars as a function of position on the sky, using the most recent PAndAS public data release \citep{McConnachie09,mcconnachie18}. Overplotted are the approximate locations of M31's disk (black oval), and rings indicating (50,100,150) kpc in projected distance from the center of M31. The fields targeted by the SPLASH survey are indicated with colored rectangles, where the color corresponds to the substructure (or lack thereof) associated with that field. In this work, we analyze a sample of 1079 M31 stars observed with 65 masks in 32 fields, spanning a large range (9-180 kpc) in projected distance from the center of M31.}
    \label{fig:m31_roadmap}
\end{figure*}

In Section~\ref{sec:data}, we describe the observation strategy and data reduction used to gather the spectra used in this work. 
To the wealth of analysis previously conducted with these data \citep[e.g.][]{gilbert12,gilbert14,gilbert18}, we contribute spectroscopic measurements of \feh\ and \alphafe\ for individual stars (Section~\ref{sec:individual_abundances}), as well as coadded abundances (Section~\ref{sec:coadded_abundances}), making use of stars that do not have secure individual abundance measurements. 
We present our measurements of the spectroscopic radial metallicity gradients in the halo of M31 in Section~\ref{sec:smooth_halo}, both for \feh\ (Section~\ref{sec:feh_gradients}), and for \alphafe\ (Section~\ref{sec:alphafe_gradients}).
We also explore the \feh--\alphafe\ abundance plane as a function of projected radius for both the smooth halo (Section~\ref{sec:alphafe_feh_smooth_halo}) and stars potentially associated with known substructure components (Section~\ref{sec:halo_substructure}). In Section~\ref{sec:halo_substructure}, we show the  \feh--\alphafe\ abundance distribution for each substructure component present in our sample.
In Section~\ref{sec:discussion}, we explore origin scenarios for the radial abundance gradients that we find, as well as the unique abundance distribution of stars in the outer halo.
Finally, Section~\ref{sec:conclusions} summarizes our findings.

\section{Observations and Data} \label{sec:data}
We use spectra of red giant branch (RGB) stars in the halo of M31 obtained via the Spectroscopic and Photometric Landscape of Andromeda's Stellar Halo (SPLASH) survey. 
The SPLASH survey was conducted over the course of approximately 10 years, collecting photometry and low and medium-resolution spectra for $\sim 10^4$ individual stars. 
We refer the reader to \citet{gilbert12,gilbert14,gilbert18} for full details on the data reduction, spectroscopic slitmask design, and survey strategy\footnote{Table~1 of \citet{gilbert12} contains a comprehensive summary of all relevant observational constraints for the data used in this study.}.

In this study, we consider 32 SPLASH fields across the M31 halo (9 $< \mathrm{R}_{\mathrm{proj}}< 180$ kpc), sourced from a total of 65 individual slitmasks.  
Figure~\ref{fig:m31_roadmap} shows the position of the slitmasks used with respect to the M31 halo, where the color of the slitmask indicates the known substructure components the field contains (if any).

\subsection{Target selection and photometry} \label{sec:target_selection}
Targets in our sample were chosen based on their magnitudes and colors. The majority of the sample was selected using imaging taken with the Mosaic Camera on the Kitt Peak National Observatory (KPNO) 4 m Mayall telescope in the Washington system $M$ and $T_2$ filters, and the narrow-band DDO51 filter  \citep{ostheimer03,beaton07}.
Fields in the innermost regions of the halo (\rproj $< 30$kpc) were designed using imaging obtained in the $g'$ and $i'$ bands using the MegaCam instrument on the Canada-France-Hawaii Telescope (CFHT) \citep{gilbert12}. For subsequent analysis, all magnitudes were transformed into the Johnson system V, I magnitudes, using the relations by \citet{majewski00} for imaging obtained in $M$ and $T_2$, and observations of Landolt photometric standard stars for imaging in $g'$ and $i'$. Targets with photometry consistent with being RGB stars at the distance of M31 were assigned a high priority to be included on the spectroscopic slitmasks \citep{gilbert18}. 

\subsection{Spectroscopy and data reduction} \label{sec:observations}
The spectroscopic slitmasks used in our sample were observed for a nominal total exposure time of 1 hour, with modifications made for particularly good or bad observing conditions. Spectra were obtained with the DEIMOS instrument on the Keck II telescope, using the 1200 line mm$^{-1}$ (1200G) grating, OG550 order blocking filter, which yields a spectral dispersion of $0.33$\AA\ pix$^{-1}$, and resolution of 1.2 \AA\ FWHM\@. The spectra cover a wavelength range of $\sim6300~\mbox{\AA} < \lambda < 9100~\mbox{\AA}$, with a resolution of $\sim6000$ at a central wavelength of $\sim7800\mbox{\AA}$. We note that 6 fields in our sample were also targeted for deep ($\sim 6$hr) observations using the low-resolution ($R\sim 3000$) 600ZD grating \citep{escala20a,escala20b}.
% and a slit width of $1''$,
Science spectra were reduced using the DEEP2 DEIMOS data reduction pipeline, developed by the DEEP2 Galaxy Redshift Survey\footnote{http://deep.ps.uci.edu/spec2d/} \citep{cooper12,newman13}. This software handles flat-fielding, sky subtraction, and extraction of one-dimensional spectra. We applied the modifications described by \citet{simon07} for increased performance with bright stellar sources. Line-of-sight velocities were measured by cross-correlating observed spectra with stellar templates provided by \citet{simon07}. Details on the stellar templates used for our sample of RGB stars can be found in Table 2 of \citet{wojno20}. We used the relative positions of the Fraunhofer A band and other telluric absorption features  to correct for the imperfect centering of a star in its slit on the slitmask, following \citet{simon07} and \citet{sohn07}.

\begin{figure*}
 	\includegraphics[width=\textwidth]{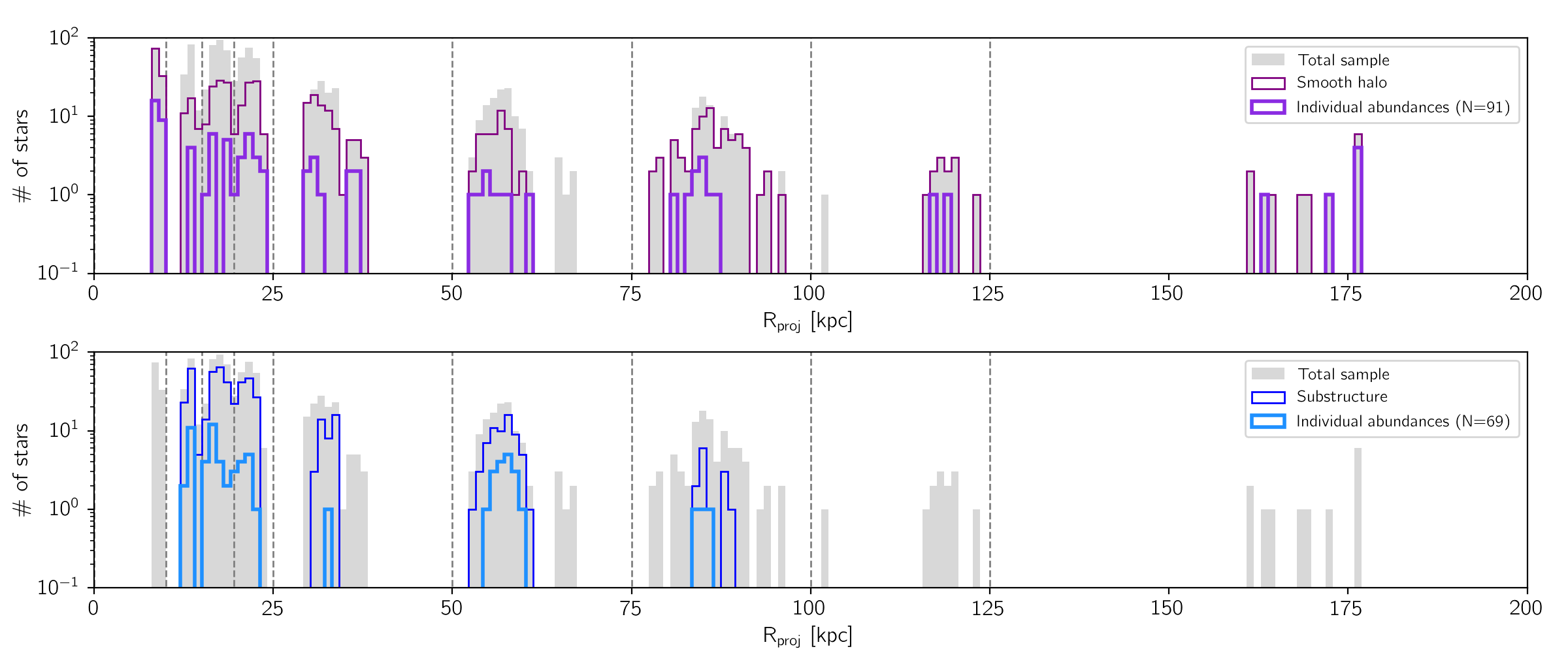}
    \caption{Histograms showing the radial distribution of our sample. The grey filled histogram represents our sample of all secure M31 stars. Top panel: the thin purple line represents the sample of smooth halo stars ($p_{\mathrm{sub}} \leq 0.2$). The thick purple line represents the sample of smooth halo stars with spectroscopic abundance measurements for both \feh\ and \alphafe\ that pass our quality criteria (Section~\ref{sec:pipeline_process}). Bottom panel: the blue line represents stars that may be associated with substructure components. The thick cyan line indicates stars with individual abundance measurements that pass the quality criteria. The vertical grey dashed lines indicate the limits of the radial bins used for binning smooth halo stars for coaddition. In total, 21\% (160 stars) of secure M31 stars in the sample have individual abundance measurements; 70\% (543 stars) are ultimately used for coadded measurements.}
    \label{fig:rproj_hist_comparison}
\end{figure*}

\subsection{M31 membership}
\label{sec:m31_membership}

We calculated the probability of M31 membership using the velocity model presented by \citet{gilbert18}.  We used the mean of the posterior probabilities of M31 membership for each star, calculated from the M31 and MW likelihoods calculated in the course of the MCMC sampling of the Gaussian mixture model \citep[Section~4.2;][]{gilbert18}.  The model likelihoods incorporated the prior probability of M31 membership based on the method established by \citet{gilbert06} to select stars more likely to be RGB stars at the distance of M31 than foreground MW dwarf stars. This method uses the following photometric/spectroscopic diagnostics: (1) position in ($M-{\mathrm{DDO51}}$) versus ($M-T_2$) color-color space, (2) position in $(I, V-I)$ color-magnitude space, (3) equivalent width of the Na I doublet ($\lambda \lambda$ 8183,8195) as a function of $(V-I)$ color, and (4) comparison between photometric and spectroscopic \feh\ measurements.  In general, the \citet{gilbert06} method also includes a diagnostic based on the line-of-sight velocity; however, this was not included in calculating the prior probability of M31 membership by \cite{gilbert18}. 

This produces an M31 membership probability ($p_{\mathrm{M31}}$) between 0 and 1, where a value of 1 indicates that a star is most likely to belong to M31. We selected M31 members by requiring  $p_{\mathrm{M31}} > 0.5$. For stars with \rproj $> 90$ kpc, we additionally utilized the original M31 membership classification presented by \citet{gilbert06}, M31CLASS\@. Based on their likelihood of belonging to M31 using the five diagnostics listed above (including line-of-sight velocity), stars are assigned integer values ($-3 < \mathrm{M31CLASS} < +3$), where a value of $+3$ indicates that the star is very likely to belong to M31. We required M31CLASS = 3 (stars must be 3 times more likely to belong to M31 than the MW and must not be bluer than the most metal-poor isochrone) for stars with \rproj $> 90$ kpc.  This additional criterion was employed for our outermost halo fields because the number of M31 halo stars at these large distances is a small fraction of the observed foreground MW contaminants.  We have shown in previous work that with this conservative sample selection, there is no indication of MW contamination in the M31 halo sample at large \rproj\ \citep{gilbert12,gilbert14}.

\subsection{Substructure classification}
\label{sec:substructure_probability}
We applied an additional classification to our sample of secure M31 stars to facilitate a comparison between the smooth halo and debris features. The velocity distributions of the most prominent tidal debris features in the SPLASH data have been characterized in previous works \citep{guhathakurta06,kalirai06b,gilbert07,gilbert09,gilbert12}. The line of sight velocity distribution of the M31 halo can be modeled as a Gaussian with a large velocity dispersion \citep{gilbert07}, and each substructure component as additional Gaussian components \citep{gilbert07,gilbert12}. These maximum-likelihood fits can then be used to determine the probability that a given star is more likely to belong to the smooth halo or a substructure component. 

We adopted substructure probability (\psub) values calculated from previous fits to the velocities of secure M31 stars in individual substructure components as well as the smooth halo, as summarized in Table 4 of \citet{gilbert18}. For this study, we defined stars with \psub $ < 0.2$ as smooth halo stars, and considered all stars with \psub $ \ge 0.2$ as potentially associated with substructure. When we examine the histogram of substructure probabilities for our sample, we find a dearth of stars around $p_{\rm sub} = 0.2$, and therefore serves as a dividing line. We note that with this cut, our smooth halo sample may have some contamination from substructure components, and vice-versa. This choice minimizes contamination of the smooth halo sample by stars associated with substructure  while enabling measurements representing as full a sample of M31's stellar halo as possible. Applying the division at \psub$ = 0.2$ yields 538 smooth halo stars, and 533 stars potentially associated with substructure. While the sample with \psub$ > 0.2$ will be dominated by stars potentially associated with substructure, 126 stars, representing 24\% of this sample, have $0.2 < $\psub$< 0.5$. This means that any abundance differences between the smooth halo and the sample of stars potentially associated with substructure are likely to be underestimated.
%switches to present tense here... double check this before submission

After applying the M31 membership criteria (Section~\ref{sec:substructure_probability}), our sample of all M31 stars consists of 1079 stars across 65 pointings, with 538 belonging to the smooth halo, and 541 that have the potential to belong to one of the various substructures in the halo of M31 (Section~\ref{sec:halo_substructure}). Figure~\ref{fig:rproj_hist_comparison} shows the total number of stars in our sample as a function of their projected distance in kpc (\rproj) from the center of M31, where the top panel shows the distribution of all stars belonging to the smooth halo, as well as those secure abundance measurements. The bottom panel shows the distribution of all stars compared to those likely associated with substructure, as well as those substructure stars that have secure abundance measurements. The projected distance from the center of M31 is calculated using a distance modulus of 24.47 (corresponding to a distance to M31 of 783 kpc; \citealt{stanek97,mcconnachie05}), in order to facilitate an accurate comparison to literature measurements of the metallicity gradient (Section~\ref{sec:abundance_gradients}). 

\section{Spectroscopic abundance measurements}
\label{sec:individual_abundances}

We used the abundance measurement pipeline presented by \citet{escala18} to measure individual \feh\ and \alphafe\ for all stars in our sample. In summary, the pipeline takes as input a 1D spectrum (and its associated uncertainty array) along with corresponding photometric estimates of \teff, \logg, and \feh\  (Section~\ref{sec:photometric_parameters}), and outputs spectroscopic effective temperature (\teff) along with \feh\ and \alphafe. It compares the observed spectrum to a grid of synthetic spectra (Section~\ref{sec:synthetic_grid}). We allowed \teff, \feh, and \alphafe\ to vary iteratively until a best-fit synthetic spectrum is found via $\chi^2$-minimization. The pipeline process for individual spectra is described in Section~\ref{sec:pipeline_process}. We then built on this methodology to allow the pipeline to handle coadded spectra (Section~\ref{sec:coadded_abundances}, \citealt{wojno20}). Further technical details of the pipeline are described by \citet{escala18}, and the relevant modifications needed to accommodate for coadded spectra are described by \citet{wojno20}.

\subsection{Photometric atmospheric stellar parameters} 
\label{sec:photometric_parameters}
We estimated photometric stellar parameters (\teff, \logg, \fehphot) by comparing the observed ($V,I$) color and $I$ magnitude to a grid of theoretical isochrones. Like \citet{wojno20}, we adopted the error-weighted mean of the best-fit Padova \citep{Girardi16}, Victoria-Regina \citep{vandenberg06}, and Yonsei-Yale \citep{demarque04} isochrone sets. For these isochrones, we assumed a stellar age of 14 Gyr and \alphafe $= 0.4$ dex for all RGB stars. We shifted the isochrones to the distance of M31 by assuming a distance modulus\footnote{We also tested the effect of assuming different distance modulus estimates: $24.47$ \citep{mcconnachie05}, and $24.38$ \citep{riess12}. We found there to be no significant impact on the final measured abundances.} of $24.63\pm0.02$ \citep{clementini11}. We determined photometric \teff, \logg, and \feh\ for each star by linearly interpolating the isochrone grid.

\subsection{Synthetic spectral grid}
\label{sec:synthetic_grid}
In each step of the fitting procedure, the observed spectrum was compared to a grid of synthetic spectra that were generated using the LTE spectral synthesis code MOOG \citep{sneden73}, with ATLAS9 stellar atmospheric models \citep{kirby11, kurucz93,sbordone04,sbordone05}. The line list was developed using the Vienna Atomic Line Database \citep[VALD;][]{kupka99}, molecular lines from \citet{kurucz92}, and a version of the hyperfine transition line list from \citet{kurucz93} tuned to match line strengths of the Sun and Arcturus \citep{kirby11,escala19b}. 

Synthetic spectra were generated for the following parameter ranges: $3500 $ K $\leq$ $T_{\mathrm{eff}}$ $\leq$ 8000 K, $0.0 \leq $ log $g \leq $ 5.0, $-5.0 \leq$ \feh\ $\leq 0.0$, and $-0.8 \leq $ \alphafe\ $\leq +1.2$, for a total of 316848 synthetic spectra \citep{kirby11}. The synthetic spectra have a wavelength spacing of $0.02$\AA\ and cover a wavelength range of $6300-9100$\AA\@. When compared to observed spectra, the synthetic spectra were interpolated and smoothed using the observed spectral resolution as a function of wavelength. For individual stars, we assumed that the observed spectral resolution ($\Delta \lambda \approx \mathrm{FWHM}/2.35$) varies as a function of wavelength, as in \citet{escala20a}.

\subsection{Chemical abundance pipeline}
\label{sec:pipeline_process}
The chemical abundance pipeline used to measure \feh\ and \alphafe\ for individual stars is described in detail by \citet{kirby08a} and \citet{escala18}, and we summarize here. Modifications made to the chemical abundance pipeline to accommodate coadded spectra are presented by \citet{wojno20}.

We removed telluric absorption features due to the Earth's atmosphere using a spectrophotometric standard star  \citep{simon07}, where we divided the observed spectra by a scaled template spectrum \citep{kirby15}. Observed spectra are shifted to the rest frame according to their line-of-sight velocities. The spectra are then continuum normalized by fitting a third-order B spline to the continuum regions \citep{kirby08a}, using a breakpoint spacing of 100 pixels.

An initial fit is performed following the initial continuum normalization. In this step, \teff\ and \feh\ are allowed to vary simultaneously. While \citet{escala18} allow for the spectral resolution ($\Delta \lambda$) to vary as well in this step, we determine $\Delta \lambda$ as a function of wavelength for each slitmask in our sample, as in \citet{escala20a}. After fits for \teff\ and \feh\ are found, we then allowed \alphafe\ to vary, keeping \teff\ and \feh\ constant. The continuum is then re-normalized, using the best-fit synthetic spectrum. This process continues iteratively until the parameter values vary by less than the following tolerance: 1 K in \teff, 0.001 dex in \feh, and 0.001 dex in \alphafe. Once the continuum refinement converged, we refitted for \feh\ and \alphafe\, and finally \feh\ again. Fit uncertainties were calculated as $\sigma_{ii}(\chi^2_{\nu})$, where  $\sigma_{ii}$ represents the diagonals of the covariance matrix of the fit, and $\chi^2_{\nu}$ is the reduced-$\chi$ statistic for a given fit parameter. We combined this fit uncertainty in quadrature with the systematic uncertainty floors determined in \citet{kirby20} of 0.101 dex for \feh, and 0.084 dex for \alphafe.

We define a star to have secure abundance measurements if all of the following conditions are met: $\sigma$\feh $ < 0.4$, $\sigma$\alphafe $ < 0.4$, the measured value is not at the edge of the grid, and the $\chi^2$ contours for the \feh\ and \alphafe\ fit parameters vary smoothly.

\subsection{Removing stars with TiO}
\label{sec:tio_stars} 
\begin{figure}
 	\includegraphics[width=\columnwidth]{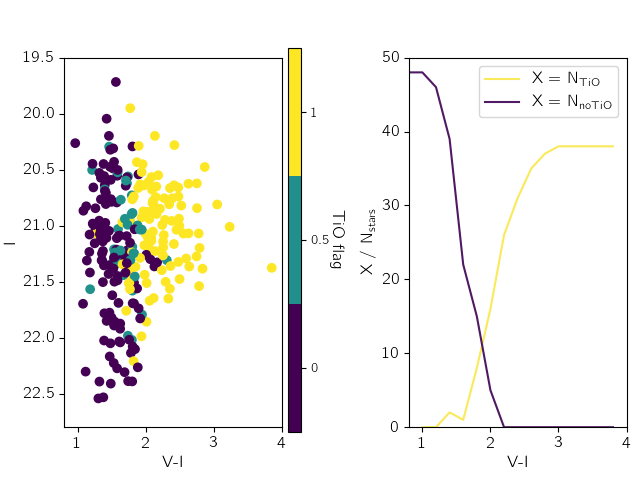}
    \caption{Left: $(V-I,I)$ color-magnitude diagram for stars with individual abundance measurements, color-coded by their assigned TiO flag (0: no TiO, 0.5: potential TiO, 1.0: clear evidence of TiO, Section~\ref{sec:tio_stars}). Right: the percent of stars identified having TiO absorption (blue) and stars without any TiO features (magenta) as a function of $(V-I)$ color. We apply a color cut at $(V-I) = 1.8$ to our sample of stars without reliable abundance measurements (Section~\ref{sec:coadded_abundances}), as this is approximately where we have the least contamination from stars with TiO absorption features in their spectra, while retaining the largest sample possible.}
    \label{fig:tio_comparison}
\end{figure}

In cool ($T < 4500$~K) giant stars, TiO absorption can be observed in the optical portion of the spectrum as large sawtooth bands typically in the wavelength range 7055-7245 \AA\  \citep{morgan1943,Jorgensen94}. This feature is not modeled in our synthetic spectra, and therefore we cannot reliably recover abundances for stars with signatures of TiO absorption. For consistency with previous works in this series using similar data sets, we remove stars with obvious TiO bands from our sample. 
For stars that pass the quality criteria on their spectroscopic measurements (see Section~\ref{sec:pipeline_process}), we examine each spectrum for the presence of TiO. Each star is then categorized as having either obvious signatures of TiO, marginal signatures of TiO, or no TiO\@. For our sample of stars with individual measurements, we find 38\% (100), 14\% (36), and 48\% (125) stars in each category, respectively. For the rest of the analysis, we remove any stars with obvious or marginal signs of TiO.

In Figure~\ref{fig:tio_comparison} we show $(V-I, I)$ color-magnitude space for our categorized stars, color-coded by the detection (or non-detection) of TiO in their spectra. Making a division at $(V-I) = 1.8$ removes the majority of stars with obvious signatures of TiO absorption, with a contamination rate (stars blue-ward of this cut but still having TiO features in their spectra) of $\sim$10\%. For stars without secure individual abundance measurements, which typically have very low S/N, we subsequently consider only those with $V-I < 1.8$ to remove the majority of stars that are likely to have TiO absorption features in their spectra. This cut preferentially removes cool, metal-rich stars from our sample, and we consider the effects of a bias on the metallicity distribution of our final sample in Section~\ref{sec:tio_bias}.

\subsection{Coadding stars} \label{sec:coadded_abundances}

For stars that do not have secure abundance measurements (i.e., they do not pass the quality criteria described in Section~\ref{sec:pipeline_process}), we combine their spectra to obtain a high enough S/N to obtain an abundance measurement. We identified groups of $\sim 5$ stars to coadd, with the aim of coadding stars that are as similar as possible in each group.  

For stars associated with the smooth halo ($p_{\rm sub} < 0.2$, Section~\ref{sec:substructure_probability}), the sample is first divided into bins according to their projected distance from the center of M31.  The boundaries of the radial bins are 10, 15, 19.5, 25, 50, 75, 100, and 125 kpc, as illustrated in Figure~\ref{fig:rproj_hist_comparison}.  For stars that may be associated with substructure ($p_{\rm sub} \ge 0.2$), the sample is divided by spectroscopic field.  For fields with multiple substructure components (e.g., multi-peaked velocity distributions), the substructure sample is further divided to ensure that only stars associated the same velocity component for a given tidal feature are combined.  Within each radial bin (smooth halo sample) or field (substructure sample), stars are then sorted according to their photometric \feh\ to obtain groups of approximately 5 stars per coadd. 

We then combine the stellar spectra in each photometric \feh\ bin according to the coaddition method described by \citet{wojno20}. First, all spectra are shifted to the rest frame using their observed line-of-sight velocities. The typical line-of-sight velocity uncertainty for stars to be coadded is $\sim6.7$ \kms.
The spectra are then continuum normalized, and rebinned such that the wavelength array of each spectrum covers the wavelength range (6300-9100 \AA) with a wavelength spacing of 0.23 \AA\ per pixel to ensure all spectra in the coadd are on the same pixel array. As $\Delta \lambda$ varies for each spectroscopic mask, and stars from different masks can be coadded, we wish to avoid assuming a ``coadded'' $\Delta \lambda$. The smoothing parameter $\Delta \lambda$ is defined as FWHM/2.35, which for the 1200G grating with a slit width of $0''.7$ is 0.45. Therefore, for coadded stars, we assume a single fixed value of $\Delta \lambda = 0.45$ (instead of $\Delta \lambda$ varying as a function of wavelength).

The continuum-normalized flux from each spectrum in a given photometric \feh\ bin is combined on a pixel-by-pixel basis, weighted by the inverse variance $(1/\sigma^2)$: 
\begin{equation}
\label{eq:coadd}
    \bar{f}_{i} = \frac{\sum^{n}_{j=1} (f_{\mathrm{pixel},ij}/\sigma{^2}_{ij})}{\sum^{n}_{j=1} 1/\sigma{^2}_{ij}}
\end{equation}
where $f_{ij}$ represents the flux of the ith pixel of the jth spectrum in a group, $\sigma{^2}_{ij}$ is the variance of $f_{ij}$, and $n$ is the number of spectra to be coadded in a given photometric \feh\ bin. The inverse variance weighted flux of the ith pixel of the coadded spectra is then given by $\bar{f}_{i}$. We also coadd the inverse variance for a group of spectra in a similar way to obtain an inverse variance for the coadded spectrum:

\begin{equation}
    \label{eq:inv_var}
    \bar{\sigma}_{\mathrm{pixel},i} = \left( \sum^{n}_{j=1} \frac{1}{\sigma{^2}_{\mathrm{pixel},ij}} \right)^{-1/2}.
\end{equation}

We then measure \feh\ and \alphafe\ following the spectroscopic synthesis and fitting procedure described by \citet{wojno20}. The method for coadded spectra is similar to that for individual spectra, with the following modifications: (1) we assume a fixed value for $\Delta \lambda$, and (2) the coadded spectrum is compared to synthetic spectra that have been coadded using the same inverse variance arrays as the observed spectra.

\begin{figure*}
 	\includegraphics[width=\textwidth]{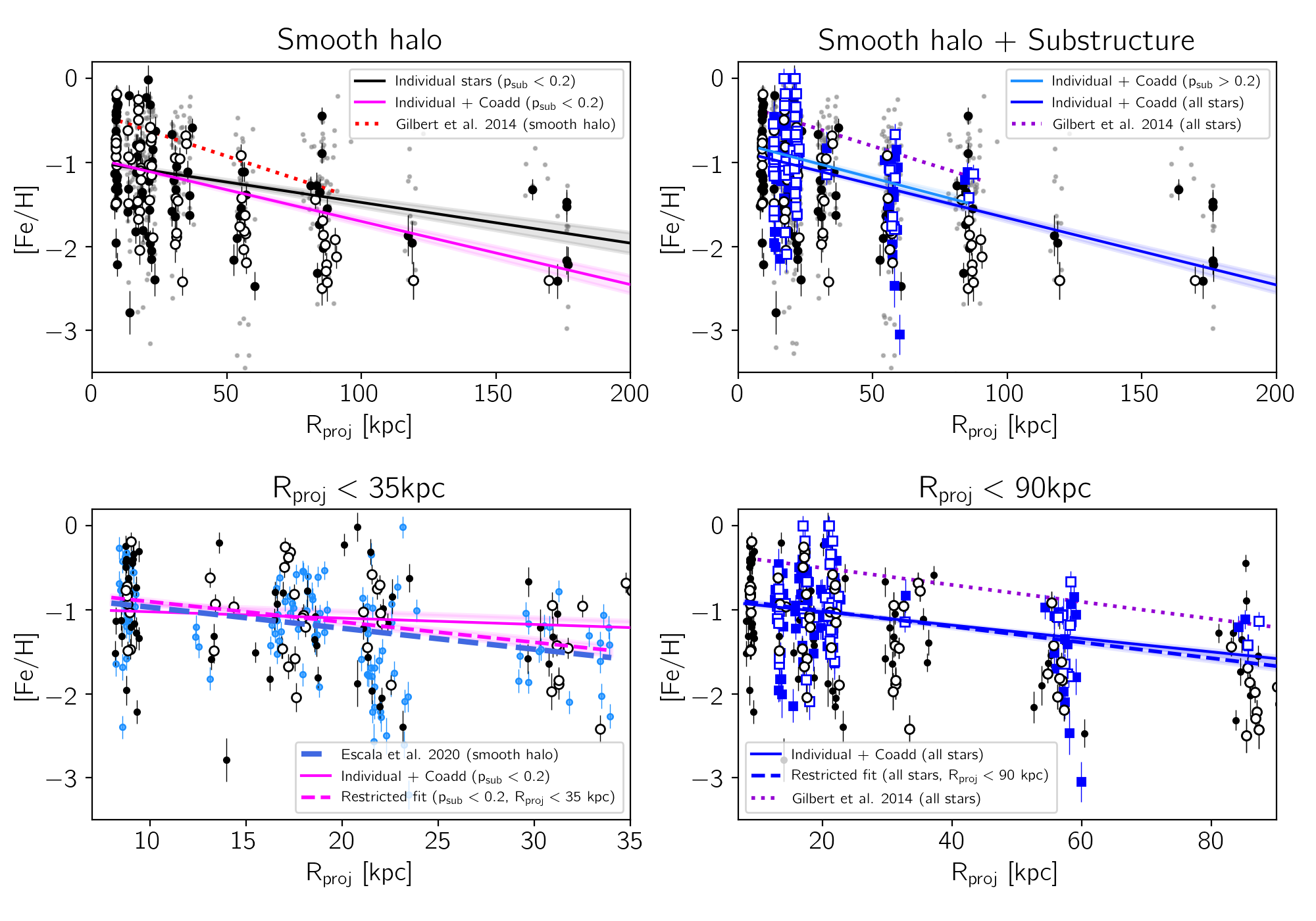}
    \caption{Spectroscopic \feh\ versus projected radius for stars with individual/coadded (filled/open circles) \feh\ measurements.  Grey points are photometric \feh\ measurements for our sample. Top left: The best-fit spectroscopic \feh\ gradient for individual smooth halo stars (Section~\ref{sec:feh_smooth_halo}, solid black), where the shaded regions indicate a 1$\sigma$ standard deviation. The red dotted line represents the gradient measured by \citet{gilbert14} for smooth halo stars, using photometric \feh. The magenta line shows the best-fit gradient calculated using both individual stars and coadds (Section~\ref{sec:coadded_abundances}). Bottom left: Individual and coadded (filled/open circles) \feh\ measurements, but for a limited radial range (\rproj$< 35$ kpc). The sample of \citet{escala20b} is shown as light blue points, and the gradient measured with their sample is shown as a dashed blue line. The solid magenta line is the same as the top-left panel. We also show the fit after limiting our sample to the same radial range (dashed magenta). Top right: Individual/coadded (filled/open dark blue squares) \feh\ measurements for stars potentially associated with substructure (Section~\ref{sec:substructure_probability}), with the best-fit gradient to individual and coadded measurements from the substructure sample (cyan), and the gradient including all stars (both the smooth halo and substructure samples; blue). The photometric \feh\ gradient from \citet{gilbert14}, fit using smooth halo and substructure stars, is shown as a dashed purple line. Bottom right: The same as above, but for a limited range (\rproj $<90$kpc). Regardless of our sample selection, we recover a statistically significant metallicity gradient in the halo of M31.}
    \label{fig:metallicity_gradient_individ}
\end{figure*}

 For each coadded spectrum, we checked the $\chi^2$ fit contours for \feh\ and \alphafe\ to confirm that the minimum is well-defined. If the fit contours were not well-defined, we removed stars with low S/N from the coadd and reassess the coadded spectrum until we achieved a good fit.  We also propagated uncertainties arising from uncertainties in the photometric measurements of \teff\ and \logg\ \citep{wojno20}. The fit uncertainty and photometric uncertainty were added in quadrature to the systematic uncertainty floors of 0.101 dex for \feh, and 0.084 dex for \alphafe\ \citep{kirby20}. 

\section{Spectroscopic abundance gradients in the M31 halo} \label{sec:smooth_halo}
We analyze trends in \feh\ and \alphafe\ as a function of \rproj\ for the smooth halo of M31, as well as stars possibly associated with tidal debris features. Our smooth halo sample consists of 91 individual stars with \feh\ and \alphafe\ measurements satisfying the quality criteria (see \ref{sec:pipeline_process}) and 64 coadds, the latter representing an additional 272 stars. Our sample of stars potentially associated with substructure consists of 69 stars with \feh\ and \alphafe\ measurements, and 53 coadds representing an additional 271 stars. In Section~\ref{sec:feh_gradients}, we explore gradients in \feh\ as a function of projected radius, for the smooth halo as well as stars associated with substructure. We explore gradients in \alphafe\ as a function of projected radius in Section~\ref{sec:alphafe_gradients}. We identify and attempt to address potential sources of bias in measurements of the \feh\ gradient arising from the presence of TiO absorption in our spectra in Section~\ref{sec:tio_bias}. Finally, we look at the \feh-\alphafe\ distribution as a function of projected radius for the smooth halo in Section~\ref{sec:alphafe_feh_smooth_halo}.
The \feh\ and \alphafe\ distributions for individual fields containing known substructure(s) are presented in Section~\ref{sec:halo_substructure}. 

\subsection{\feh\ metallicity gradients}
\label{sec:feh_gradients}

The halo of M31 is known to have a significant radial metallicity gradient \citep{kalirai06, Koch08, gilbert14, escala20a}. 
To measure the spectroscopic metallicity gradient, we assume it follows a linear model parameterized by an angle $\theta$, and the perpendicular distance of the line from the origin $b_{\perp}$. We prefer this parameterization over the standard ($m$,$b$) because it avoids the preference for shallow slopes when flat priors are used \citep{hogg10}. Uncertainties on both \feh\ and \alphafe\ are taken into account when comparing our measured abundances to the model, and we assume flat priors on both $\theta$ and $b_{\perp}$. To fit this simple model, we use emcee, a python implementation of a Markov Chain Monte Carlo Ensemble sampler \citep{foreman-mackey13}, following the same method described by \citet{wojno20}. Using emcee, we generate a distribution of linear models that fit our data, taking the 50th percentile of the marginalized posterior distribution as our measured gradient, and the average of the 16th and 84th percentiles as the quoted uncertainty on the gradient. The measured gradients for all subsamples described in this section are given in Table~\ref{tab:abundance_gradients}.

\subsubsection{Smooth halo stars}
\label{sec:feh_smooth_halo}

Figure~\ref{fig:metallicity_gradient_individ} shows \feh\ as a function of projected radius from the center of M31 for our sample of smooth halo stars with either individual or coadded spectroscopic \feh\ and \alphafe\ abundances.
For the individual and coadded measurements, which cover a radial range \rproj\ of 8-177 kpc, we measure a slope of \gradientfehicsm\ dex kpc$^{-1}$ (intercept at \rproj$=0.0$ of $-0.991\pm0.015$ dex), indicated by the solid magenta line the top left panel in Figure~\ref{fig:metallicity_gradient_individ}. We also show the metallicity gradient fit using only stars with individual \feh\ measurements (black line), where we measure a value of $-0.0048\pm0.0003$ dex kpc$^{-1}$. 

Considering the difference in the distributions between the individual and coadded measurements, the spectra that are coadded typically have low S/N, and/or are more metal-poor and therefore have weaker lines. Therefore, it is reasonable that the metallicity gradient measured from individual stars is shallower because it leaves out a significant number of metal-poor (\feh\ $< -1.5$ dex) stars at more distant projected radii (\rproj\ $> 50$ kpc). We consider the combined individual and coadded abundances to be the most comprehensive sample for measuring the radial metallicity gradient of the smooth halo.

For comparison, we also show the photometric \feh\ for secure M31 smooth halo stars (grey points) in addition to the slope measured by \citet{gilbert14} using photometric \feh\ for a sample of secure M31 smooth halo stars out to a projected radius of 90 kpc. We note that the sample used to determine the photometric \feh\ gradient includes  stars with signs of TiO absorption, and used different criteria to define stars more likely associated with the smooth halo than substructure components. We find that the slope of the gradient measured using both the individual and coadded abundance measurements agrees within $2\sigma$ with the value of the gradient measured from photometric \feh, albeit with an offset on the normalization. If we restrict our sample of individual and coadded measurements to the same radial range as the \citet{gilbert14} measurements (\rproj $< 90$ kpc), the gradient is unchanged. 

In the bottom left panel of Figure~\ref{fig:metallicity_gradient_individ}, we show the inner 35 kpc of our smooth halo sample. In addition to our sample (black points), we show the spectroscopic \feh\ abundances from \citet{escala20b}, where they used deep ($\sim6$ hour, 600ZD grating) observations of fields in the inner halo of M31 to obtain spectroscopic abundances for a sample of individual stars significantly overlapping our own (6 in common out of 10 spectroscopic fields in our sample with \rproj$< 35$ kpc).
\citeauthor{escala20b} found a radial \feh\ gradient of $-0.025\pm0.002$ dex kpc$^{-1}$, shown in the bottom left panel of Figure~\ref{fig:metallicity_gradient_individ} as a dashed blue line. To facilitate a more accurate comparison to the measurements from \citet{escala20b}, we re-measure the radial \feh\ gradient using all individual and coadded smooth halo measurements within \rproj $< 35$ kpc, finding a value of $-0.018\pm0.003$ dex kpc$^{-1}$ (dashed magenta line), consistent within $\sim 2\sigma$. 

\subsubsection{Substructure}
The right-side panels of Figure~\ref{fig:metallicity_gradient_individ} show stars potentially associated with substructure (blue), where individual abundance measurements are indicated with filled points, and coadds are open circles. For individual and coadded \feh\ measurements, we measure a gradient of \gradientfehicss\ dex kpc$^{-1}$, indicated with a solid cyan line.
This gradient is fit to a smaller radial extent compared to the smooth halo, as the most distant substructure identified in SPLASH is at $\sim 90$ kpc. We also measure a gradient of \gradientfehicsms\ dex kpc$^{-1}$ for the combined sample of all M31 stars (both the smooth halo and substructure samples), using both individual and coadded measurements.

In the bottom right panel of Figure~\ref{fig:metallicity_gradient_individ}, we show our sample limited to a radial range of \rproj $< 90$ kpc. The solid blue line is the same fit from the top right panel, for individual and coadded stars in both the smooth halo and substructure samples. We also fit this gradient using a restricted radial range (\rproj $< 90$), finding a value of \gradientfehicsmslimited (dashed blue line), for comparison to the gradient measured using both smooth halo and substructure from \citet{gilbert14} (purple dotted line). Compared to the gradient measured by \citet{gilbert14}, we find that our measured gradient for all stars in this restricted radial range is consistent within $\sim 2\sigma$, with an offset of approximately 0.5 dex with respect to the normalization. 

\subsection{\alphafe\ abundance gradients}
\label{sec:alphafe_gradients}

\begin{figure*}
 	\includegraphics[width=\textwidth]{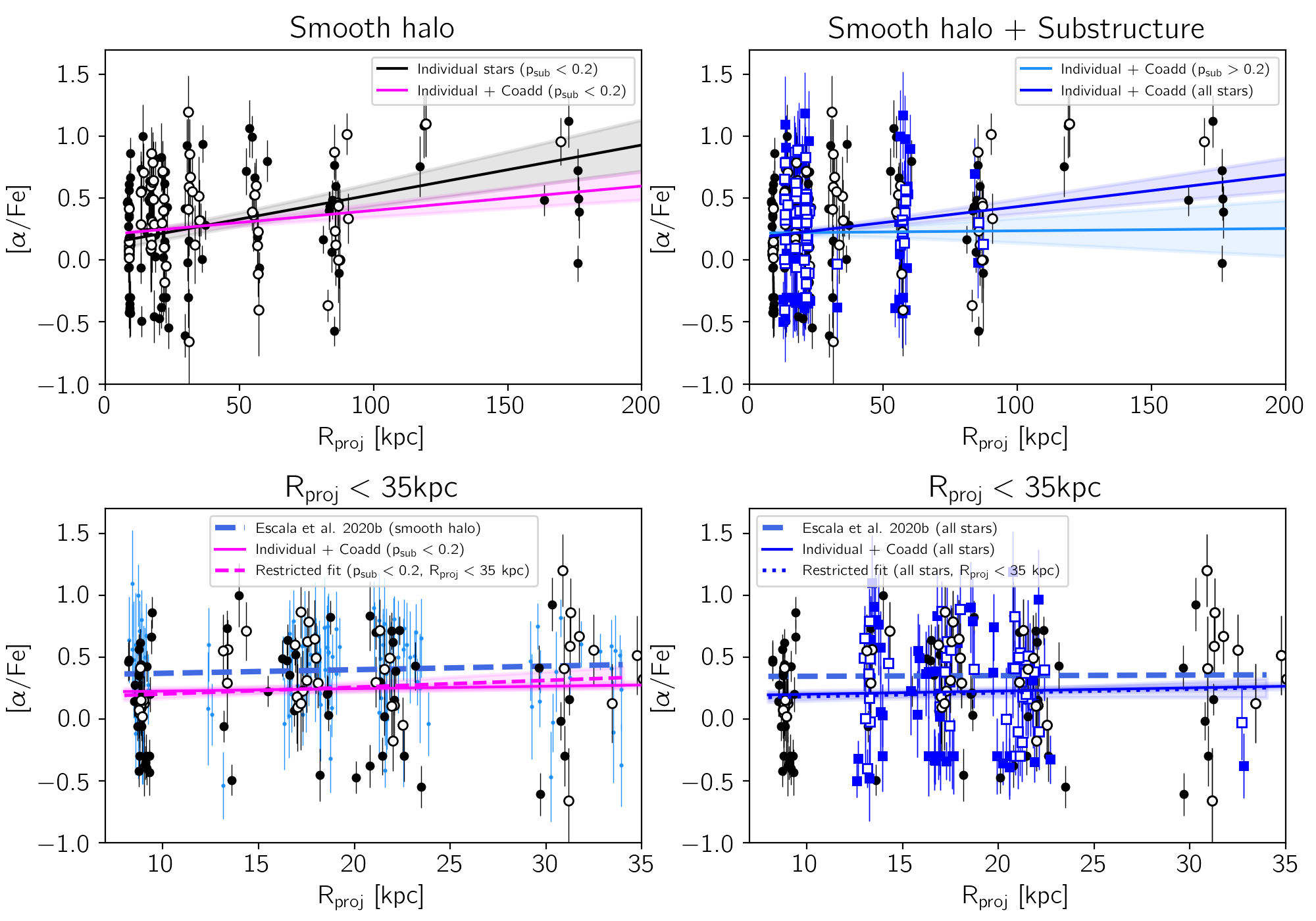}
    \caption{The same as Figure~\ref{fig:metallicity_gradient_individ}, but for the abundance gradient with respect to \alphafe. Top left: The best-fit spectroscopic \alphafe\ gradient fit to individual and coadded abundance measurements is shown as a solid magenta line, with the shaded regions indicating a 1$\sigma$ standard deviation. The gradient fit using only individual measurements is shown as a black line. Bottom left: Zoomed in version of the top panel, now showing spectroscopic measurements from \citet{escala20b} (blue points), with their gradient (blue dashed line). Using the same radial range as in \citet{escala20b}, we find a gradient of \gradientalphaicsmlimited\ dex kpc$^{-1}$ (magenta dashed line). Top right: The best-fit gradient for individual and coadded stars associated with substructure is shown as the light blue line, with the gradient measured for all stars (individual and coadded, smooth halo and substructure) shown in dark blue. Bottom right: Zoomed in version of the top right panel, showing the best-fit gradient of all stars restricted to the same radial range as in \citet{escala20b} (\gradientalphaicsmslimited\ dex kpc$^{-1}$, blue dotted line).}
    \label{fig:alphafe_gradient}
\end{figure*}

In Figure~\ref{fig:alphafe_gradient}, we show \alphafe\ as a function of projected radius, for the smooth halo sample (black points) as well as the sample of stars possibly associated with substructure (blue points). Using individual (filled points) and coadded measurements (open points) for smooth halo stars, we find a best-fit gradient of \gradientalphaicsm\ dex kpc$^{-1}$, shown as a solid magenta line, with the shaded regions indicating $1\sigma$ uncertainties. This is consistent with the gradient measured from individual \alphafe\ measurements only (black line, \gradientalphaism\ dex kpc$^{-1}$). When we consider only stars possibly associated with substructure in the top right panel of Figure~\ref{fig:alphafe_gradient}, we measure a value of \gradientalphaiss\ dex kpc$^{-1}$, consistent with no gradient (cyan line). For the full sample of individual and coadded measurements for all stars (smooth halo and substructure samples), we measure a gradient of \gradientalphaicsms\ dex kpc$^{-1}$ (blue line).

The bottom left panel of Figure~\ref{fig:alphafe_gradient} is analogous to the bottom left panel of Figure~\ref{fig:metallicity_gradient_individ}, where we compare our sample to the sample presented in \citet{escala20b} (light blue points). \citeauthor{escala20b} measure an \alphafe\ gradient of $+0.0029 \pm 0.0027$ dex kpc$^{-1}$ for smooth halo stars $8-35$ kpc in projected radius, shown as the dashed light blue line. This value is consistent with our value for the smooth halo using both individual and coadded abundance measurements (\gradientalphaicsm). If we restrict our smooth halo sample to the same radial range as the \citet{escala20b} sample ($8-35$ kpc), we measure a slightly steeper gradient of \gradientalphaicsmlimited\ dex kpc$^{-1}$ (dashed magenta line). In the bottom right plot, we compare the gradient measured for all individual stars and coadds (solid blue line) and the gradient from \citet{escala20b} to the gradient measured using all stars within 35 kpc ($+0.0030\pm0.0017$ dex kpc$^{-1}$ dex kpc, dashed blue line). The measured gradient for all stars in this restricted radial range falls between the \citet{escala20b} measurement and the gradient found for our full sample (all stars, including individual and coadded measurements), and is consistent within 2$\sigma$ with the latter. 

There are multiple potential sources for the differences between our measured gradients and those measured by \citeauthor{escala20b}.  Measuring reliable \alphafe\ abundances with shallow 1200G spectra is typically more difficult compared to the deep 600ZD spectra used in the \citet{escala20b} sample, due to both the higher S/N of the deep 600ZD spectra, and the expanded spectral range towards the blue ($4100-9000$\AA), which provides significantly more lines for measuring $\alpha$-elements. Our sample has a larger number of stars with \alphafe\ $\sim -0.5$ at small projected radii (\rproj $< 25$kpc), which may serve to ``tilt" the gradient towards the positive direction.  In addition, as the coadd technique increases the ability to recover abundance measurements for metal-poor stars, and there is a general relationship between \alphafe\ and \feh, the addition of measurements from coadded spectra will tend to add more $\alpha$-enhanced stars to the sample of abundance measurements.  This may also result in a more positive measured gradient, as the fraction of metal-poor stars is higher at larger projected radius. In summary, while evidence for a gradient in \alphafe\ as a function of \rproj\ for the inner regions of the M31 halo is tentative at best, we find that there is a significant radial gradient in \alphafe\ over the entire radial range that we consider for this study.

\begin{deluxetable*}{lllll}
\label{tab:abundance_gradients}
\tablewidth{\textwidth}
\tablenum{1}
\tablecaption{Abundance gradients measured for different M31 halo subsamples. The number in parentheses in the N$_{\mathrm{coadds}}$ column refers to the number of stars that went into the coadds.}
\tablehead{\colhead{Halo Component} & \colhead{N$_{\mathrm{stars}}$} & \colhead{N$_{\mathrm{coadds}}$} &\colhead{d\feh/d\rproj (dex kpc$^{-1}$)} & \colhead{d\alphafe/d\rproj (dex kpc$^{-1}$)} }

\startdata
Smooth Halo & 91 & 0 & $-0.0048\pm0.0003$ & \gradientalphaism\\
Smooth Halo & 91 & 64 (272) & $-0.0075\pm0.0003$ & \gradientalphaicsm\\
Substructure & 69 & 0 & $-0.007\pm0.0007$ & \gradientalphaiss\\
Substructure &69 & 53 (271) &$-0.005\pm0.0003$ & \gradientalphaicss\\
Smooth Halo + Substructure & 160 & 117 (543) & $-0.005\pm0.0003$ & \gradientalphaicsms\\ 
\enddata
\end{deluxetable*}

\subsection{Effect of excluding stars with TiO}
\label{sec:tio_bias}
We have removed stars likely to have TiO absorption from our spectral synthesis abundance measurements (Section~\ref{sec:tio_stars}), as the synthetic spectra used in our spectroscopic abundance pipeline do not model TiO absorption, and therefore we cannot reliably measure abundances for stars with significant TiO absorption features. 
Our secure M31 sample contains 1079 stars, and we remove 308 stars ($\sim29\%$) that are either observed or are likely to have TiO absorption features. The removal of stars with TiO absorption features removes predominantly cool, metal-rich stars.  As these TiO stars make up a significant portion of our M31 halo sample, in this section we attempt to quantify any potential bias on the radial metallicity gradient measurement due to the removal of these stars.

To do this, we use the photometric metallicities of all smooth halo stars (including those with likely TiO absorption) to find the average \fehphot\ for a given radial bin, and compare this to the average \fehphot\ of the radial bin after we remove stars with TiO. We define a TiO correction factor, \feh$_{\mathrm{TiO corr}}$ as the difference in the mean \fehphot\ between the two averages for each radial bin. 

\begin{figure}
 	\includegraphics[width=\columnwidth]{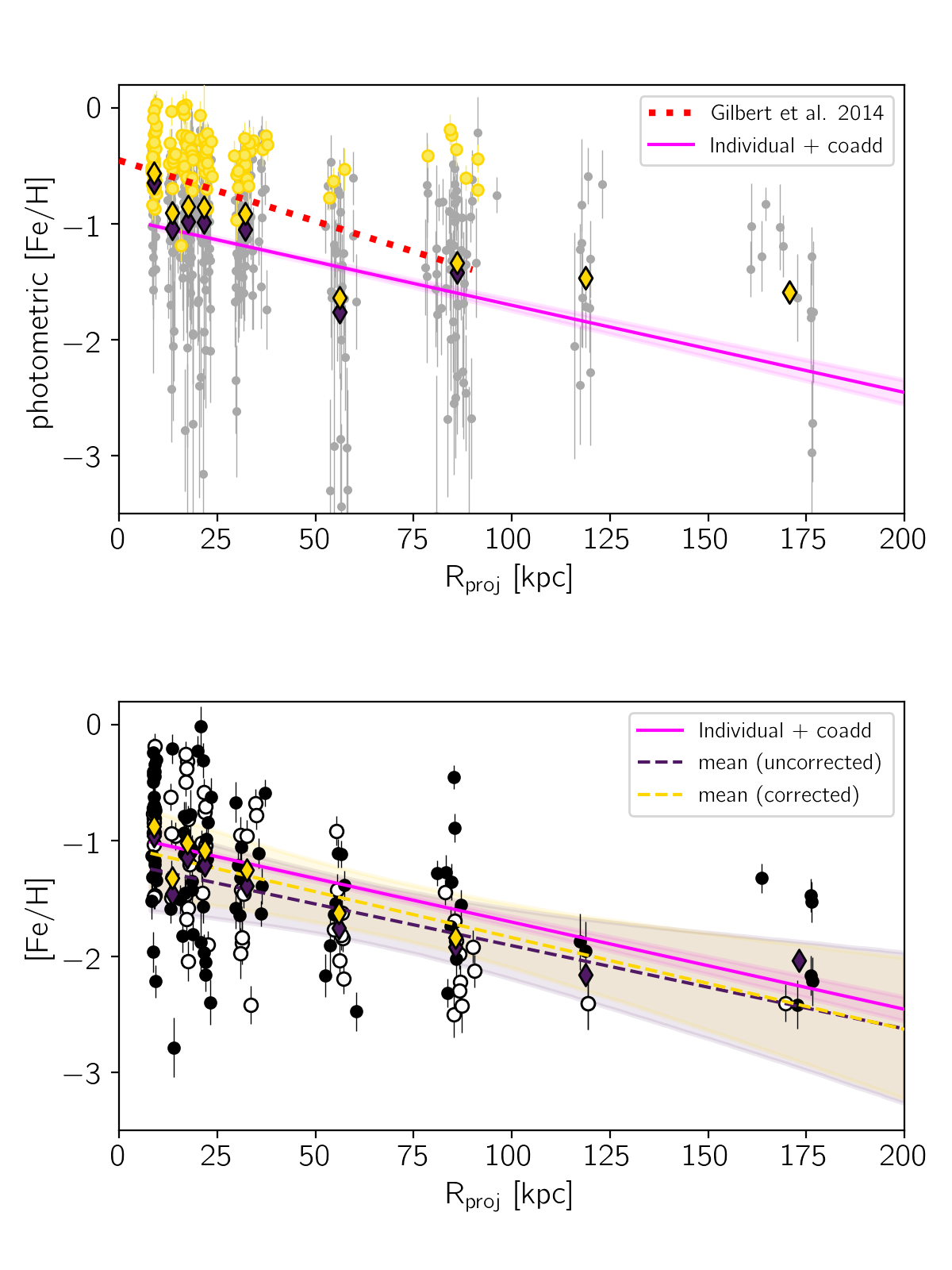}
    \caption{Top: Photometric \feh\ measurements as a function of their projected radius for all secure M31 smooth halo stars (grey). The yellow points indicate stars likely to have TiO in their spectra (Section~\ref{sec:tio_stars}). The outlined diamonds indicate the mean photometric \feh\ for each radial bin, where yellow indicates the mean \feh\ for all stars in that bin, and dark purple is the mean \feh\ with TiO stars removed. The best-fit spectroscopic \feh\ gradient (Section~\ref{sec:feh_gradients}) using our sample of individual and coadded \feh\ measurements in the smooth halo is shown as a solid magenta line, with the shaded regions indicating a 1$\sigma$ standard deviation. The dotted red line is the metallicity gradient measured in \citet{gilbert14}, using photometric \feh. Bottom: Spectroscopic \feh\ as a function of  projected radius for stars with individual spectral synthesis-based \feh\ measurements (black) and coadded \feh\ measurements (open points). We measure the slope from both the uncorrected (dark purple) and corrected (yellow) means, and find no statistically significant difference between the slope before and after accounting for the impact of the removal of TiO stars (Section~\ref{sec:tio_bias}).}
    \label{fig:metallicity_gradient_tio_comparison}
\end{figure}

In Figure~\ref{fig:metallicity_gradient_tio_comparison}, we show the estimated bias in our \feh\ gradient measurement due to removing stars likely to have TiO in their spectra. The top panel shows photometric \feh\ as a function of projected radius for secure M31 smooth halo stars, with smooth halo stars that do not pass our TiO cuts shown in yellow. 
The overplotted diamonds represent the weighted mean photometric metallicity for a given radial bin, for all smooth halo stars (yellow) and for stars unlikely to have TiO in their spectra following the criteria in Section~\ref{sec:tio_stars} (dark purple). The mean \rproj\ is calculated using the position of all points in a given radial bin. 
The TiO correction factor, \feh$_{\mathrm{TiO corr}}$, is defined as the difference between these means, and is of the order of $\sim 0.1$ dex for any given radial bin. This correction factor estimates the shift of the mean metallicity if we were to include stars with TiO absorption in their spectra. The solid magenta line shows the spectroscopic \feh\ gradient measured using both the individual and coadded abundance measurements for the smooth halo sample (Section~\ref{sec:feh_smooth_halo}). The red dotted line is the gradient presented in \citet{gilbert14}, measured using photometric metallicities of stars in the smooth halo of M31 (see their Figure~10). We include this measurement for comparison as their sample selection is similar to our sample selection (including TiO stars; Section~\ref{sec:m31_membership}), although the gradient is fit over a smaller radial range (\rproj\ $< 90$ kpc). Their measured gradient ($-0.0105\pm0.0013$ dex kpc$^{-1}$) is consistent (within 2$\sigma$) with what we measure from both individual and coadded spectroscopic measurements of stars in the smooth halo, with a vertical offset of approximately 0.5 dex.

In the bottom panel of 
Figure~\ref{fig:metallicity_gradient_tio_comparison}, we show the sample of individual smooth halo stars with spectroscopic \feh\ measurements (filled points) and coadded \feh\ measurements (open points). The mean spectroscopic \feh\ for each radial bin is indicated with dark purple diamonds. The yellow diamonds represent the weighted mean \feh\ plus \feh$_{\mathrm{TiO corr}}$ for a given radial bin. We remeasure the radial gradient using only these mean measurements, resulting in the ``uncorrected" (dark purple) and ``corrected" (yellow) fits indicated with dashed lines. The shaded regions indicate $1\sigma$ uncertainties. The slope of the gradient that we measure from the ``corrected'' measurements is consistent ($-0.0079\pm0.0020$) with the gradient measured from the ``uncorrected" means ($-0.0072\pm0.0021$), and is encapsulated within the large uncertainties of the fits. We estimate that the effect on the radial metallicity gradient due to removing TiO stars from our sample produces an offset on the normalization of approximately the order of the TiO correction factor ($\sim 0.1$ dex), and does not significantly affect the measured slope. Since there is significant dispersion in the \alphafe\ values as a function of \feh\ (Section~\ref{sec:alphafe_feh_smooth_halo}), we cannot reasonably estimate the potential bias on the \alphafe\ gradient due to removing stars with TiO.

\subsection{\feh--\alphafe\ distribution as a function of radius}
\label{sec:alphafe_feh_smooth_halo}

\begin{figure*}
 	\includegraphics[width=\textwidth]{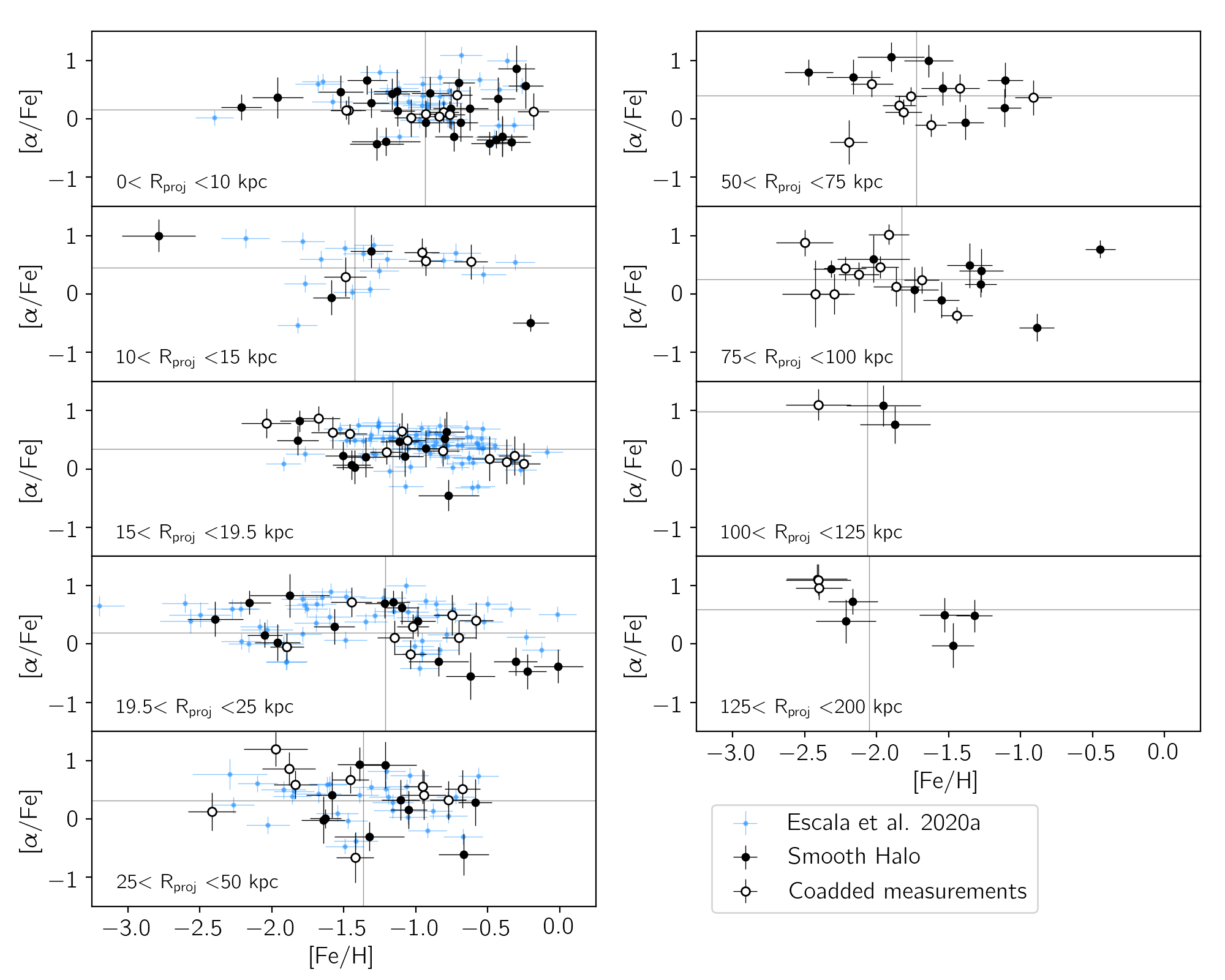}
    \caption{\alphafe--\feh\ as a function of projected radius for our nine radial bins. The corresponding radial range for a given panel is indicated in the bottom left. In each panel we show the sample of smooth halo stars with individual abundance measurements (black points), coadded smooth halo stars (open circles), and the sample from \citet{escala20b} (blue points). The mean \feh\ and \alphafe\ for each radial bin for our sample including both individual and coadded measurements are shown as solid grey lines. We find that the mean \feh\ gradually shifts to lower metallicities as a function of increasing projected radius, and that outermost radial bins have a higher mean \alphafe\ compared to the inner bins.}
    \label{fig:feh_alphafe_smooth_halo}
\end{figure*}

We combine both the \feh\ and \alphafe\ spectroscopic abundance information in Figure~\ref{fig:feh_alphafe_smooth_halo}, where we show the \feh--\alphafe\ distribution as a function of radius for our nine radial bins. We show both the individual (black points) and coadded (open circles) abundance measurements for smooth halo stars. For comparison, we also show the spectroscopic sample from \citet{escala20b} (blue points). 

The mean \feh\ shifts as a function of projected radius, from primarily metal-rich populations in the inner regions, to relatively metal-poor populations in the most distant regions of the smooth halo. Given the aspect ratio of the plots, the \alphafe\ abundance gradient is more difficult to detect in the 2D abundance space by eye, however, there are distinct differences if we compare the more centrally concentrated bins (\rproj $< 20$ kpc) to the outermost bins.

We find that both our individual and coadded abundance measurements follow the abundance distribution of the \citet{escala20b} sample, for the regions where our samples overlap. The spread of the distribution varies significantly for each radial bin. The outer most bins (\rproj $> 50$kpc) have slightly smaller dispersions in \feh\ and \alphafe\, but these bins are more susceptible to small number statistics, and therefore we do not detect any correlation between dispersion in \feh\ or \alphafe\ with increasing projected distance. 
For bins larger than 50 kpc, we present some of the first spectroscopic abundance measurements for stars at large projected radii. Prior to this work, only 9 stars in M31's outer halo ($43 < $ \rproj\ $< 165$ kpc) had spectroscopic \feh\ and \alphafe\ measurements \citep{vargas14,gilbert20}. We measure spectroscopic \feh\ and \alphafe\ abundances for 45 halo stars with \rproj $> 50$ kpc, providing an increase of 600\% in outer halo stars with \alphafe\ and \feh\ measurements. In addition, we have 33 coadded measurements of outer halo stars, representing the averaged abundances of an additional 174 individual stars. 

\section{Substructure}
\label{sec:halo_substructure}
In this section we explore the chemical abundances of stars that may belong to various substructures in M31's halo. The analysis was done on a field-by-field basis, where multiple masks/pointings were merged for the final sample for that field \citep[e.g.,][]{gilbert12}. 
To obtain coadded measurements for stars likely associated with substructure, they are first grouped by their observed field. For example, a0\_1, a0\_2, and a0\_3 are three masks observed in the same GSS field, and thus we consider candidates for coaddition from those three masks combined. As for the smooth halo sample, we sort stars according to their photometric \feh\ to obtain groupings of approximately five stars per coadd. For fields that contain multiple substructures/debris features, we coadd stars according to their line-of-sight velocity measurements as well as photometric \feh\ (see Figures~\ref{fig:substr_comparison_gss}-\ref{fig:substr_comparison_seshelf}).

As discussed in Section~\ref{sec:substructure_probability}, we intentionally chose a division between smooth halo and substructure samples ($p_{\rm sub} = 0.2$) that minimizes contamination by substructure in the smooth halo sample.  While it is not possible to isolate a pure substructure sample from line-of-sight velocities alone, the choice of $p_{\rm sub} \ge 0.2$ for the substructure sample does increase the fraction of stars associated with the smooth halo included in the substructure sample compared to using a higher $p_{\rm sub}$ value.  
It has been shown in previous works \citep{escala20b} that the abundance distributions of the smooth halo and substructure components overlap considerably in M31's inner halo. The inclusion of a higher fraction of smooth halo stars in the substructure sample will tend to increase the apparent overlap in chemical space, and thus the true differences between smooth halo and substructure can be expected to be larger than seen here. 

In this work, our fields target or overlap the following named substructures in the halo of M31: the Giant Stellar Stream (GSS) \citep{ibata01}, a Kinematically Cold Component (KCC) associated with the GSS \citep{gilbert09, gilbert19a}, Stream C \citep{ibata07}, and the Southeast Shelf \citep{gilbert07}. We also only have coadded measurements for fields f135 and A220, located at 17 and 85 kpc in \rproj, respectively. These fields have previously been identified as having substructure \citep{gilbert12}, however, they are not associated with the named substructures listed above. The locations of these fields with substructure are shown in Figure~\ref{fig:m31_roadmap}, overplotted on the stellar density map of M31 from PAndAS \citep{mcconnachie18} to illustrate the position and extent of the relevant substructure(s). 

\subsection{\feh--\alphafe\ distribution as a function of radius}
\begin{figure*}
 	\includegraphics[width=\textwidth]{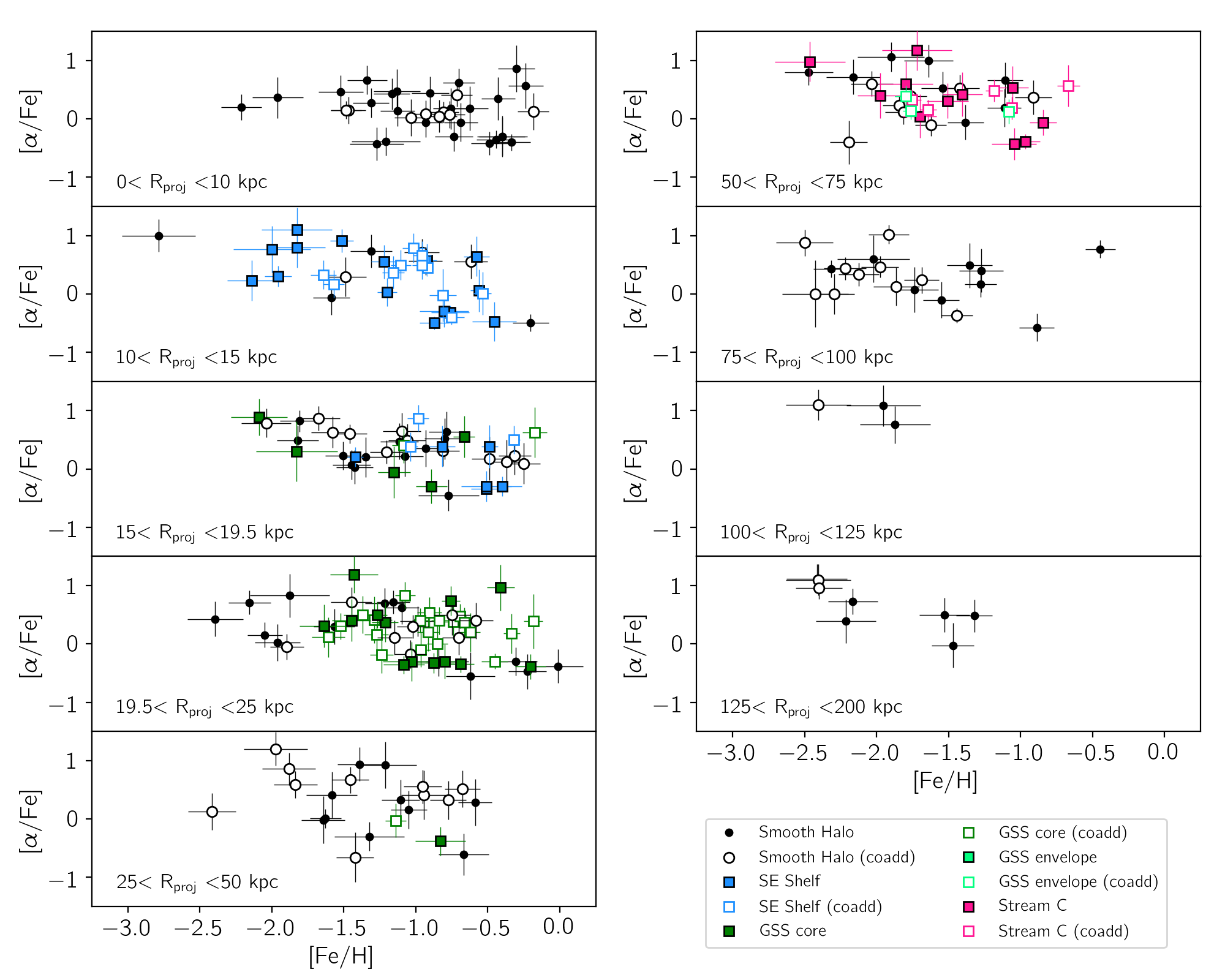}
    \caption{The same as Figure~\ref{fig:feh_alphafe_smooth_halo}, but including the individual and coadded abundance measurements for stars possibly associated with substructure (Section~\ref{sec:halo_substructure}). The individual and coadded measurements for the smooth halo (black) are included for comparison. Individual stars potentially associated with substructure are shown color-coded by their respective substructure component (blue: SE shelf, green: GSS core, light green: GSS envelope, magenta: stream C), where the coadded measurements for a given substructure are indicated in the same color as open points. There are no substructure components detected in the sparsely populated outer halo fields (\rproj $ > 90 $kpc). 
    We find that the abundance distributions of stars in the substructure sample generally vary similarly as stars in the smooth halo sample, for any given \rproj\ bin.} 
    \label{fig:feh_alphafe_substructure}
\end{figure*}

In Figure~\ref{fig:feh_alphafe_substructure}, we show the  distribution of stars in \feh--\alphafe\ space for both our smooth halo and substructure samples. We also show coadded abundance measurements for both the smooth halo and each of the substructure components in our sample. The substructure coadds are color-coded according to the substructure with which they are associated.  We use the same colors for each substructure/field for the rest of this section. 

In a number of radial bins, we find that the distributions of stars possibly associated with substructure components vary similarly as compared to the distribution of smooth halo stars, in the sense that stars at small \rproj\ are on average more metal-rich than stars at large \rproj. 
In addition, we find that stars potentially associated with any substructure in the M31 halo are on average more metal-rich (higher \feh), and have lower \alphafe\ (Figures~\ref{fig:substr_comparison_gss}-\ref{fig:substr_comparison_seshelf}), compared the smooth halo stars in the same radial bin. 
However, overall the substructure components also tend to follow the same radial \feh\ gradient as the smooth halo. Indeed, we measure nearly the same halo \feh\ gradient if we include all substructure, compared to the \feh\ gradient measured for the smooth halo only. 
Some of these substructures (GSS, SE shelf) likely originate from the same merger event \citep{escala21}, and we find that qualitatively, the distributions of these two components are more similar to each other than the underlying M31 smooth halo. In the following sections, we explore the abundance distributions for each of the substructure components separately. 

\subsection{Giant Stellar Stream (GSS)}
\label{sec:substructure_gss}
The GSS is immediately identifiable in stellar density maps of the halo of M31 \citep{ibata01} as a radial structure extending approximately 6\degree\ ($\sim80$~kpc) to the south of M31's center, to the west of the southern minor axis. The core of the GSS is metal-rich and has a high surface brightness, while photometric studies have shown the envelope is relatively more metal-poor and has a lower surface brightness \citep{ibata01,mcconnachie03,mcconnachie18}. The stellar density sharply declines along the eastern edge of the GSS, and decreases gradually across the western envelope.  

Our observations target fields on the core (a3, f207, H13s; Section~\ref{sec:substructure_gss_core}) as well as the envelope of the GSS (a13; Section~\ref{sec:substructure_gss_envelope}). The mean velocity of stream members varies as a function of projected radius, from $-524.9$~\kms\ at 17~kpc to $-301.6$~\kms\ at 58~kpc, and for this work we adopt the values for each kinematic component as a function of radius compiled in Table 4 of \citet{gilbert18}. 
%mention the KCC 

\subsubsection{GSS core}
\label{sec:substructure_gss_core}
\begin{figure*}
 	\includegraphics[width=\textwidth]{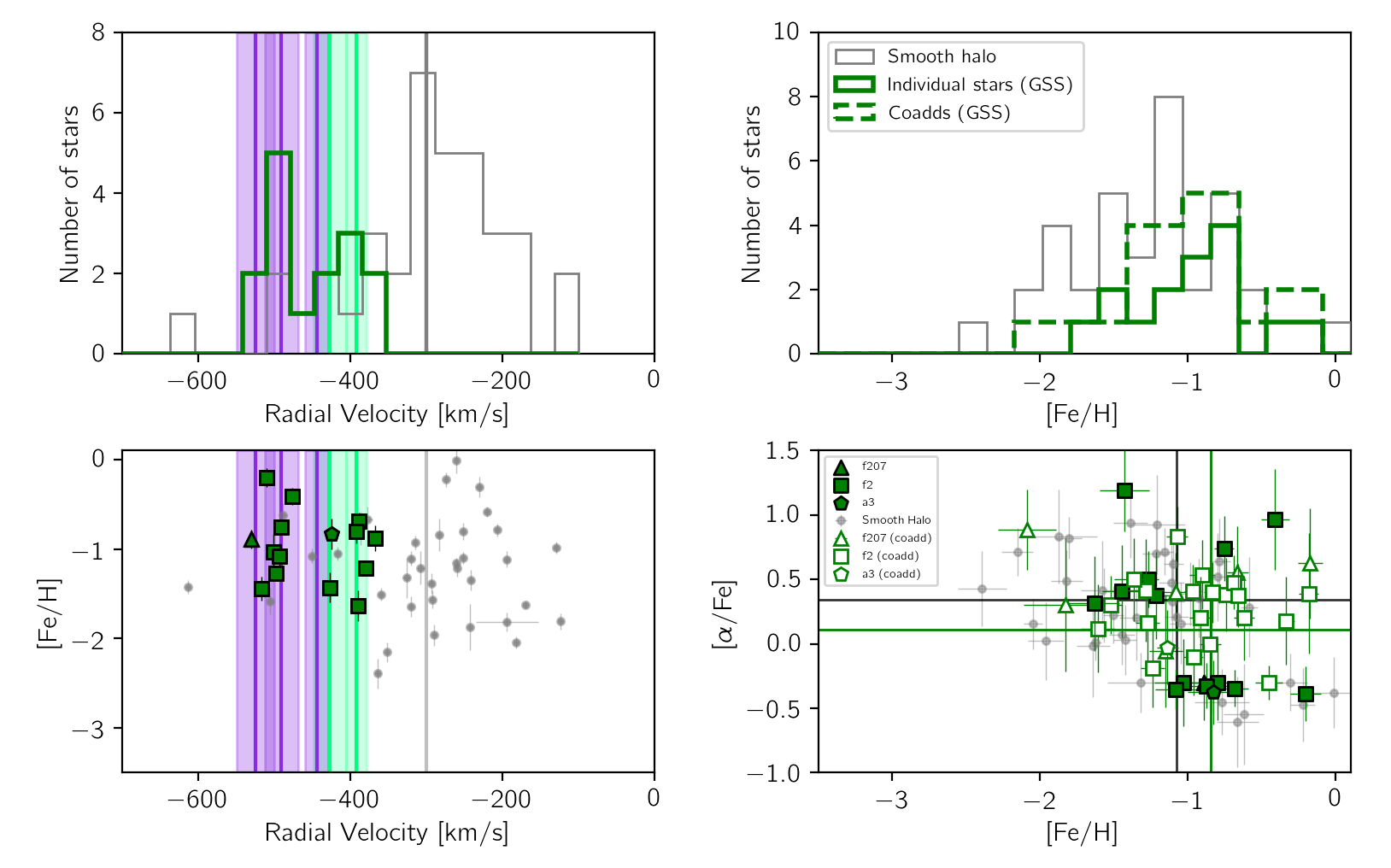}
    \caption{Top left: Histogram of the radial velocities for smooth halo stars with secure abundances in the radial range $15 <$ \rproj $< 50$ kpc (grey) and stars associated with substructure in fields on the GSS core (f207, f2, a3; green). The vertical lines represent the mean line-of-sight velocity for the GSS (purple), 
    and the secondary kinematically cold component (light green, Section~\ref{sec:substructure_gss}) for each field shown \citep[][see their Table 4]{gilbert18}. The shaded regions indicate the corresponding velocity dispersion. 
    The systemic velocity ($-300.0$ \kms) of M31 is in grey. Top right: histograms of spectroscopic \feh\ for the smooth halo (grey), individual (15 stars; solid green) and coadded stars potentially associated with substructure in these fields (27 coadds representing 136 stars; dashed green). Bottom left: Spectroscopic \feh\ as a function of line-of-sight velocity for substructure stars (outlined green symbols) and smooth halo stars (grey). We show the same mean and dispersion velocity for the GSS and KCC as in the top left panel. 
    Bottom right: \feh\--\alphafe\ measurements for individual substructure stars (open green symbols), as well as coadded groupings of stars (filled green symbols). 
    We also show the \feh\--\alphafe\ distribution of smooth halo stars in grey. Solid lines show the mean \feh\ and \alphafe\ for the entire substructure sample (green), and the smooth halo (grey).}
    \label{fig:substr_comparison_gss}
\end{figure*}

In Figure~\ref{fig:substr_comparison_gss}, we show the (a) line-of-sight velocity histogram, (b) spectroscopic \feh\ histogram, (c) line-of-sight velocity vs.\ \feh, and (d) \feh\ vs.\ \alphafe\ for stars likely belonging to substructure (outlined points), and those belonging to our M31 smooth halo sample (grey). Field f207 is 17 kpc in projected radius from the center of M31. \citet{gilbert19a} provides a thorough overview of this field, where they use deeper spectroscopic observations to investigate the three kinematical components in the field: the GSS, a kinematically cold component (KCC), and a kinematically hot halo component. Field H13s is located at 21 kpc from the center of M31, roughly southeast along the minor axis, and spatially overlaps with the deeper spectroscopic observations of field S, presented in \citet{escala20a}. This field contains substructure associated with both the GSS and KCC. Field a3 is located at 33 kpc in projected radius, and is the most distant field along the core of the GSS. Deep ($\sim$ 6 hour) spectroscopic observations of this field were published in \citet{escala20b}, and the substructure component was subsequently analyzed in \citet{escala21}. For these three fields, we obtain abundance measurements for 15 individual stars, and 27 coadds, representing an additional 136 stars.

\begin{figure*}
 	\includegraphics[width=\textwidth]{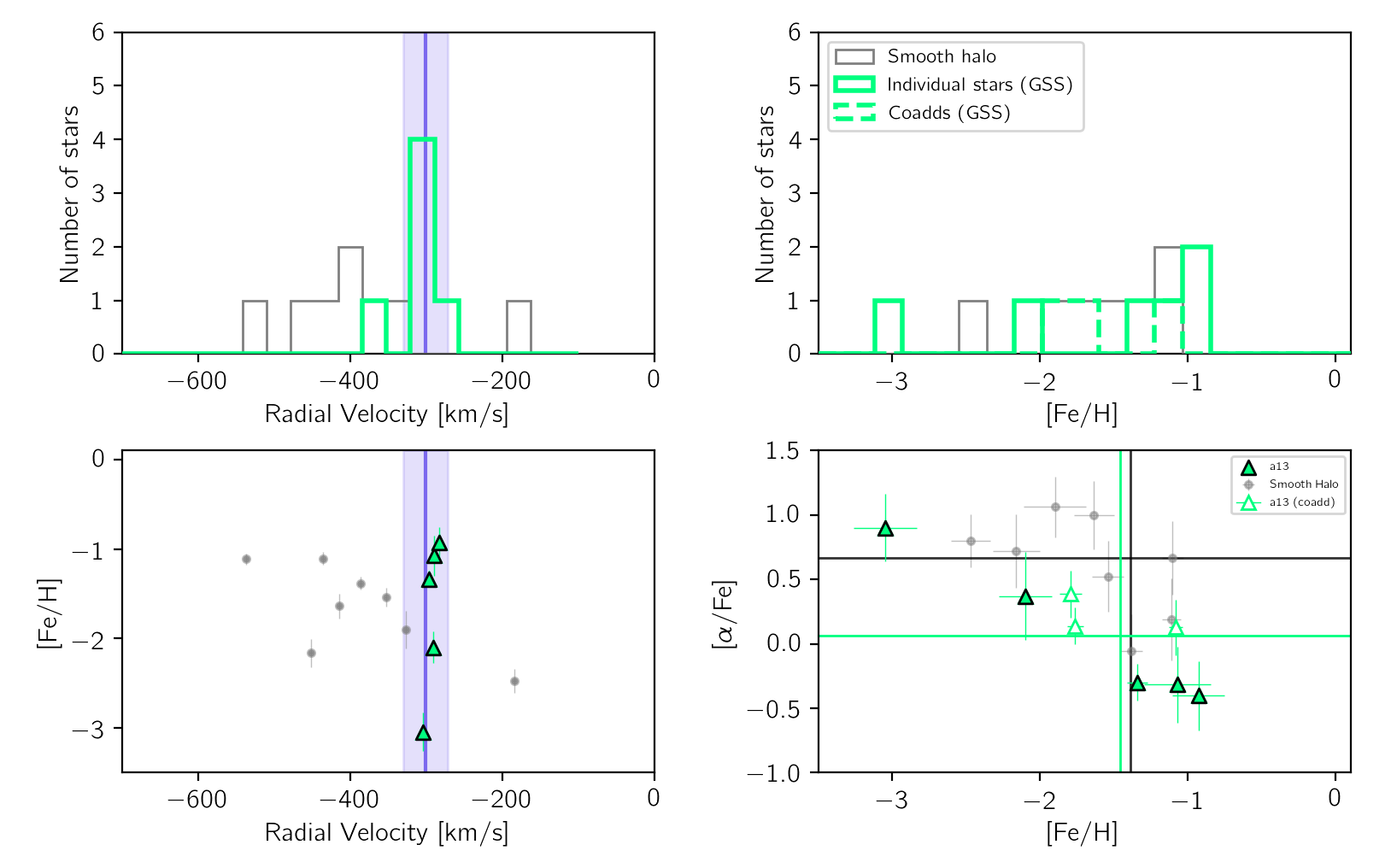}
    \caption{The same as Figure~\ref{fig:substr_comparison_gss}, but for field a13, located on the envelope of the GSS. Stars likely to belong to the GSS envelope are indicated with light green triangles, and smooth halo stars from the same radial bin as field a13 ($50 <$\rproj$<75$~kpc) are indicated with grey points. In the left side panels, we show the mean velocity $-301.6$ \kms\ ($\sigma = 29.2$ \kms) \citep{gilbert18} of the GSS envelope at the location of this field (purple line), where the shaded region indicates the velocity dispersion. At a distance of 58 kpc, the mean \feh\ of this field (\meanfehGSSenv) agrees with the smooth halo (\meanfehGSSenvsmooth), however, we find there to be a difference in the mean \alphafe. The mean \alphafe\ of this field (\meanalphafeGSSenv) is significantly lower than that of the smooth halo (\meanalphafeGSSenvsmooth), but is similar to the mean \alphafe\ of the GSS core fields (\meanalphafeGSS, Figure~\ref{fig:substr_comparison_gss}).}
    \label{fig:substr_comparison_gss_envelope}
\end{figure*}

We find that the \feh\--\alphafe\ distribution of the substructure sample from both individual and coadded spectroscopic measurements (Figure \ref{fig:substr_comparison_gss}) is similar, but slightly more metal-rich, compared to the smooth halo of M31 in the same radial range. We measure a mean \feh\ of \meanfehGSSsmooth\ ($\sigma$\feh$= 0.49$) and a mean \alphafe\ of \meanalphafeGSSsmooth\ ($\sigma$\alphafe$= 0.39$) for the smooth halo in the radial range spanned by these fields ($10 < $\rproj$< 50$kpc). For all individual and coadded stars potentially associated with the GSS, we find a mean \feh\ of \meanfehGSS\ ($\sigma$\feh$= 0.35$) and a mean \alphafe\ of \meanalphafeGSS\ ($\sigma$\alphafe$= 0.43$).

This value differs slightly from the mean \feh\ reported in \citet{escala21} of $-1.03\pm0.07$, where they used a sample of stars that overlaps the sample presented in this work, using deeper ($\sim 6$ hour) observations made with the 600ZD grating on DEIMOS. This grating allows for expanded wavelength coverage at the blue end at the expense of lower resolution. Additional validation of the agreement between stars observed with 600ZD and 1200G grating has been presented in \citet{wojno20} and \citet{escala20a}. However, we note that \citet{escala21} applies a different criterion to obtain their sample of stars likely associated with substructure (\psub $> 0.5$), while our sample of stars possibly associated with substructure is more likely to be contaminated by smooth halo stars (\psub $\ge 0.2$).

\subsubsection{GSS envelope}
\label{sec:substructure_gss_envelope}

Field a13 is located at approximately 58 kpc in projected distance from the center of M31, on the western side of the envelope of the GSS. Photometric studies have found the envelope to have a lower surface brightness and metallicity compared to the GSS core \citep{ibata07}. Using SPLASH data to identify GSS stars, \citet{gilbert09} found a photometric metallicity difference between the GSS core (fields f207, H13s, and a3) and envelope (a13) of $\sim$0.7 dex. 
\citet{escala21} used an overlapping sample of M31 RGB stars observed with the 600ZD grating on DEIMOS to explore abundance gradients in the GSS. They measure a difference in the spectroscopic \feh\ of $\sim0.1$ dex between the core and envelope of the GSS. For field a13, we obtain abundance measurements for 5 individual stars, and 3 coadds representing an additional 15 stars. 

\begin{figure*}
 	\includegraphics[width=\textwidth]{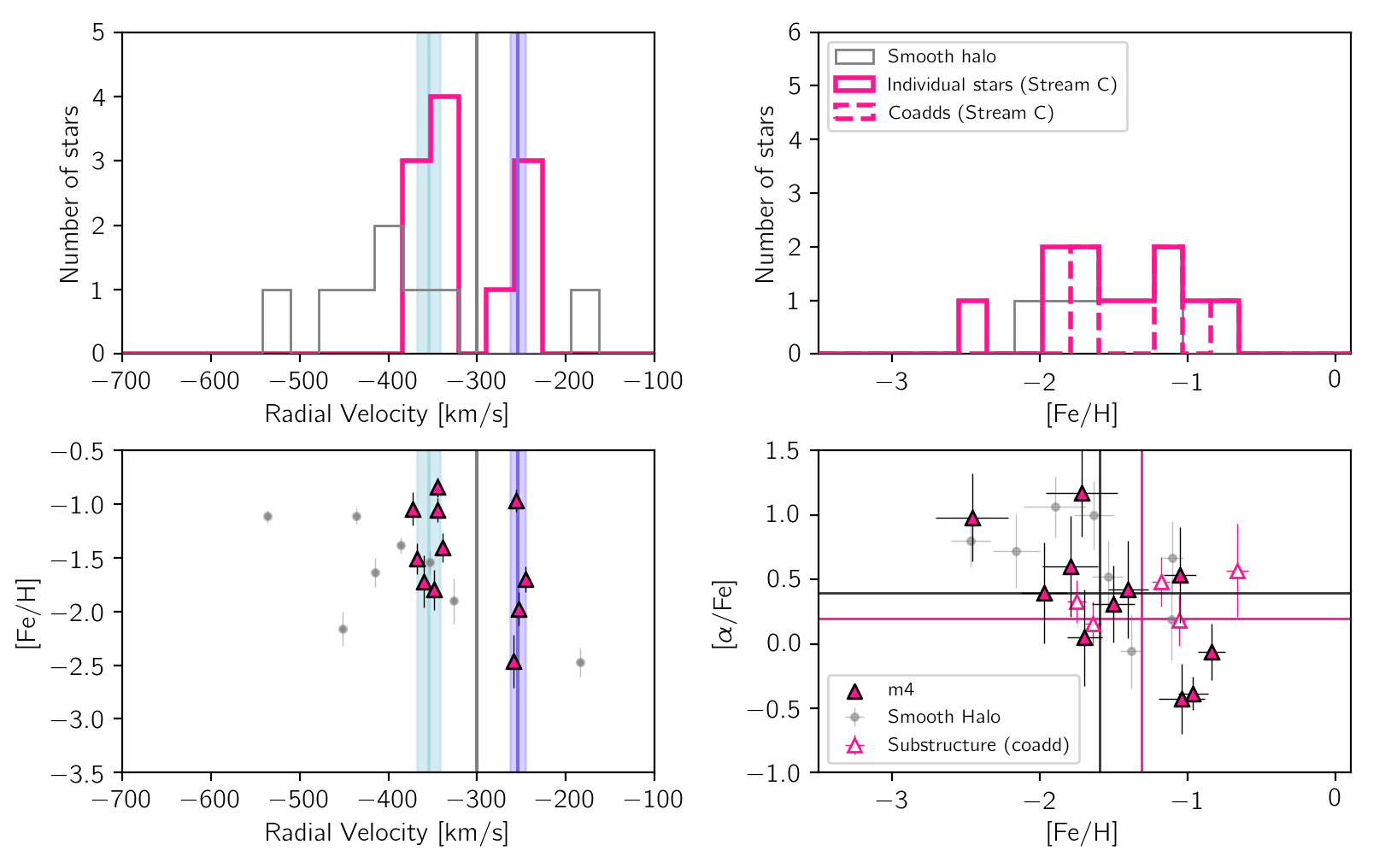}
    \caption{The same as Figures~\ref{fig:substr_comparison_gss} and \ref{fig:substr_comparison_gss_envelope}, but for the sample of stars possibly associated with Stream C (field m4) and smooth halo stars from the same radial bin as the m4 field ($50 <$\rproj$<75$~kpc). The systemic velocity of the M31 halo is indicated with a solid grey line in the left column panels, and the mean velocities of the kinematically cold substructures in this field \citep{gilbert18} are shown as blue and purple vertical lines, where the shaded region indicates the velocity dispersion.}
    \label{fig:substr_comparison_streamc}
\end{figure*}

Figure~\ref{fig:substr_comparison_gss_envelope} shows stars in a13 possibly associated with substructure (light green) compared to the smooth halo sample from the same radial bin as field a13 (grey; $50 <$ \rproj\ $<75$~kpc). We find that stars in the GSS envelope and smooth halo samples have a similar mean \feh\, but the mean \alphafe\ of the GSS envelope sample is significantly lower ($\sim 0.6$ dex). For the GSS envelope sample, we measure a mean \feh\ of \meanfehGSSenv, (\sigmafeh\ $ = 0.39$), and a mean \alphafe\ of \meanalphafeGSSenv\ (\sigmafeh\ $ = 0.36$). In comparison, for the smooth halo sample in the same radial range we measure a mean \feh\ of \meanfehGSSenvsmooth, (\sigmafeh\ $ = 0.39$), and a mean \alphafe\ of \meanalphafeGSSenvsmooth\ (\sigmafeh\ $ = 0.34$).
The mean \alphafe\ (\meanalphafeGSSenv) for this field agrees more closely with the GSS core fields (\meanalphafeGSS), despite a difference of approximately 40 kpc in projected radius. We note that the \meanalphafe\ that we measure for the GSS core and the GSS envelope samples are significantly lower compared to the values presented in \citet{escala21}, using deep 600ZD observations of a similar set of fields.  This may be due, at least in part, to the differing $p_{\rm sub}$ thresholds used in the two analyses.

\begin{figure*}
 	\includegraphics[width=\textwidth]{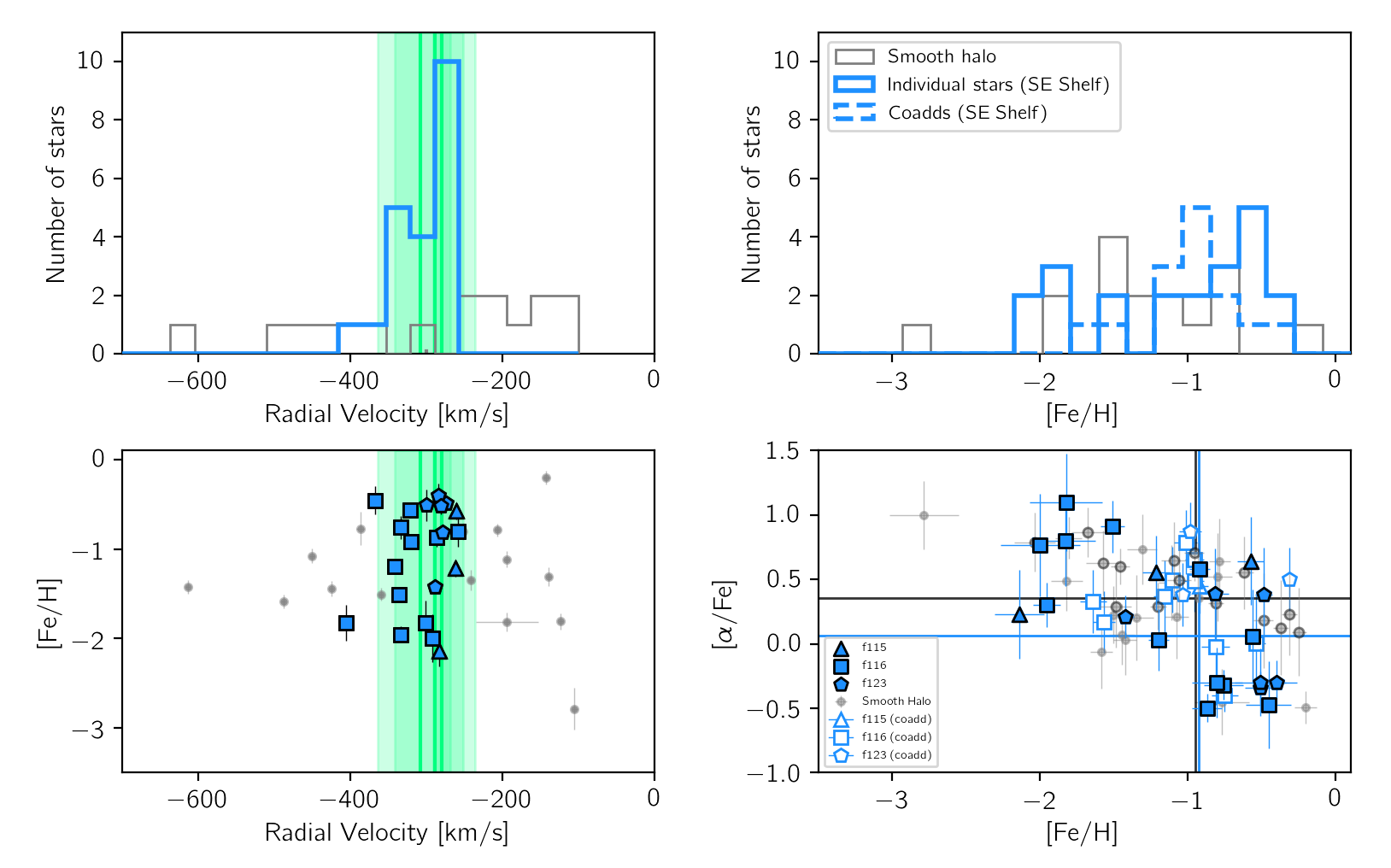}
    \caption{The same as Figure~\ref{fig:substr_comparison_gss}, but for stars possibly associated with the SE Stream (fields f115, f116, and f123).}
    \label{fig:substr_comparison_seshelf}
\end{figure*}

\subsection{Stream C}

Field m4 is located on the minor axis of M31. This field is known to contain multiple kinematical components, and spatially overlaps `Stream C', a debris stream perpendicular to the GSS. This stream was first detected using deep photometric data \citep{ibata07}, and was targeted for spectroscopic follow-up \citep{chapman08}. \citeauthor{chapman08} find that the stream is composed of two distinct kinematic structures that overlap spatially: a primary metal-rich (\feh\ $\sim -0.6$ dex) component with a systemic velocity of $-349$ \kms, and a secondary metal-poor (\feh $\sim -1.3$ dex) component with a systemic velocity of $-286$ \kms. Both components were determined to have similar velocity dispersion, $\sigma_ v \sim 5$ km s$^{-1}$. These properties were confirmed by \citet{gilbert09} using the m4 field, located at a slightly different location on Stream C than the \citet{chapman08} spectroscopic sample. 

Our sample of stars in m4 have a mean projected radius of 58 kpc. We obtain abundance measurements for 11 individual stars, and 5 coadds, representing an additional 26 stars.
Figure~\ref{fig:substr_comparison_streamc} shows the line-of-sight velocity histogram, metallicity (\feh) histogram, \feh\ vs.\ line-of-sight velocity distribution, and \feh-\alphafe\ distribution for stars possibly associated with substructure in m4, as well as stars in the smooth halo sample in the same radial bin as field m4 ($50 <$\rproj$<75$~kpc). Stars in this field have a bi-modal velocity distribution, with peaks around $-350$ and $-250$ \kms\  \citep{gilbert18}. The individual spectroscopic metallicities of stars potentially associated with the two kinematically cold components follow the same pattern found in earlier work: the more negative peak at $-354.8$~\kms\ is on average more metal-rich than the peak at $-254.1$~\kms. In this work, we measure a mean \feh\ of $-1.18\pm0.02$ for stars associated with the more negative peak in the velocity distribution, and a mean \feh\ of $-1.49\pm0.03$ for stars associated with the peak at $-254.1$~\kms. Compared to the smooth halo, we find that the stars possibly associated with stream C are slightly more metal-rich (\meanfeh $=$\meanfehStC, \sigmafeh $ = 0.36$) compared to the smooth halo (\meanfeh $=$\meanfehStCsmooth,  \sigmafeh $ = 0.36$), and have a lower mean \alphafe\  (\meanalphafeStC, \sigmaalphafe $=0.38$) compared to the smooth halo (\meanalphafe $ = $\meanalphafeStCsmooth, \sigmaalphafe $=0.34$). 

\subsection{SE Shelf}
The southeast (SE) shelf substructure was first presented in \citet{gilbert07}, where they describe it as a kinematically distinct cold component of the velocity distribution. Using photometrically derived metallicity estimates, they find that the stars likely to belong to this substructure are more metal-rich than smooth halo stars at the same radial range. The presence of this shelf feature has been predicted by N-body models for the GSS progenitor \citep{fardal07}, where the SE shelf is a shell formed by the fourth pericentric passage of tidal debris from the GSS progenitor \citep{fardal06,fardal07}.

\citet{escala20a}, using deep 600ZD spectroscopic observations of M31 inner halo stars (\rproj$\sim 12$ kpc), compared the abundances of stars associated with the GSS to those in the SE shelf. They found that the SE shelf was $\gtrsim 0.1$ dex more metal poor than the GSS at the same range of projected radii. They indicate that this may be evidence for the presence of a radial metallicity gradient in the GSS progenitor, if the GSS and SE shelf are indeed related.
\citet{escala21} added another field to this analysis, at a larger distance in projected radius (\rproj$\sim 18$ kpc) to be able to reliably disentangle the substructure component from the background halo population via line-of-sight velocity. With deep 600ZD and 1200G spectroscopy, they compared the spectroscopic metallicity and abundance distributions between the GSS and the SE shelf with this expanded sample, finding that the mean measured abundances agree within 1$\sigma$. They find that both distributions are consistent with being drawn from the same distribution, further indicating they originate from the same merger event. 

Our sample of stars possibly associated with this substructure is drawn from fields f115, f116, and f123. These fields are located at 13, 15, and 17 kpc, respectively, in projected distance from the center of M31. We obtain abundance measurements for 21 individual stars and 14 coadds, representing an additional 72 stars. In Figure~\ref{fig:substr_comparison_seshelf}, we show the comparison between our SE shelf sample (blue symbols), and the smooth halo sample in the same radial range as the SE shelf fields ($10 <$ \rproj $<19.5$ kpc, grey). We find that stars that are potentially associated with the SE shelf substructure have a very similar mean \feh\ ($-0.93\pm0.01$, \sigmafeh\ $ = 0.41$) compared to the smooth halo (\meanfeh$ = -0.94\pm0.01$, \sigmafeh\ $ = 0.49$). In addition, we measure a mean \alphafe\ for our SE shelf sample of $0.06\pm0.02$ (\sigmaalphafe $=0.46$), which is lower than the mean \alphafe\ for the smooth halo (\meanalphafe $= 0.36\pm0.02$, \sigmaalphafe$=0.42$). Compared to the GSS core and envelope fields, the SE shelf has a mean \feh\ more similar to the GSS core, but the mean \alphafe\ is consistent within 2$\sigma$ to both the GSS core and GSS envelope fields. 

\section{Discussion}\label{sec:discussion}
In this work, we find abundance gradients in both \feh\ and \alphafe\ as a function of projected radius from the center of M31. These gradients are measured using spectroscopic abundance measurements for individual M31 RGB stars with high quality spectra, as well as coadded abundance measurements for groups of low S/N stars. 
In addition, we also explore the  \feh--\alphafe\ abundance distribution as a function of projected radius, for both the smooth halo, as well as multiple tidal debris components present in the M31 halo sample. 

\subsection{The outer halo of M31} 
\begin{figure}
 	\includegraphics[width=\columnwidth]{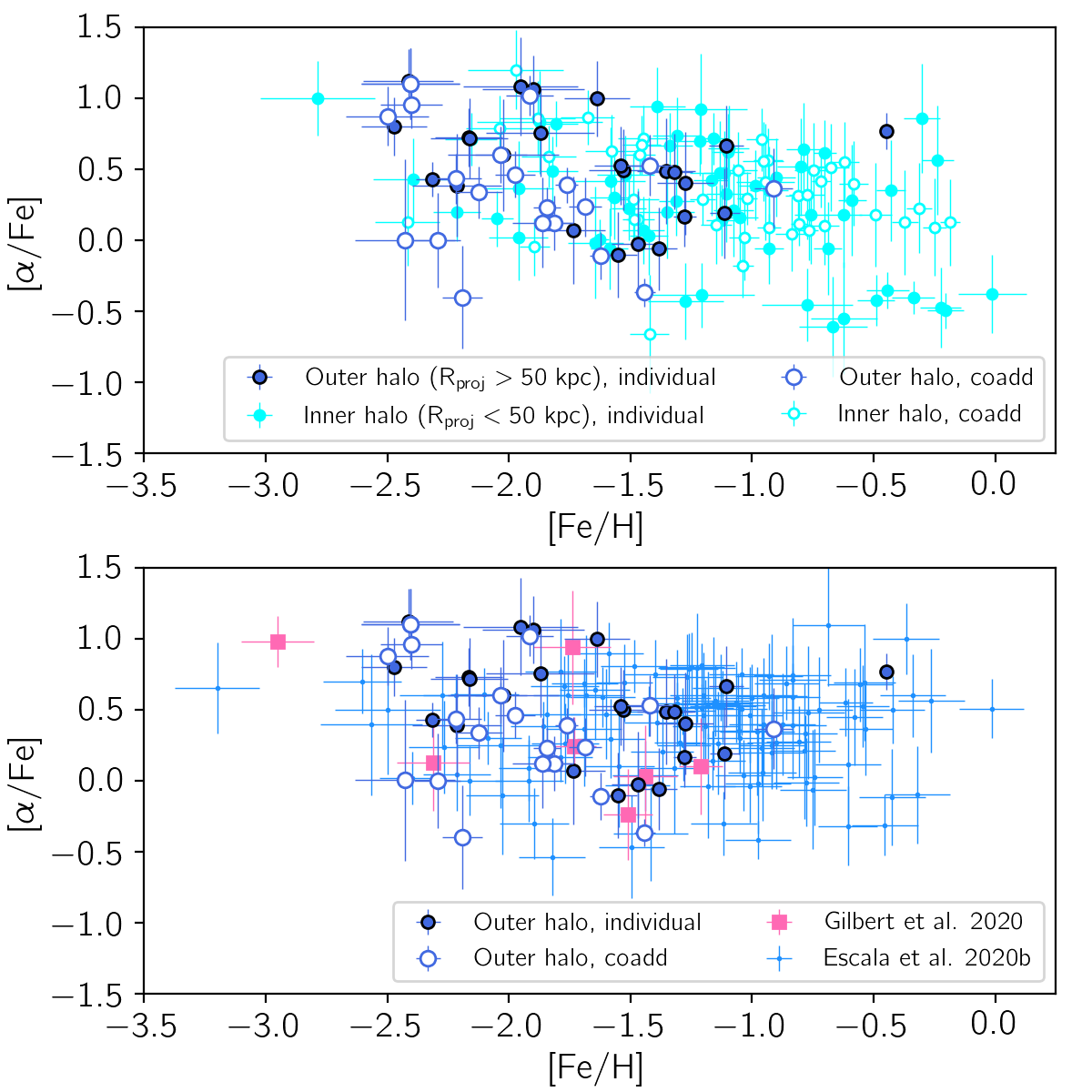}
    \caption{Top: Distribution of our outer halo (\rproj$> 50 $kpc; dark blue) and inner halo (\rproj$< 50 $kpc; cyan)  individual (filled) and coadded (open) abundance measurements. Bottom: Our outer halo sample (dark blue) overplotted on the \feh--\alphafe\ distributions from \citet{gilbert20} (outer halo, pink), and \citet{escala20b} (inner halo, light blue). The abundances of our outer halo sample are consistent with those from the sample of outer halo stars from \citet{gilbert20}, and the abundances of our inner halo sample are consistent with those from the deep spectroscopic data presented in \citet{escala20b}. While the abundances of the outer and inner halo samples overlap significantly, we find that stars in the outer halo are almost exclusively more metal-poor than \feh\ $\sim -1.0$.} 
    \label{fig:outer_halo_comparison}
\end{figure}

It is well-established that halos of massive galaxies form primarily through hierarchical accretion, i.e., are built up through the tidal disruption of less massive satellite galaxies \citep[e.g.,][]{Searle78,Bullock05,font11,conroy19}. In the halo of M31, the GSS provides the most significant evidence of a recent/ongoing accretion event, where the satellite debris is not yet phase mixed, and is therefore visible as streams and shells \citep{ibata01,ferguson02,fardal07,fardal08}. However, a significant population of stars in the outer halo of M31 are \textit{not} associated with known substructure, which in this work we call the ``smooth halo". Of particular interest are the smooth halo stars found at large projected radii from the center of M31 (\rproj\ $> 50$ kpc). In this section, we will attempt to provide constraints on the origin of these outer smooth halo stars.

In Figure~\ref{fig:outer_halo_comparison}, we show the \feh--\alphafe\ abundance distribution for all individual and coadded abundance measurements of the smooth halo sample. In the top panel, outer (\rproj\ $> 50$ kpc) halo stars are indicated with dark blue points, and inner halo stars are indicated with cyan points. We find that for the smooth halo, stars in the outer halo are almost exclusively more metal-poor than \feh\ $\sim -1.0$, with a mean metallicity \meanfeh\ $= -1.94\pm0.20$ dex and dispersion \sigmafeh\ $ = 0.41$ dex. We measure a mean \alphafe\ abundance for these stars of \meanalphafe\ $ = 0.43\pm0.12$ dex, with a dispersion of \sigmaalphafe$ = 0.40$ dex. In comparison, the inner (\rproj\ $< 50$ kpc) halo sample has a significantly higher mean metallicity, with a slightly larger dispersion  (\meanfeh\ $ = -1.27\pm0.21$ dex,  \sigmafeh\ $ = 0.61$). We find that the inner halo is less enhanced in \alphafe, with a similar dispersion in \alphafe\ as the outer halo (\meanalphafe\ $= 0.29\pm0.08$,  \sigmaalphafe\ $ = 0.39$).  

\subsubsection{Comparison with other studies of the M31 halo}

We show the distribution of our outer halo sample compared to other recent, similar samples of the M31 halo in the bottom panel of Figure~\ref{fig:outer_halo_comparison}. The outer halo sample presented in \citet{gilbert20} is shown in pink, where their sample spans a radial range $46< $\rproj $< 110$ kpc. Their sample does not contain any stars likely associated with known substructure, and therefore should represent the smooth outer halo. For comparison, we also show the inner M31 halo ($8 < $\rproj$< 35$ kpc) sample presented in \citet{escala20b}, where we have applied the criterion that the probability of a star being associated with substructure is less than 0.5, to ensure a comparable ``smooth halo'' sample. The abundance measurements for both the \citet{gilbert20} and \citet{escala20b} samples were obtained using the same spectral synthesis technique used for this paper, and should therefore be comparable in the sense that the abundance measurements are on the same scale.

Our smooth halo stars closely match the \feh--\alphafe\ distribution space spanned by the \citet{escala20b} and \citet{gilbert20} samples, for the inner and outer halo, respectively. For the inner halo, \citet{escala20b} measure a mean metallicity of \meanfeh\ $ -1.08\pm0.04$ dex, and \meanalphafe\ $= 0.40\pm0.03$. These measurements are statistically consistent within 1$\sigma$ with our inner halo sample (\meanfeh\ $=-1.27\pm0.21$ dex, \meanalphafe\ $ = 0.43\pm0.12$ dex). For the outer halo, \citet{gilbert20} measured a mean \feh\ of $ = -1.92\pm0.13$ dex, and a mean \alphafe\ of $ = 0.30\pm0.16$ dex. Again, we find these values to be consistent with what we measure for the smooth outer halo sample (\meanfeh\ $= -1.94\pm0.20$, \meanalphafe\ $ = 0.43\pm0.12$ dex), although we do find a slightly higher \meanalphafe. Between the inner \citep{escala20b} and outer \citep{gilbert20} smooth halo samples, there is evidence for a systematic difference in the mean \feh\ between the inner and outer halo. As our sample is consistent with both of these previously published results, we also recover a significant difference in the mean \feh\ between the inner and outer smooth halo, with a magnitude of approximately 0.7 dex.

\subsubsection{Comparison with M31 dSphs}

\begin{figure}
 	\includegraphics[width=\columnwidth]{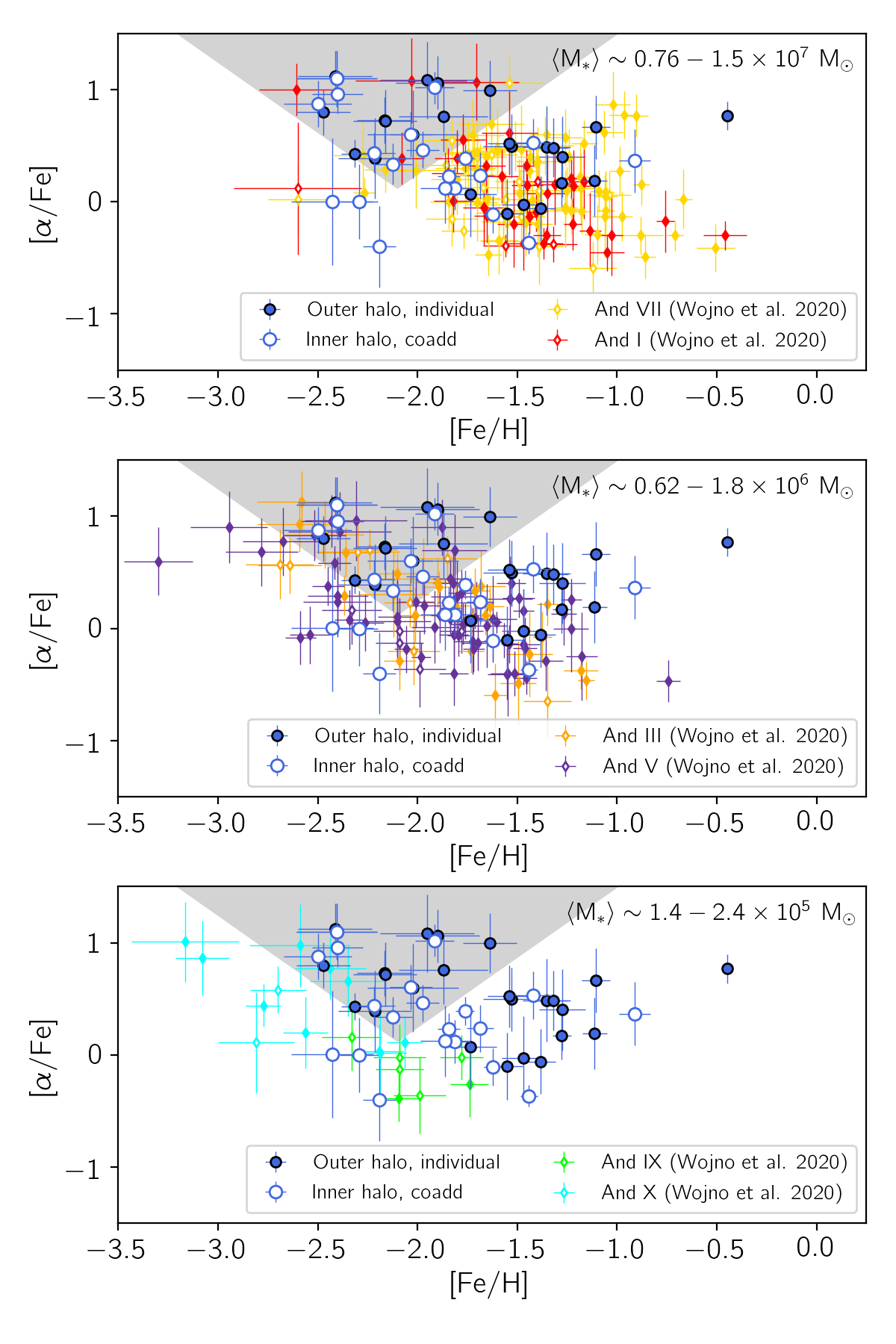}
    \caption{The \feh--\alphafe  distribution of our outer halo (\rproj$> 50 $kpc; dark blue) individual (filled) and coadded (open) abundance measurements, compared to that for six dSphs of M31 in three mass ranges, where the mass range is given in the upper right of each panel. The shaded area indicates the region used to compare the fraction of metal-poor, $\alpha$-enhanced in the outer halo with the fraction of stars in dSphs with these abundances.
    Top: The M31 outer halo sample overplotted with abundance measurements for two of the largest dSphs presented in \citet{wojno20}: And VII (gold) and And I (red), for individual (filled diamonds) and coadded (open diamonds) abundance measurements. Middle: The outer halo sample overplotted with the abundance measurements of two dSphs in the middle of the mass range from \citet{wojno20}, And III (orange) and And V (purple). Bottom: The outer halo sample overplotted with the \feh--\alphafe\ distributions from the two least massive dSphs presented in \citet{wojno20}: And IX (lime) and And X (cyan). The abundance distribution of our outer halo sample overlaps the most with the distributions of the more massive satellites (top and middle panels). Interestingly, the metal-poor, high-$\alpha$ region of our outer halo sample is not well-represented by any one of the dSphs, indicating that these stars may originate from multiple accretion events.}
    \label{fig:outer_halo_comparison_dsphs}
\end{figure}

\citet{gilbert20} found that the abundance distribution of their sample of outer halo stars best resembled the abundance distributions of the largest M31 dSphs, And I and And VII. In  Figure~\ref{fig:outer_halo_comparison_dsphs}, we show our outer halo sample compared to the individual (filled diamonds) and coadded (open diamonds) abundance measurements for dSphs in three mass ranges from \citet{wojno20}. The mass range for the dSphs shown is given in the upper right for each panel, respectively, where all mass estimates are sourced from \citet{kirby13} and \citet{kirby20} (see Table~7 of \citet{wojno20} for more details). In the top panel, we show our outer halo abundance distribution compared to two of the most massive M31 dSphs, And VII (gold, M$_{\sun} \sim 7.6\times10^6$) and And I (red, M$_{\sun} \sim 1.5\times10^7$).  In the middle panel, we show the outer halo abundance distribution compared to And III (orange, $M_{\sun} \sim 6.2\times10^5$) and And V (orange, $M_{\sun} \sim 1.8\times10^6$), which represent the middle of the mass range of dSphs presented in \citet{wojno20}. Finally, in the bottom panel, we show the outer halo abundance distribution compared to And IX (lime, $M_{\sun} \sim 1.4\times10^5$) and And X (cyan, $M_{\sun} \sim 2.4\times10^5$), two of the least massive dSphs where we have abundance measurements for individual stars. 
Compared to inner halo samples from both this work and \citet{escala20b} (Figure~\ref{fig:outer_halo_comparison}), we find that our outer halo sample is more consistent with the abundance distribution of the more massive dSphs of M31 (top and middle panels of Figure~\ref{fig:outer_halo_comparison_dsphs}). However, the abundance distributions for the more massive dSphs are slightly more metal-rich and lower in average \alphafe, and none of the M31 dSphs with measured \feh--\alphafe\ abundance distributions to date populate the parameter space occupied by the most metal-poor and $\alpha$-enhanced outer halo stars. 

To better quantify how well this region of parameter space is populated by a sample of stars from dSphs in a given range, we calculate the fraction of stars with abundances that put them in the shaded grey region in each panel of Figure~\ref{fig:outer_halo_comparison_dsphs}, compared to the entire abundance space. We find $\sim46\%$ of our outer halo sample occupies this space, compared to $\sim10\%$ for the inner halo sample. For the most massive galaxies (top panel, And VII and And I), we find that $\sim8\%$ of the dSph sample has abundance measurements that fall within the shaded region. For the lower mass dSphs, we find $\sim18\%$ and $\sim16\%$ of the dSph abundance distributions fall within the shaded region for the middle (And III, And V) and bottom (And IX, And X) panels, respectively. While the lower mass satellites contain a relatively larger fraction of stars that occupy the same region as the most metal-poor, $\alpha$-enhanced outer halo stars, they still represent a small minority of the overall abundance distribution. 

\subsection{Abundance gradients in the halo of M31}
\label{sec:abundance_gradients}
Generally, negative abundance gradients form when star formation efficiency is higher in dense inner regions, and lower in sparsely populated outer regions. However, stellar halos are built up primarily of accreted components \citep[e.g.,][]{Bullock05,conroy19}, where these progenitors likely have multiple generations of star formation, and may even have abundance gradients of their own. The inner halos of galaxies can contain a mix of in-situ and accreted stars, while the outer regions are overwhelmingly comprised of accreted satellites \citep{conroy19,Zolotov09}. 

\subsubsection{\feh\ gradients}
M31 is known to have a significant negative radial metallicity gradient \citep{kalirai06,Koch08,ibata14,gilbert14} with a magnitude of approximately $-0.01$ dex kpc$^{-1}$. This gradient persists with the inclusion of stars associated with substructure \citep{gilbert14}. \citet{gilbert14} propose that this large-scale gradient indicates that M31 accreted at least one relatively massive progenitor. In addition, \citet{gilbert14} find significant variation in the metallicity distribution between fields in the outer halo, and suggest that multiple smaller progenitors contributed to the observed metallicity distribution at large projected radii.

We measure a radial gradient of $-0.0075\pm0.0001$ dex kpc$^{-1}$, consistent with results in the literature measured using photometric \feh\ estimates. We note there is an offset with respect to the normalization compared to the \feh\ gradient measured using photometric \feh\ from \citet{gilbert14}, and we verify that this offset is consistent with the difference between the photometric and spectroscopic \feh\ measurements. The persistence of a large-scale metallicity gradient corresponds with a scenario where the bulk of the halo was built up from at least one massive satellite \citep{cooper10}, although we acknowledge that there may be significant contributions from less massive satellites. Additionally, if relatively more massive satellites with higher metallicities fall into the central regions of the halo, and less massive satellites interact with the less dense regions of the halo \citep{cooper10,tissera12}, we would expect to see metal-poor, \alphafe-high stars predominantly in the outer halo. 

\subsubsection{\alphafe\ gradients}

Like \feh, gradients in \alphafe\ abundances provide important clues to the star formation history of a galaxy. The ratio of $\alpha$-elements to Fe in a given stellar population is directly linked to the star formation efficiency and supernovae rates (for a through review for this abundance space in the MW, see \citealt {Matteucci21}), as $\alpha$-elements are mainly produced in core-collapse SNe by massive, short-lived stars, and low- and intermediate- mass stars explode as Type Ia SNe, producing much higher yields of Fe than core-collapse SNe. 

In this work, we find a positive gradient (\gradientalphaicsm) in \alphafe\ as a function of projected radius for smooth halo stars when we fit over the full radial range of the sample (see upper left panel of Figure~\ref{fig:alphafe_gradient}). If we consider only stars possibly associated with known substructures, we do not measure a significant gradient as a function of projected radius (\gradientalphaicss; \rproj$<90$~kpc).  However, if all M31 stars are included in the fit, the measured \alphafe\ gradient remains statistically significant (\gradientalphaicsms, upper right panel of Figure~\ref{fig:alphafe_gradient}).  The strength and significance of the measured gradient in \alphafe\ is affected by the stars, and the radial range, included in the fit. 

While gradients in \alphafe\ as a function of projected radius have been studied primarily in the star-forming disks of galaxies, studies of the MW halo indicate the presence of a negative gradient in \alphafe\ \citep[e.g.,][]{fulbright02,Nissen10,hawkins15,fernandez-alvar17}. In addition, \citet{conroy19} found a flat vertical gradient in \alphafe\ for the MW, using a sample of stars spanning a radial range out to 100 kpc, and vertical distance $|z|<10$ kpc. However, these studies typically cover a limited range in radius from the center of the MW compared to the range in projected radii for our stars in M31, and therefore we turn to simulations for a more comprehensive comparison.

Using simulations of MW and M31-mass galaxies,  \citet{font11} found a slightly positive radial gradient in \alphafe. An updated version of this analysis using the ARTEMIS simulations \citep{font20} found  similar negative \feh\ gradients for both the in-situ and accreted halo components, with constant \alphafe\ as a function of radius (i.e., no gradient), although there is some indication of a positive trend in the outermost regions of the simulated galaxies. \citet{font11} ascribe the presence of a positive gradient in \alphafe\ as a function of \rproj\ as due to the relative importance of SNe Ia (resulting from intermediate-mass stars) in the inner regions of the galaxy, where SNe II (resulting from massive stars) produce metals at all radii. They found that increasing the rate of SNe Ia in their simulated galaxies increases the steepness of the gradient in \alphafe\ as a function of \rproj.
The lack of a radial gradient in \alphafe\ can be the result of early accretion across the entire halo, where both the in-situ and accreted components have had similar enrichment histories. On the other hand, a negative gradient implies low-\alphafe\ stars in the outer halo compared to the inner halo, which may result from the accretion of satellites with a star formation efficiency such that they have been sufficiently enriched to higher values of \feh\, and therefore  lower values of \alphafe. 

If we compare the results of these simulations with the most analogous sample in this work (i.e., the gradient fit to all M31 stars, \gradientalphaicsms), we find a 
positive gradient in \alphafe\ as a function of \rproj. 
However, we note that if we consider only stars at \rproj $ < 90$ kpc, we find that the trend in \alphafe\ as a function of \rproj\ is consistent within $2\sigma$ with being flat ($0.001\pm0.0005$ dex kpc$^{-1}$).
In our sample, stars at larger projected radii (\rproj\ $> 100$kpc) have higher \alphafe, which increases the mesaured gradient when the entire radial range is included. The location of these stars in \feh--\alphafe\ space is most similar to the ``in-situ" or ``canonical halo" population of MW stars \citep[e.g.,][]{hawkins15,hayes18}, however, at \rproj $> 100$kpc, these stars would have had to migrate a significant distance outwards from the central regions of M31. Therefore, we conclude that these stars are likely to be related to accretion events. 

\section{Conclusions}
\label{sec:conclusions}
In this work, we present newly-determined spectroscopic \feh\ and \alphafe\ for 160 stars observed in the halo of M31. We classify stars as belonging to either the smooth halo (91 stars), or potentially associated with substructure components, i.e. tidal debris from an accreted population (69 stars). Where we cannot measure abundances for individual stars reliably, we group spectra of different stars together and coadd them. We obtain a total of 118 coadded measurements (62 smooth halo, 56 substructure stars), from a pool of 611 stars without reliable individual abundances. These coadded abundance measurements have been shown to reflect the weighted average of the component stars \citep{wojno20} reliably. At large projected radii (\rproj$> 50$ kpc), we obtain measurements for 25 individual stars (as well as 21 coadds from a pool of 111 stars) in the smooth halo, and 20 individual stars (as well as 12 coadds from a pool of 63 stars) possibly associated with substructure components. Prior to this work, only 9 stars in the outer halo of M31 ($43 < $\rproj$< 165$ kpc) had \feh\ and \alphafe\ measurements available in the literature \citep{vargas14,gilbert20}. We therefore greatly expand on the number of spectroscopic abundance measurements in the M31 halo using this technique, especially at large radii where confirmed M31 halo stars are sparse. 

We measure a radial \feh\ gradient of \gradientfehicsm\ for the smooth halo. This gradient is consistent with previous results using photometric \feh\ \citep{gilbert14}, but over a larger radial range. We find a similar gradient for the full M31 halo sample (\gradientfehicsms), including stars possibly associated with substructure components. Additionally, we find evidence for a significant \alphafe\ gradient in the smooth halo (\gradientalphaicsm) over the entire radial range of our sample ($9 < $\rproj$< 180$ kpc), using both individual and coadded abundance measurements. We do not find evidence for a gradient when considering only stars possibly associated with substructure. The finding for our smooth halo sample is in contrast to \citet{escala20b}, where they find no significant \alphafe\ gradient using abundance measurements for individual stars in the radial range $9 < $\rproj$< 35$ kpc. 

Following these gradients, we also find that the distribution of stars in \feh-\alphafe\ abundance space changes as a function of projected radius from the center of M31. The inner regions of the halo are typically more metal-rich, and less enhanced in \alphafe\ compared to the outer regions. We find that the spread in both \feh\ and \alphafe\ does not depend on projected radius.
Finally, we compare the \feh-\alphafe\ abundance distribution of stars possibly associated with tidal debris features with the abundance distribution of the underlying smooth halo of M31. We find that for the majority of substructure features, the mean \feh\ and \alphafe\ for stars potentially associated with substructure differs from the mean \feh\ and \alphafe\ for the smooth halo stars in the same radial range.
This is expected, as the progenitors of the substructure populations were likely accreted more recently than those of the phase-mixed inner halo (built up from more ancient accreted progenitors), and likely had a different chemical evolution history.

Stars associated with the GSS substructure, likely caused by a major (4:1 mass ratio) merger a few Gyr ago \citep{dsouza18,hammer18,quirk20}, are on average more metal-rich, and have a slightly lower mean \alphafe\ compared to the smooth halo at the same \rproj\ distance. The observed decline in \alphafe\ as a function of \feh\ also points to the progenitor of the GSS having an extended star formation history.
We see a similar effect in our sample of stars associated with the envelope of the GSS, where the \alphafe\ values of the substructure stars are on average much lower than the smooth halo of M31. 

When we consider only stars at large projected radii (``outer halo'' stars, Section~\ref{sec:discussion}), we find they are almost exclusively more metal poor than \feh $< -1.0$, and are enhanced with respect to \alphafe. We find that our sample of inner halo stars is on average more metal-rich, with a slightly broader distribution in \feh\ compared to the outer halo stars. If we compare our sample of stars in the outer halo, we find, as in \citet{gilbert20}, that the abundance distribution of smooth outer halo stars is  consistent with the abundance distributions of the more massive dSphs of M31 (e.g. And I, III, V, VII). However, we find that the abundance distribution of the smooth outer halo stars is not fully represented by any one of the surviving dSphs, implying that the stars in the metal-poor, \alphafe-enhanced region of the parameter space may have been accreted through multiple separate merger events.

While it is still difficult to accurately reconstruct the chemical evolution pathways within the progenitors of current-day substructure, this work shows that the halo of M31 is incredibly complex not only kinematically, but also chemically. Obtaining additional spectroscopic abundance measurements covering the entirety of the M31 halo is outside of the reach of the current generation of multi-object spectroscopic instruments (although see \citet{dey22} for a recent spectroscopic study of M31 halo stars with $R\sim2000-5000$), but may be achievable in the next generation of large scale spectroscopic surveys, such as the Subaru Prime Focus Spectrograph (PFS). 

\acknowledgments
This material is based upon work supported by the National Science
Foundation under Grant Nos.\ AST-1614569 (JLW, KMG) and AST-1614081 (ENK).  ENK
gratefully acknowledges support from a Cottrell Scholar award
administered by the Research Corporation for Science Advancement as
well as funding from generous donors to the California Institute of
Technology.  IE acknowledges generous support from a Carnegie-Princeton Fellowship awarded by the Carnegie Observatories and Princeton University.
%RLB and SRM thank NSF grants AST-0307842, AST-0607726, AST-1009882, and AST-1413269. 

The data presented herein were obtained at the W. M. Keck Observatory, which is operated as a scientific partnership among the California Institute of Technology, the University of California and the National Aeronautics and Space Administration. The Observatory was made possible by the generous financial support of the W. M. Keck Foundation.

The authors wish to recognize and acknowledge the very significant cultural role and reverence that the summit of Mauna Kea has always had within the indigenous Hawaiian community.  We are most fortunate to have the opportunity to conduct observations from this mountain.

This research made use of Astropy, a community-developed core Python package for Astronomy \citep{astropy2013,astropy2018}\footnote{http://www.astropy.org}. 
 
\facilities{Keck:II (DEIMOS)}
\software{Astropy \citep{astropy2013},
Matplotlib \citep{matplotlib}, numpy \citep{numpy}, scipy \citep{scipy}, emcee \citep{foreman-mackey13}}
\bibliography{mybib}{}

\begin{thebibliography}{}
\expandafter\ifx\csname natexlab\endcsname\relax\def\natexlab#1{#1}\fi
\providecommand{\url}[1]{\href{#1}{#1}}
\providecommand{\dodoi}[1]{doi:~\href{http://doi.org/#1}{\nolinkurl{#1}}}
\providecommand{\doeprint}[1]{\href{http://ascl.net/#1}{\nolinkurl{http://ascl.net/#1}}}
\providecommand{\doarXiv}[1]{\href{https://arxiv.org/abs/#1}{\nolinkurl{https://arxiv.org/abs/#1}}}

\bibitem[{{Abdurro'uf} {et~al.}(2022){Abdurro'uf}, {Accetta}, {Aerts}, {Silva
  Aguirre}, {Ahumada}, {Ajgaonkar}, {Filiz Ak}, {Alam}, {Allende Prieto},
  {Almeida}, {Anders}, {Anderson}, {Andrews}, {Anguiano}, {Aquino-Ort{\'\i}z},
  {Arag{\'o}n-Salamanca}, {Argudo-Fern{\'a}ndez}, {Ata}, {Aubert},
  {Avila-Reese}, {Badenes}, {Barb{\'a}}, {Barger}, {Barrera-Ballesteros},
  {Beaton}, {Beers}, {Belfiore}, {Bender}, {Bernardi}, {Bershady}, {Beutler},
  {Bidin}, {Bird}, {Bizyaev}, {Blanc}, {Blanton}, {Boardman}, {Bolton},
  {Boquien}, {Borissova}, {Bovy}, {Brandt}, {Brown}, {Brownstein}, {Brusa},
  {Buchner}, {Bundy}, {Burchett}, {Bureau}, {Burgasser}, {Cabang}, {Campbell},
  {Cappellari}, {Carlberg}, {Wanderley}, {Carrera}, {Cash}, {Chen}, {Chen},
  {Cherinka}, {Chiappini}, {Choi}, {Chojnowski}, {Chung}, {Clerc}, {Cohen},
  {Comerford}, {Comparat}, {da Costa}, {Covey}, {Crane}, {Cruz-Gonzalez},
  {Culhane}, {Cunha}, {Dai}, {Damke}, {Darling}, {Davidson}, {Davies},
  {Dawson}, {De Lee}, {Diamond-Stanic}, {Cano-D{\'\i}az}, {S{\'a}nchez},
  {Donor}, {Duckworth}, {Dwelly}, {Eisenstein}, {Elsworth}, {Emsellem},
  {Eracleous}, {Escoffier}, {Fan}, {Farr}, {Feng}, {Fern{\'a}ndez-Trincado},
  {Feuillet}, {Filipp}, {Fillingham}, {Frinchaboy}, {Fromenteau}, {Galbany},
  {Garc{\'\i}a}, {Garc{\'\i}a-Hern{\'a}ndez}, {Ge}, {Geisler}, {Gelfand},
  {G{\'e}ron}, {Gibson}, {Goddy}, {Godoy-Rivera}, {Grabowski}, {Green},
  {Greener}, {Grier}, {Griffith}, {Guo}, {Guy}, {Hadjara}, {Harding},
  {Hasselquist}, {Hayes}, {Hearty}, {Hern{\'a}ndez}, {Hill}, {Hogg},
  {Holtzman}, {Horta}, {Hsieh}, {Hsu}, {Hsu}, {Huber}, {Huertas-Company},
  {Hutchinson}, {Hwang}, {Ibarra-Medel}, {Chitham}, {Ilha}, {Imig}, {Jaekle},
  {Jayasinghe}, {Ji}, {Johnson}, {Jones}, {J{\"o}nsson}, {Katkov}, {Khalatyan},
  {Kinemuchi}, {Kisku}, {Knapen}, {Kneib}, {Kollmeier}, {Kong}, {Kounkel},
  {Kreckel}, {Krishnarao}, {Lacerna}, {Lane}, {Langgin}, {Lavender}, {Law},
  {Lazarz}, {Leung}, {Leung}, {Lewis}, {Li}, {Li}, {Lian}, {Liang}, {Lin},
  {Lin}, {Lin}, {Lintott}, {Long}, {Longa-Pe{\~n}a}, {L{\'o}pez-Cob{\'a}},
  {Lu}, {Lundgren}, {Luo}, {Mackereth}, {de la Macorra}, {Mahadevan},
  {Majewski}, {Manchado}, {Mandeville}, {Maraston}, {Margalef-Bentabol},
  {Masseron}, {Masters}, {Mathur}, {McDermid}, {Mckay}, {Merloni},
  {Merrifield}, {Meszaros}, {Miglio}, {Di Mille}, {Minniti}, {Minsley},
  {Monachesi}, {Moon}, {Mosser}, {Mulchaey}, {Muna}, {Mu{\~n}oz}, {Myers},
  {Myers}, {Nadathur}, {Nair}, {Nandra}, {Neumann}, {Newman}, {Nidever},
  {Nikakhtar}, {Nitschelm}, {O'Connell}, {Garma-Oehmichen}, {Luan Souza de
  Oliveira}, {Olney}, {Oravetz}, {Ortigoza-Urdaneta}, {Osorio}, {Otter},
  {Pace}, {Padilla}, {Pan}, {Pan}, {Parikh}, {Parker}, {Peirani}, {Pe{\~n}a
  Ram{\'\i}rez}, {Penny}, {Percival}, {Perez-Fournon}, {Pinsonneault},
  {Poidevin}, {Poovelil}, {Price-Whelan}, {B{\'a}rbara de Andrade Queiroz},
  {Raddick}, {Ray}, {Rembold}, {Riddle}, {Riffel}, {Riffel}, {Rix}, {Robin},
  {Rodr{\'\i}guez-Puebla}, {Roman-Lopes}, {Rom{\'a}n-Z{\'u}{\~n}iga}, {Rose},
  {Ross}, {Rossi}, {Rubin}, {Salvato}, {S{\'a}nchez}, {S{\'a}nchez-Gallego},
  {Sanderson}, {Santana Rojas}, {Sarceno}, {Sarmiento}, {Sayres}, {Sazonova},
  {Schaefer}, {Schiavon}, {Schlegel}, {Schneider}, {Schultheis}, {Schwope},
  {Serenelli}, {Serna}, {Shao}, {Shapiro}, {Sharma}, {Shen}, {Shetrone}, {Shu},
  {Simon}, {Skrutskie}, {Smethurst}, {Smith}, {Sobeck}, {Spoo}, {Sprague},
  {Stark}, {Stassun}, {Steinmetz}, {Stello}, {Stone-Martinez},
  {Storchi-Bergmann}, {Stringfellow}, {Stutz}, {Su}, {Taghizadeh-Popp},
  {Talbot}, {Tayar}, {Telles}, {Teske}, {Thakar}, {Theissen}, {Tkachenko},
  {Thomas}, {Tojeiro}, {Hernandez Toledo}, {Troup}, {Trump}, {Trussler},
  {Turner}, {Tuttle}, {Unda-Sanzana}, {V{\'a}zquez-Mata}, {Valentini},
  {Valenzuela}, {Vargas-Gonz{\'a}lez}, {Vargas-Maga{\~n}a}, {Alfaro},
  {Villanova}, {Vincenzo}, {Wake}, {Warfield}, {Washington}, {Weaver},
  {Weijmans}, {Weinberg}, {Weiss}, {Westfall}, {Wild}, {Wilde}, {Wilson},
  {Wilson}, {Wilson}, {Wolf}, {Wood-Vasey}, {Yan}, {Zamora}, {Zasowski},
  {Zhang}, {Zhao}, {Zheng}, {Zheng}, \& {Zhu}}]{apogee_dr17}
{Abdurro'uf}, {Accetta}, K., {Aerts}, C., {et~al.} 2022, \apjs, 259, 35,
  \dodoi{10.3847/1538-4365/ac4414}

\bibitem[{{Astropy Collaboration} {et~al.}(2013){Astropy Collaboration},
  {Robitaille}, {Tollerud}, {Greenfield}, {Droettboom}, {Bray}, {Aldcroft},
  {Davis}, {Ginsburg}, {Price-Whelan}, {Kerzendorf}, {Conley}, {Crighton},
  {Barbary}, {Muna}, {Ferguson}, {Grollier}, {Parikh}, {Nair}, {Unther},
  {Deil}, {Woillez}, {Conseil}, {Kramer}, {Turner}, {Singer}, {Fox}, {Weaver},
  {Zabalza}, {Edwards}, {Azalee Bostroem}, {Burke}, {Casey}, {Crawford},
  {Dencheva}, {Ely}, {Jenness}, {Labrie}, {Lim}, {Pierfederici}, {Pontzen},
  {Ptak}, {Refsdal}, {Servillat}, \& {Streicher}}]{astropy2013}
{Astropy Collaboration}, {Robitaille}, T.~P., {Tollerud}, E.~J., {et~al.} 2013,
  \aap, 558, A33, \dodoi{10.1051/0004-6361/201322068}

\bibitem[{{Astropy Collaboration} {et~al.}(2018){Astropy Collaboration},
  {Price-Whelan}, {Sip{\H{o}}cz}, {G{\"u}nther}, {Lim}, {Crawford}, {Conseil},
  {Shupe}, {Craig}, {Dencheva}, {Ginsburg}, {Vand erPlas}, {Bradley},
  {P{\'e}rez-Su{\'a}rez}, {de Val-Borro}, {Aldcroft}, {Cruz}, {Robitaille},
  {Tollerud}, {Ardelean}, {Babej}, {Bach}, {Bachetti}, {Bakanov}, {Bamford},
  {Barentsen}, {Barmby}, {Baumbach}, {Berry}, {Biscani}, {Boquien}, {Bostroem},
  {Bouma}, {Brammer}, {Bray}, {Breytenbach}, {Buddelmeijer}, {Burke},
  {Calderone}, {Cano Rodr{\'\i}guez}, {Cara}, {Cardoso}, {Cheedella}, {Copin},
  {Corrales}, {Crichton}, {D'Avella}, {Deil}, {Depagne}, {Dietrich}, {Donath},
  {Droettboom}, {Earl}, {Erben}, {Fabbro}, {Ferreira}, {Finethy}, {Fox},
  {Garrison}, {Gibbons}, {Goldstein}, {Gommers}, {Greco}, {Greenfield},
  {Groener}, {Grollier}, {Hagen}, {Hirst}, {Homeier}, {Horton}, {Hosseinzadeh},
  {Hu}, {Hunkeler}, {Ivezi{\'c}}, {Jain}, {Jenness}, {Kanarek}, {Kendrew},
  {Kern}, {Kerzendorf}, {Khvalko}, {King}, {Kirkby}, {Kulkarni}, {Kumar},
  {Lee}, {Lenz}, {Littlefair}, {Ma}, {Macleod}, {Mastropietro}, {McCully},
  {Montagnac}, {Morris}, {Mueller}, {Mumford}, {Muna}, {Murphy}, {Nelson},
  {Nguyen}, {Ninan}, {N{\"o}the}, {Ogaz}, {Oh}, {Parejko}, {Parley}, {Pascual},
  {Patil}, {Patil}, {Plunkett}, {Prochaska}, {Rastogi}, {Reddy Janga},
  {Sabater}, {Sakurikar}, {Seifert}, {Sherbert}, {Sherwood-Taylor}, {Shih},
  {Sick}, {Silbiger}, {Singanamalla}, {Singer}, {Sladen}, {Sooley},
  {Sornarajah}, {Streicher}, {Teuben}, {Thomas}, {Tremblay}, {Turner},
  {Terr{\'o}n}, {van Kerkwijk}, {de la Vega}, {Watkins}, {Weaver}, {Whitmore},
  {Woillez}, {Zabalza}, \& {Astropy Contributors}}]{astropy2018}
{Astropy Collaboration}, {Price-Whelan}, A.~M., {Sip{\H{o}}cz}, B.~M., {et~al.}
  2018, \aj, 156, 123, \dodoi{10.3847/1538-3881/aabc4f}

\bibitem[{{Beaton} {et~al.}(2007){Beaton}, {Majewski}, {Patterson},
  {Guhathakurta}, {Gilbert}, {Kalirai}, {Kirby}, \& {Ostheimer}}]{beaton07}
{Beaton}, R., {Majewski}, S., {Patterson}, R., {et~al.} 2007, in American
  Astronomical Society Meeting Abstracts, Vol. 211, American Astronomical
  Society Meeting Abstracts, 104.17

\bibitem[{{Bellazzini} {et~al.}(2003){Bellazzini}, {Cacciari}, {Federici},
  {Fusi Pecci}, \& {Rich}}]{bellazzini03}
{Bellazzini}, M., {Cacciari}, C., {Federici}, L., {Fusi Pecci}, F., \& {Rich},
  M. 2003, \aap, 405, 867, \dodoi{10.1051/0004-6361:20030455}

\bibitem[{{Bullock} \& {Johnston}(2005)}]{Bullock05}
{Bullock}, J.~S., \& {Johnston}, K.~V. 2005, \apj, 635, 931,
  \dodoi{10.1086/497422}

\bibitem[{{Bullock} {et~al.}(2001){Bullock}, {Kravtsov}, \&
  {Weinberg}}]{bullock01}
{Bullock}, J.~S., {Kravtsov}, A.~V., \& {Weinberg}, D.~H. 2001, \apj, 548, 33,
  \dodoi{10.1086/318681}

\bibitem[{{Chapman} {et~al.}(2006){Chapman}, {Ibata}, {Lewis}, {Ferguson},
  {Irwin}, {McConnachie}, \& {Tanvir}}]{chapman06}
{Chapman}, S.~C., {Ibata}, R., {Lewis}, G.~F., {et~al.} 2006, \apj, 653, 255,
  \dodoi{10.1086/508599}

\bibitem[{{Chapman} {et~al.}(2008){Chapman}, {Ibata}, {Irwin}, {Koch},
  {Letarte}, {Martin}, {Collins}, {Lewis}, {McConnachie}, {Pe{\~n}arrubia},
  {Rich}, {Trethewey}, {Ferguson}, {Huxor}, \& {Tanvir}}]{chapman08}
{Chapman}, S.~C., {Ibata}, R., {Irwin}, M., {et~al.} 2008, \mnras, 390, 1437,
  \dodoi{10.1111/j.1365-2966.2008.13703.x}

\bibitem[{{Clementini} {et~al.}(2011){Clementini}, {Contreras Ramos},
  {Federici}, {Macario}, {Beccari}, {Testa}, {Cignoni}, {Marconi}, {Ripepi},
  {Tosi}, {Bellazzini}, {Fusi Pecci}, {Diolaiti}, {Cacciari}, {Marano},
  {Giallongo}, {Ragazzoni}, {Di Paola}, {Gallozzi}, \&
  {Smareglia}}]{clementini11}
{Clementini}, G., {Contreras Ramos}, R., {Federici}, L., {et~al.} 2011, \apj,
  743, 19, \dodoi{10.1088/0004-637X/743/1/19}

\bibitem[{{Cole} {et~al.}(1994){Cole}, {Aragon-Salamanca}, {Frenk}, {Navarro},
  \& {Zepf}}]{cole94}
{Cole}, S., {Aragon-Salamanca}, A., {Frenk}, C.~S., {Navarro}, J.~F., \&
  {Zepf}, S.~E. 1994, \mnras, 271, 781, \dodoi{10.1093/mnras/271.4.781}

\bibitem[{{Conroy} {et~al.}(2019){Conroy}, {Naidu}, {Zaritsky}, {Bonaca},
  {Cargile}, {Johnson}, \& {Caldwell}}]{conroy19}
{Conroy}, C., {Naidu}, R.~P., {Zaritsky}, D., {et~al.} 2019, \apj, 887, 237,
  \dodoi{10.3847/1538-4357/ab5710}

\bibitem[{{Cooper} {et~al.}(2010){Cooper}, {Cole}, {Frenk}, {White}, {Helly},
  {Benson}, {De Lucia}, {Helmi}, {Jenkins}, {Navarro}, {Springel}, \&
  {Wang}}]{cooper10}
{Cooper}, A.~P., {Cole}, S., {Frenk}, C.~S., {et~al.} 2010, \mnras, 406, 744,
  \dodoi{10.1111/j.1365-2966.2010.16740.x}

\bibitem[{{Cooper} {et~al.}(2012){Cooper}, {Newman}, {Davis}, {Finkbeiner}, \&
  {Gerke}}]{cooper12}
{Cooper}, M.~C., {Newman}, J.~A., {Davis}, M., {Finkbeiner}, D.~P., \& {Gerke},
  B.~F. 2012, {spec2d: DEEP2 DEIMOS Spectral Pipeline}, Astrophysics Source
  Code Library.
\newblock \doeprint{1203.003}

\bibitem[{{Demarque} {et~al.}(2004){Demarque}, {Woo}, {Kim}, \&
  {Yi}}]{demarque04}
{Demarque}, P., {Woo}, J.-H., {Kim}, Y.-C., \& {Yi}, S.~K. 2004, \apjs, 155,
  667, \dodoi{10.1086/424966}

\bibitem[{{Dey} {et~al.}(2022){Dey}, {Najita}, {Koposov}, {Josephy-Zack},
  {Maxemin}, {Bell}, {Poppett}, {Patel}, {Beraldo e Silva}, {Raichoor},
  {Schlegel}, {Lang}, {Meisner}, {Myers}, {Aguilar}, {Ahlen}, {Allende Prieto},
  {Brooks}, {Cooper}, {Dawson}, {de la Macorra}, {Doel}, {Font-Ribera},
  {Garcia-Bellido}, {Gontcho}, {Guy}, {Honscheid}, {Kehoe}, {Kisner}, {Kremin},
  {Landriau}, {Le Guillou}, {Levi}, {Li}, {Martini}, {Miquel}, {Moustakas},
  {Nie}, {Palanque-Delabrouille}, {Prada}, {Schlafly}, {Sharples}, {Tarle},
  {Ting}, {Tyas}, {Valluri}, {Wechsler}, \& {Zou}}]{dey22}
{Dey}, A., {Najita}, J.~R., {Koposov}, S.~E., {et~al.} 2022, arXiv e-prints,
  arXiv:2208.11683.
\newblock \doarXiv{2208.11683}

\bibitem[{{Dorman} {et~al.}(2012){Dorman}, {Guhathakurta}, {Fardal}, {Lang},
  {Geha}, {Howley}, {Kalirai}, {Bullock}, {Cuillandre}, {Dalcanton}, {Gilbert},
  {Seth}, {Tollerud}, {Williams}, \& {Yniguez}}]{dorman12}
{Dorman}, C.~E., {Guhathakurta}, P., {Fardal}, M.~A., {et~al.} 2012, \apj, 752,
  147, \dodoi{10.1088/0004-637X/752/2/147}

\bibitem[{{D'Souza} \& {Bell}(2018)}]{dsouza18}
{D'Souza}, R., \& {Bell}, E.~F. 2018, Nature Astronomy, 2, 737,
  \dodoi{10.1038/s41550-018-0533-x}

\bibitem[{{Durrell} {et~al.}(1994){Durrell}, {Harris}, \&
  {Pritchet}}]{durrell94}
{Durrell}, P.~R., {Harris}, W.~E., \& {Pritchet}, C.~J. 1994, \aj, 108, 2114,
  \dodoi{10.1086/117223}

\bibitem[{{Durrell} {et~al.}(2001){Durrell}, {Harris}, \&
  {Pritchet}}]{durrell01}
---. 2001, \aj, 121, 2557, \dodoi{10.1086/320403}

\bibitem[{{Durrell} {et~al.}(2004){Durrell}, {Harris}, \&
  {Pritchet}}]{durrell04}
---. 2004, \aj, 128, 260, \dodoi{10.1086/421746}

\bibitem[{{Eggen} {et~al.}(1962){Eggen}, {Lynden-Bell}, \& {Sandage}}]{Eggen62}
{Eggen}, O.~J., {Lynden-Bell}, D., \& {Sandage}, A.~R. 1962, \apj, 136, 748,
  \dodoi{10.1086/147433}

\bibitem[{{Escala} {et~al.}(2022){Escala}, {Gilbert}, {Fardal}, {Guhathakurta},
  {Sanderson}, {Kalirai}, \& {Mobasher}}]{escala22}
{Escala}, I., {Gilbert}, K.~M., {Fardal}, M., {et~al.} 2022, arXiv e-prints,
  arXiv:2203.16675.
\newblock \doarXiv{2203.16675}

\bibitem[{{Escala} {et~al.}(2019{\natexlab{a}}){Escala}, {Gilbert}, {Kirby},
  {Wojno}, {Cunningham}, \& {Guhathakurta}}]{escala19b}
{Escala}, I., {Gilbert}, K.~M., {Kirby}, E.~N., {et~al.} 2019{\natexlab{a}},
  arXiv e-prints, arXiv:1909.00006.
\newblock \doarXiv{1909.00006}

\bibitem[{{Escala} {et~al.}(2020{\natexlab{a}}){Escala}, {Gilbert}, {Kirby},
  {Wojno}, {Cunningham}, \& {Guhathakurta}}]{escala20a}
---. 2020{\natexlab{a}}, \apj, 889, 177, \dodoi{10.3847/1538-4357/ab6659}

\bibitem[{{Escala} {et~al.}(2021){Escala}, {Gilbert}, {Wojno}, {Kirby}, \&
  {Guhathakurta}}]{escala21}
{Escala}, I., {Gilbert}, K.~M., {Wojno}, J., {Kirby}, E.~N., \& {Guhathakurta},
  P. 2021, \aj, 162, 45, \dodoi{10.3847/1538-3881/abfec4}

\bibitem[{{Escala} {et~al.}(2019{\natexlab{b}}){Escala}, {Kirby}, {Gilbert},
  {Cunningham}, \& {Wojno}}]{escala18}
{Escala}, I., {Kirby}, E.~N., {Gilbert}, K.~M., {Cunningham}, E.~C., \&
  {Wojno}, J. 2019{\natexlab{b}}, \apj, 878, 42,
  \dodoi{10.3847/1538-4357/ab1eac}

\bibitem[{{Escala} {et~al.}(2020{\natexlab{b}}){Escala}, {Kirby}, {Gilbert},
  {Wojno}, {Cunningham}, \& {Guhathakurta}}]{escala20b}
{Escala}, I., {Kirby}, E.~N., {Gilbert}, K.~M., {et~al.} 2020{\natexlab{b}},
  \apj, 902, 51, \dodoi{10.3847/1538-4357/abb474}

\bibitem[{{Fardal} {et~al.}(2006){Fardal}, {Babul}, {Geehan}, \&
  {Guhathakurta}}]{fardal06}
{Fardal}, M.~A., {Babul}, A., {Geehan}, J.~J., \& {Guhathakurta}, P. 2006,
  \mnras, 366, 1012, \dodoi{10.1111/j.1365-2966.2005.09864.x}

\bibitem[{{Fardal} {et~al.}(2008){Fardal}, {Babul}, {Guhathakurta}, {Gilbert},
  \& {Dodge}}]{fardal08}
{Fardal}, M.~A., {Babul}, A., {Guhathakurta}, P., {Gilbert}, K.~M., \& {Dodge},
  C. 2008, \apjl, 682, L33, \dodoi{10.1086/590386}

\bibitem[{{Fardal} {et~al.}(2007){Fardal}, {Guhathakurta}, {Babul}, \&
  {McConnachie}}]{fardal07}
{Fardal}, M.~A., {Guhathakurta}, P., {Babul}, A., \& {McConnachie}, A.~W. 2007,
  \mnras, 380, 15, \dodoi{10.1111/j.1365-2966.2007.11929.x}

\bibitem[{{Fardal} {et~al.}(2012){Fardal}, {Guhathakurta}, {Gilbert},
  {Tollerud}, {Kalirai}, {Tanaka}, {Beaton}, {Chiba}, {Komiyama}, \&
  {Iye}}]{fardal12}
{Fardal}, M.~A., {Guhathakurta}, P., {Gilbert}, K.~M., {et~al.} 2012, \mnras,
  423, 3134, \dodoi{10.1111/j.1365-2966.2012.21094.x}

\bibitem[{{Fardal} {et~al.}(2013){Fardal}, {Weinberg}, {Babul}, {Irwin},
  {Guhathakurta}, {Gilbert}, {Ferguson}, {Ibata}, {Lewis}, {Tanvir}, \&
  {Huxor}}]{fardal13}
{Fardal}, M.~A., {Weinberg}, M.~D., {Babul}, A., {et~al.} 2013, \mnras, 434,
  2779, \dodoi{10.1093/mnras/stt1121}

\bibitem[{{Ferguson} {et~al.}(2002){Ferguson}, {Irwin}, {Ibata}, {Lewis}, \&
  {Tanvir}}]{ferguson02}
{Ferguson}, A. M.~N., {Irwin}, M.~J., {Ibata}, R.~A., {Lewis}, G.~F., \&
  {Tanvir}, N.~R. 2002, \aj, 124, 1452, \dodoi{10.1086/342019}

\bibitem[{{Fern{\'a}ndez-Alvar} {et~al.}(2017){Fern{\'a}ndez-Alvar}, {Carigi},
  {Allende Prieto}, {Hayden}, {Beers}, {Fern{\'a}ndez-Trincado}, {Meza},
  {Schultheis}, {Santiago}, {Queiroz}, {Anders}, {da Costa}, \&
  {Chiappini}}]{fernandez-alvar17}
{Fern{\'a}ndez-Alvar}, E., {Carigi}, L., {Allende Prieto}, C., {et~al.} 2017,
  \mnras, 465, 1586, \dodoi{10.1093/mnras/stw2861}

\bibitem[{{Font} {et~al.}(2011){Font}, {McCarthy}, {Crain}, {Theuns}, {Schaye},
  {Wiersma}, \& {Dalla Vecchia}}]{font11}
{Font}, A.~S., {McCarthy}, I.~G., {Crain}, R.~A., {et~al.} 2011, \mnras, 416,
  2802, \dodoi{10.1111/j.1365-2966.2011.19227.x}

\bibitem[{{Font} {et~al.}(2020){Font}, {McCarthy}, {Poole-Mckenzie},
  {Stafford}, {Brown}, {Schaye}, {Crain}, {Theuns}, \& {Schaller}}]{font20}
{Font}, A.~S., {McCarthy}, I.~G., {Poole-Mckenzie}, R., {et~al.} 2020, \mnras,
  498, 1765, \dodoi{10.1093/mnras/staa2463}

\bibitem[{{Foreman-Mackey} {et~al.}(2013){Foreman-Mackey}, {Hogg}, {Lang}, \&
  {Goodman}}]{foreman-mackey13}
{Foreman-Mackey}, D., {Hogg}, D.~W., {Lang}, D., \& {Goodman}, J. 2013, \pasp,
  125, 306, \dodoi{10.1086/670067}

\bibitem[{{Fulbright}(2002)}]{fulbright02}
{Fulbright}, J.~P. 2002, \aj, 123, 404, \dodoi{10.1086/324630}

\bibitem[{{Gaia Collaboration} {et~al.}(2022){Gaia Collaboration}, {Vallenari},
  {Brown}, {Prusti}, {de Bruijne}, {Arenou}, {Babusiaux}, {Biermann},
  {Creevey}, {Ducourant}, {Evans}, {Eyer}, {Guerra}, {Hutton}, {Jordi},
  {Klioner}, {Lammers}, {Lindegren}, {Luri}, {Mignard}, {Panem}, {Pourbaix},
  {Randich}, {Sartoretti}, {Soubiran}, {Tanga}, {Walton}, {Bailer-Jones},
  {Bastian}, {Drimmel}, {Jansen}, {Katz}, {Lattanzi}, {van Leeuwen}, {Bakker},
  {Cacciari}, {Casta{\~n}eda}, {De Angeli}, {Fabricius}, {Fouesneau},
  {Fr{\'e}mat}, {Galluccio}, {Guerrier}, {Heiter}, {Masana}, {Messineo},
  {Mowlavi}, {Nicolas}, {Nienartowicz}, {Pailler}, {Panuzzo}, {Riclet}, {Roux},
  {Seabroke}, {Sordo{\o}rcit}, {Th{\'e}venin}, {Gracia-Abril}, {Portell},
  {Teyssier}, {Altmann}, {Andrae}, {Audard}, {Bellas-Velidis}, {Benson},
  {Berthier}, {Blomme}, {Burgess}, {Busonero}, {Busso}, {C{\'a}novas}, {Carry},
  {Cellino}, {Cheek}, {Clementini}, {Damerdji}, {Davidson}, {de Teodoro},
  {Nu{\~n}ez Campos}, {Delchambre}, {Dell'Oro}, {Esquej},
  {Fern{\'a}ndez-Hern{\'a}ndez}, {Fraile}, {Garabato}, {Garc{\'\i}a-Lario},
  {Gosset}, {Haigron}, {Halbwachs}, {Hambly}, {Harrison}, {Hern{\'a}ndez},
  {Hestroffer}, {Hodgkin}, {Holl}, {Jan{\ss}en}, {Jevardat de Fombelle},
  {Jordan}, {Krone-Martins}, {Lanzafame}, {L{\"o}ffler}, {Marchal}, {Marrese},
  {Moitinho}, {Muinonen}, {Osborne}, {Pancino}, {Pauwels}, {Recio-Blanco},
  {Reyl{\'e}}, {Riello}, {Rimoldini}, {Roegiers}, {Rybizki}, {Sarro}, {Siopis},
  {Smith}, {Sozzetti}, {Utrilla}, {van Leeuwen}, {Abbas}, {{\'A}brah{\'a}m},
  {Abreu Aramburu}, {Aerts}, {Aguado}, {Ajaj}, {Aldea-Montero}, {Altavilla},
  {{\'A}lvarez}, {Alves}, {Anders}, {Anderson}, {Anglada Varela}, {Antoja},
  {Baines}, {Baker}, {Balaguer-N{\'u}{\~n}ez}, {Balbinot}, {Balog}, {Barache},
  {Barbato}, {Barros}, {Barstow}, {Bartolom{\'e}}, {Bassilana}, {Bauchet},
  {Becciani}, {Bellazzini}, {Berihuete}, {Bernet}, {Bertone}, {Bianchi},
  {Binnenfeld}, {Blanco-Cuaresma}, {Blazere}, {Boch}, {Bombrun}, {Bossini},
  {Bouquillon}, {Bragaglia}, {Bramante}, {Breedt}, {Bressan}, {Brouillet},
  {Brugaletta}, {Bucciarelli}, {Burlacu}, {Butkevich}, {Buzzi}, {Caffau},
  {Cancelliere}, {Cantat-Gaudin}, {Carballo}, {Carlucci}, {Carnerero},
  {Carrasco}, {Casamiquela}, {Castellani}, {Castro-Ginard}, {Chaoul},
  {Charlot}, {Chemin}, {Chiaramida}, {Chiavassa}, {Chornay}, {Comoretto},
  {Contursi}, {Cooper}, {Cornez}, {Cowell}, {Crifo}, {Cropper}, {Crosta},
  {Crowley}, {Dafonte}, {Dapergolas}, {David}, {David}, {de Laverny}, {De
  Luise}, {De March}, {De Ridder}, {de Souza}, {de Torres}, {del Peloso}, {del
  Pozo}, {Delbo}, {Delgado}, {Delisle}, {Demouchy}, {Dharmawardena}, {Di
  Matteo}, {Diakite}, {Diener}, {Distefano}, {Dolding}, {Edvardsson}, {Enke},
  {Fabre}, {Fabrizio}, {Faigler}, {Fedorets}, {Fernique}, {Fienga}, {Figueras},
  {Fournier}, {Fouron}, {Fragkoudi}, {Gai}, {Garcia-Gutierrez},
  {Garcia-Reinaldos}, {Garc{\'\i}a-Torres}, {Garofalo}, {Gavel}, {Gavras},
  {Gerlach}, {Geyer}, {Giacobbe}, {Gilmore}, {Girona}, {Giuffrida}, {Gomel},
  {Gomez}, {Gonz{\'a}lez-N{\'u}{\~n}ez}, {Gonz{\'a}lez-Santamar{\'\i}a},
  {Gonz{\'a}lez-Vidal}, {Granvik}, {Guillout}, {Guiraud},
  {Guti{\'e}rrez-S{\'a}nchez}, {Guy}, {Hatzidimitriou}, {Hauser}, {Haywood},
  {Helmer}, {Helmi}, {Sarmiento}, {Hidalgo}, {Hilger}, {H{\l}adczuk}, {Hobbs},
  {Holland}, {Huckle}, {Jardine}, {Jasniewicz}, {Jean-Antoine Piccolo},
  {Jim{\'e}nez-Arranz}, {Jorissen}, {Juaristi Campillo}, {Julbe}, {Karbevska},
  {Kervella}, {Khanna}, {Kontizas}, {Kordopatis}, {Korn}, {K{\'o}sp{\'a}l},
  {Kostrzewa-Rutkowska}, {Kruszy{\'n}ska}, {Kun}, {Laizeau}, {Lambert},
  {Lanza}, {Lasne}, {Le Campion}, {Lebreton}, {Lebzelter}, {Leccia}, {Leclerc},
  {Lecoeur-Taibi}, {Liao}, {Licata}, {Lindstr{\o}m}, {Lister}, {Livanou},
  {Lobel}, {Lorca}, {Loup}, {Madrero Pardo}, {Magdaleno Romeo}, {Managau},
  {Mann}, {Manteiga}, {Marchant}, {Marconi}, {Marcos}, {Marcos Santos},
  {Mar{\'\i}n Pina}, {Marinoni}, {Marocco}, {Marshall}, {Polo},
  {Mart{\'\i}n-Fleitas}, {Marton}, {Mary}, {Masip}, {Massari},
  {Mastrobuono-Battisti}, {Mazeh}, {McMillan}, {Messina}, {Michalik}, {Millar},
  {Mints}, {Molina}, {Molinaro}, {Moln{\'a}r}, {Monari}, {Mongui{\'o}},
  {Montegriffo}, {Montero}, {Mor}, {Mora}, {Morbidelli}, {Morel}, {Morris},
  {Muraveva}, {Murphy}, {Musella}, {Nagy}, {Noval}, {Oca{\~n}a}, {Ogden},
  {Ordenovic}, {Osinde}, {Pagani}, {Pagano}, {Palaversa}, {Palicio},
  {Pallas-Quintela}, {Panahi}, {Payne-Wardenaar}, {Pe{\~n}alosa Esteller},
  {Penttil{\"a}}, {Pichon}, {Piersimoni}, {Pineau}, {Plachy}, {Plum}, {Poggio},
  {Pr{\v{s}}a}, {Pulone}, {Racero}, {Ragaini}, {Rainer}, {Raiteri}, {Rambaux},
  {Ramos}, {Ramos-Lerate}, {Re Fiorentin}, {Regibo}, {Richards}, {Rios Diaz},
  {Ripepi}, {Riva}, {Rix}, {Rixon}, {Robichon}, {Robin}, {Robin}, {Roelens},
  {Rogues}, {Rohrbasser}, {Romero-G{\'o}mez}, {Rowell}, {Royer}, {Ruz Mieres},
  {Rybicki}, {Sadowski}, {S{\'a}ez N{\'u}{\~n}ez}, {Sagrist{\`a} Sell{\'e}s},
  {Sahlmann}, {Salguero}, {Samaras}, {Sanchez Gimenez}, {Sanna},
  {Santove{\~n}a}, {Sarasso}, {Schultheis}, {Sciacca}, {Segol}, {Segovia},
  {S{\'e}gransan}, {Semeux}, {Shahaf}, {Siddiqui}, {Siebert}, {Siltala},
  {Silvelo}, {Slezak}, {Slezak}, {Smart}, {Snaith}, {Solano}, {Solitro},
  {Souami}, {Souchay}, {Spagna}, {Spina}, {Spoto}, {Steele},
  {Steidelm{\"u}ller}, {Stephenson}, {S{\"u}veges}, {Surdej}, {Szabados},
  {Szegedi-Elek}, {Taris}, {Taylo}, {Teixeira}, {Tolomei}, {Tonello}, {Torra},
  {Torra}, {Torralba Elipe}, {Trabucchi}, {Tsounis}, {Turon}, {Ulla}, {Unger},
  {Vaillant}, {van Dillen}, {van Reeven}, {Vanel}, {Vecchiato}, {Viala},
  {Vicente}, {Voutsinas}, {Weiler}, {Wevers}, {Wyrzykowski}, {Yoldas}, {Yvard},
  {Zhao}, {Zorec}, {Zucker}, \& {Zwitter}}]{gaiadr3}
{Gaia Collaboration}, {Vallenari}, A., {Brown}, A.~G.~A., {et~al.} 2022, arXiv
  e-prints, arXiv:2208.00211.
\newblock \doarXiv{2208.00211}

\bibitem[{{Gilbert} {et~al.}(2019){Gilbert}, {Kirby}, {Escala}, {Wojno},
  {Kalirai}, \& {Guhathakurta}}]{gilbert19a}
{Gilbert}, K.~M., {Kirby}, E.~N., {Escala}, I., {et~al.} 2019, \apj, 883, 128,
  \dodoi{10.3847/1538-4357/ab3807}

\bibitem[{{Gilbert} {et~al.}(2020){Gilbert}, {Wojno}, {Kirby}, {Escala},
  {Beaton}, {Guhathakurta}, \& {Majewski}}]{gilbert20}
{Gilbert}, K.~M., {Wojno}, J., {Kirby}, E.~N., {et~al.} 2020, \aj, 160, 41,
  \dodoi{10.3847/1538-3881/ab9602}

\bibitem[{{Gilbert} {et~al.}(2006){Gilbert}, {Guhathakurta}, {Kalirai}, {Rich},
  {Majewski}, {Ostheimer}, {Reitzel}, {Cenarro}, {Cooper}, {Luine}, \&
  {Patterson}}]{gilbert06}
{Gilbert}, K.~M., {Guhathakurta}, P., {Kalirai}, J.~S., {et~al.} 2006, \apj,
  652, 1188, \dodoi{10.1086/508643}

\bibitem[{{Gilbert} {et~al.}(2007){Gilbert}, {Fardal}, {Kalirai},
  {Guhathakurta}, {Geha}, {Isler}, {Majewski}, {Ostheimer}, {Patterson},
  {Reitzel}, {Kirby}, \& {Cooper}}]{gilbert07}
{Gilbert}, K.~M., {Fardal}, M., {Kalirai}, J.~S., {et~al.} 2007, \apj, 668,
  245, \dodoi{10.1086/521094}

\bibitem[{{Gilbert} {et~al.}(2009){Gilbert}, {Guhathakurta}, {Kollipara},
  {Beaton}, {Geha}, {Kalirai}, {Kirby}, {Majewski}, \& {Patterson}}]{gilbert09}
{Gilbert}, K.~M., {Guhathakurta}, P., {Kollipara}, P., {et~al.} 2009, \apj,
  705, 1275, \dodoi{10.1088/0004-637X/705/2/1275}

\bibitem[{{Gilbert} {et~al.}(2012){Gilbert}, {Guhathakurta}, {Beaton},
  {Bullock}, {Geha}, {Kalirai}, {Kirby}, {Majewski}, {Ostheimer}, {Patterson},
  {Tollerud}, {Tanaka}, \& {Chiba}}]{gilbert12}
{Gilbert}, K.~M., {Guhathakurta}, P., {Beaton}, R.~L., {et~al.} 2012, \apj,
  760, 76, \dodoi{10.1088/0004-637X/760/1/76}

\bibitem[{{Gilbert} {et~al.}(2014){Gilbert}, {Kalirai}, {Guhathakurta},
  {Beaton}, {Geha}, {Kirby}, {Majewski}, {Patterson}, {Tollerud}, {Bullock},
  {Tanaka}, \& {Chiba}}]{gilbert14}
{Gilbert}, K.~M., {Kalirai}, J.~S., {Guhathakurta}, P., {et~al.} 2014, \apj,
  796, 76, \dodoi{10.1088/0004-637X/796/2/76}

\bibitem[{{Gilbert} {et~al.}(2018){Gilbert}, {Tollerud}, {Beaton},
  {Guhathakurta}, {Bullock}, {Chiba}, {Kalirai}, {Kirby}, {Majewski}, \&
  {Tanaka}}]{gilbert18}
{Gilbert}, K.~M., {Tollerud}, E., {Beaton}, R.~L., {et~al.} 2018, \apj, 852,
  128, \dodoi{10.3847/1538-4357/aa9f26}

\bibitem[{{Girardi}(2016)}]{Girardi16}
{Girardi}, L. 2016, \araa, 54, 95, \dodoi{10.1146/annurev-astro-081915-023354}

\bibitem[{{Grand} {et~al.}(2017){Grand}, {G{\'o}mez}, {Marinacci}, {Pakmor},
  {Springel}, {Campbell}, {Frenk}, {Jenkins}, \& {White}}]{grand17}
{Grand}, R. J.~J., {G{\'o}mez}, F.~A., {Marinacci}, F., {et~al.} 2017, \mnras,
  467, 179, \dodoi{10.1093/mnras/stx071}

\bibitem[{{Guhathakurta} {et~al.}(2005){Guhathakurta}, {Ostheimer}, {Gilbert},
  {Rich}, {Majewski}, {Kalirai}, {Reitzel}, \& {Patterson}}]{guhathakurta05}
{Guhathakurta}, P., {Ostheimer}, J.~C., {Gilbert}, K.~M., {et~al.} 2005, arXiv
  e-prints, astro.
\newblock \doarXiv{astro-ph/0502366}

\bibitem[{{Guhathakurta} {et~al.}(2006){Guhathakurta}, {Rich}, {Reitzel},
  {Cooper}, {Gilbert}, {Majewski}, {Ostheimer}, {Geha}, {Johnston}, \&
  {Patterson}}]{guhathakurta06}
{Guhathakurta}, P., {Rich}, R.~M., {Reitzel}, D.~B., {et~al.} 2006, \aj, 131,
  2497, \dodoi{10.1086/499562}

\bibitem[{{Hammer} {et~al.}(2018){Hammer}, {Yang}, {Wang}, {Ibata}, {Flores},
  \& {Puech}}]{hammer18}
{Hammer}, F., {Yang}, Y.~B., {Wang}, J.~L., {et~al.} 2018, \mnras, 475, 2754,
  \dodoi{10.1093/mnras/stx3343}

\bibitem[{{Hawkins} {et~al.}(2015){Hawkins}, {Jofr{\'e}}, {Masseron}, \&
  {Gilmore}}]{hawkins15}
{Hawkins}, K., {Jofr{\'e}}, P., {Masseron}, T., \& {Gilmore}, G. 2015, \mnras,
  453, 758, \dodoi{10.1093/mnras/stv1586}

\bibitem[{{Hayes} {et~al.}(2018){Hayes}, {Majewski}, {Shetrone},
  {Fern{\'a}ndez-Alvar}, {Allende Prieto}, {Schuster}, {Carigi}, {Cunha},
  {Smith}, {Sobeck}, {Almeida}, {Beers}, {Carrera}, {Fern{\'a}ndez-Trincado},
  {Garc{\'\i}a-Hern{\'a}ndez}, {Geisler}, {Lane}, {Lucatello}, {Matthews},
  {Minniti}, {Nitschelm}, {Tang}, {Tissera}, \& {Zamora}}]{hayes18}
{Hayes}, C.~R., {Majewski}, S.~R., {Shetrone}, M., {et~al.} 2018, \apj, 852,
  49, \dodoi{10.3847/1538-4357/aa9cec}

\bibitem[{{Helmi}(2008)}]{helmi08}
{Helmi}, A. 2008, \aapr, 15, 145, \dodoi{10.1007/s00159-008-0009-6}

\bibitem[{{Hogg} {et~al.}(2010){Hogg}, {Bovy}, \& {Lang}}]{hogg10}
{Hogg}, D.~W., {Bovy}, J., \& {Lang}, D. 2010, arXiv e-prints, arXiv:1008.4686.
\newblock \doarXiv{1008.4686}

\bibitem[{{Hunter}(2007)}]{matplotlib}
{Hunter}, J.~D. 2007, Computing in Science and Engineering, 9, 90,
  \dodoi{10.1109/MCSE.2007.55}

\bibitem[{{Ibata} {et~al.}(2001){Ibata}, {Irwin}, {Lewis}, {Ferguson}, \&
  {Tanvir}}]{ibata01}
{Ibata}, R., {Irwin}, M., {Lewis}, G., {Ferguson}, A. M.~N., \& {Tanvir}, N.
  2001, Nature, 412, 49.
\newblock \doarXiv{astro-ph/0107090}

\bibitem[{{Ibata} {et~al.}(2007){Ibata}, {Martin}, {Irwin}, {Chapman},
  {Ferguson}, {Lewis}, \& {McConnachie}}]{ibata07}
{Ibata}, R., {Martin}, N.~F., {Irwin}, M., {et~al.} 2007, \apj, 671, 1591,
  \dodoi{10.1086/522574}

\bibitem[{{Ibata} {et~al.}(2014){Ibata}, {Lewis}, {McConnachie}, {Martin},
  {Irwin}, {Ferguson}, {Babul}, {Bernard}, {Chapman}, {Collins}, {Fardal},
  {Mackey}, {Navarro}, {Pe{\~n}arrubia}, {Rich}, {Tanvir}, \&
  {Widrow}}]{ibata14}
{Ibata}, R.~A., {Lewis}, G.~F., {McConnachie}, A.~W., {et~al.} 2014, \apj, 780,
  128, \dodoi{10.1088/0004-637X/780/2/128}

\bibitem[{{Irwin} {et~al.}(2005){Irwin}, {Ferguson}, {Ibata}, {Lewis}, \&
  {Tanvir}}]{irwin05}
{Irwin}, M.~J., {Ferguson}, A. M.~N., {Ibata}, R.~A., {Lewis}, G.~F., \&
  {Tanvir}, N.~R. 2005, \apjl, 628, L105, \dodoi{10.1086/432718}

\bibitem[{{Johnston} {et~al.}(1996){Johnston}, {Hernquist}, \&
  {Bolte}}]{johnston96}
{Johnston}, K.~V., {Hernquist}, L., \& {Bolte}, M. 1996, \apj, 465, 278,
  \dodoi{10.1086/177418}

\bibitem[{{Jorgensen}(1994)}]{Jorgensen94}
{Jorgensen}, U.~G. 1994, \aap, 284, 179

\bibitem[{{Kalirai} {et~al.}(2006{\natexlab{a}}){Kalirai}, {Guhathakurta},
  {Gilbert}, {Reitzel}, {Majewski}, {Rich}, \& {Cooper}}]{kalirai06b}
{Kalirai}, J.~S., {Guhathakurta}, P., {Gilbert}, K.~M., {et~al.}
  2006{\natexlab{a}}, \apj, 641, 268, \dodoi{10.1086/498700}

\bibitem[{{Kalirai} {et~al.}(2006{\natexlab{b}}){Kalirai}, {Gilbert},
  {Guhathakurta}, {Majewski}, {Ostheimer}, {Rich}, {Cooper}, {Reitzel}, \&
  {Patterson}}]{kalirai06}
{Kalirai}, J.~S., {Gilbert}, K.~M., {Guhathakurta}, P., {et~al.}
  2006{\natexlab{b}}, \apj, 648, 389, \dodoi{10.1086/505697}

\bibitem[{{Kirby}(2011)}]{kirby11}
{Kirby}, E.~N. 2011, \pasp, 123, 531, \dodoi{10.1086/660019}

\bibitem[{{Kirby} {et~al.}(2013){Kirby}, {Cohen}, {Guhathakurta}, {Cheng},
  {Bullock}, \& {Gallazzi}}]{kirby13}
{Kirby}, E.~N., {Cohen}, J.~G., {Guhathakurta}, P., {et~al.} 2013, \apj, 779,
  102, \dodoi{10.1088/0004-637X/779/2/102}

\bibitem[{{Kirby} {et~al.}(2020){Kirby}, {Gilbert}, {Escala}, {Wojno},
  {Guhathakurta}, {Majewski}, \& {Beaton}}]{kirby20}
{Kirby}, E.~N., {Gilbert}, K.~M., {Escala}, I., {et~al.} 2020, \aj, 159, 46,
  \dodoi{10.3847/1538-3881/ab5f0f}

\bibitem[{{Kirby} {et~al.}(2008){Kirby}, {Guhathakurta}, \&
  {Sneden}}]{kirby08a}
{Kirby}, E.~N., {Guhathakurta}, P., \& {Sneden}, C. 2008, \apj, 682, 1217,
  \dodoi{10.1086/589627}

\bibitem[{{Kirby} {et~al.}(2015){Kirby}, {Simon}, \& {Cohen}}]{kirby15}
{Kirby}, E.~N., {Simon}, J.~D., \& {Cohen}, J.~G. 2015, \apj, 810, 56,
  \dodoi{10.1088/0004-637X/810/1/56}

\bibitem[{{Kobayashi} \& {Nakasato}(2011)}]{kobayashi11}
{Kobayashi}, C., \& {Nakasato}, N. 2011, \apj, 729, 16,
  \dodoi{10.1088/0004-637X/729/1/16}

\bibitem[{{Koch} {et~al.}(2008){Koch}, {Grebel}, {Gilmore}, {Wyse}, {Kleyna},
  {Harbeck}, {Wilkinson}, \& {Wyn Evans}}]{Koch08}
{Koch}, A., {Grebel}, E.~K., {Gilmore}, G.~F., {et~al.} 2008, \aj, 135, 1580,
  \dodoi{10.1088/0004-6256/135/4/1580}

\bibitem[{{Kupka} {et~al.}(1999){Kupka}, {Piskunov}, {Ryabchikova}, {Stempels},
  \& {Weiss}}]{kupka99}
{Kupka}, F., {Piskunov}, N., {Ryabchikova}, T.~A., {Stempels}, H.~C., \&
  {Weiss}, W.~W. 1999, \aaps, 138, 119, \dodoi{10.1051/aas:1999267}

\bibitem[{{Kurucz}(1992)}]{kurucz92}
{Kurucz}, R.~L. 1992, \rmxaa, 23

\bibitem[{{Kurucz}(1993)}]{kurucz93}
---. 1993, Physica Scripta Volume T, 47, 110,
  \dodoi{10.1088/0031-8949/1993/T47/017}

\bibitem[{{Mackey} {et~al.}(2019){Mackey}, {Lewis}, {Brewer}, {Ferguson},
  {Veljanoski}, {Huxor}, {Collins}, {C{\^o}t{\'e}}, {Ibata}, {Irwin}, {Martin},
  {McConnachie}, {Pe{\~n}arrubia}, {Tanvir}, \& {Wan}}]{mackey19}
{Mackey}, D., {Lewis}, G.~F., {Brewer}, B.~J., {et~al.} 2019, \nat, 574, 69,
  \dodoi{10.1038/s41586-019-1597-1}

\bibitem[{{Majewski} {et~al.}(2000){Majewski}, {Ostheimer}, {Kunkel}, \&
  {Patterson}}]{majewski00}
{Majewski}, S.~R., {Ostheimer}, J.~C., {Kunkel}, W.~E., \& {Patterson}, R.~J.
  2000, \aj, 120, 2550, \dodoi{10.1086/316836}

\bibitem[{{Matteucci}(2021)}]{Matteucci21}
{Matteucci}, F. 2021, \aapr, 29, 5, \dodoi{10.1007/s00159-021-00133-8}

\bibitem[{{McConnachie}(2012)}]{mcconnachie12}
{McConnachie}, A.~W. 2012, \aj, 144, 4, \dodoi{10.1088/0004-6256/144/1/4}

\bibitem[{{McConnachie} {et~al.}(2005){McConnachie}, {Irwin}, {Ferguson},
  {Ibata}, {Lewis}, \& {Tanvir}}]{mcconnachie05}
{McConnachie}, A.~W., {Irwin}, M.~J., {Ferguson}, A.~M.~N., {et~al.} 2005,
  \mnras, 356, 979, \dodoi{10.1111/j.1365-2966.2004.08514.x}

\bibitem[{{McConnachie} {et~al.}(2003){McConnachie}, {Irwin}, {Ibata},
  {Ferguson}, {Lewis}, \& {Tanvir}}]{mcconnachie03}
{McConnachie}, A.~W., {Irwin}, M.~J., {Ibata}, R.~A., {et~al.} 2003, \mnras,
  343, 1335, \dodoi{10.1046/j.1365-8711.2003.06785.x}

\bibitem[{{McConnachie} {et~al.}(2009){McConnachie}, {Irwin}, {Ibata},
  {Dubinski}, {Widrow}, {Martin}, {C{\^o}t{\'e}}, {Dotter}, {Navarro},
  {Ferguson}, {Puzia}, {Lewis}, {Babul}, {Barmby}, {Bienaym{\'e}}, {Chapman},
  {Cockcroft}, {Collins}, {Fardal}, {Harris}, {Huxor}, {Mackey},
  {Pe{\~n}arrubia}, {Rich}, {Richer}, {Siebert}, {Tanvir}, {Valls-Gabaud}, \&
  {Venn}}]{McConnachie09}
---. 2009, \nat, 461, 66, \dodoi{10.1038/nature08327}

\bibitem[{{McConnachie} {et~al.}(2018){McConnachie}, {Ibata}, {Martin},
  {Ferguson}, {Collins}, {Gwyn}, {Irwin}, {Lewis}, {Mackey}, {Davidge},
  {Arias}, {Conn}, {C{\^o}t{\'e}}, {Crnojevic}, {Huxor}, {Penarrubia},
  {Spengler}, {Tanvir}, {Valls-Gabaud}, {Babul}, {Barmby}, {Bate}, {Bernard},
  {Chapman}, {Dotter}, {Harris}, {McMonigal}, {Navarro}, {Puzia}, {Rich},
  {Thomas}, \& {Widrow}}]{mcconnachie18}
{McConnachie}, A.~W., {Ibata}, R., {Martin}, N., {et~al.} 2018, \apj, 868, 55,
  \dodoi{10.3847/1538-4357/aae8e7}

\bibitem[{{McWilliam}(1997)}]{mcwilliam97}
{McWilliam}, A. 1997, \araa, 35, 503, \dodoi{10.1146/annurev.astro.35.1.503}

\bibitem[{{Monachesi} {et~al.}(2019){Monachesi}, {G{\'o}mez}, {Grand},
  {Simpson}, {Kauffmann}, {Bustamante}, {Marinacci}, {Pakmor}, {Springel},
  {Frenk}, {White}, \& {Tissera}}]{monachesi19}
{Monachesi}, A., {G{\'o}mez}, F.~A., {Grand}, R. J.~J., {et~al.} 2019, \mnras,
  485, 2589, \dodoi{10.1093/mnras/stz538}

\bibitem[{{Morgan} {et~al.}(1943){Morgan}, {Keenan}, \& {Kellman}}]{morgan1943}
{Morgan}, W.~W., {Keenan}, P.~C., \& {Kellman}, E. 1943, {An atlas of stellar
  spectra, with an outline of spectral classification}

\bibitem[{{Mould} \& {Kristian}(1986)}]{mould86}
{Mould}, J., \& {Kristian}, J. 1986, \apj, 305, 591, \dodoi{10.1086/164273}

\bibitem[{{Newman} {et~al.}(2013){Newman}, {Cooper}, {Davis}, {Faber}, {Coil},
  {Guhathakurta}, {Koo}, {Phillips}, {Conroy}, {Dutton}, {Finkbeiner}, {Gerke},
  {Rosario}, {Weiner}, {Willmer}, {Yan}, {Harker}, {Kassin}, {Konidaris},
  {Lai}, {Madgwick}, {Noeske}, {Wirth}, {Connolly}, {Kaiser}, {Kirby},
  {Lemaux}, {Lin}, {Lotz}, {Luppino}, {Marinoni}, {Matthews}, {Metevier}, \&
  {Schiavon}}]{newman13}
{Newman}, J.~A., {Cooper}, M.~C., {Davis}, M., {et~al.} 2013, \apjs, 208, 5,
  \dodoi{10.1088/0067-0049/208/1/5}

\bibitem[{{Nissen} \& {Schuster}(2010)}]{Nissen10}
{Nissen}, P.~E., \& {Schuster}, W.~J. 2010, \aap, 511, L10,
  \dodoi{10.1051/0004-6361/200913877}

\bibitem[{{Ostheimer}(2003)}]{ostheimer03}
{Ostheimer}, James~Craig, J. 2003, PhD thesis, UNIVERSITY OF VIRGINIA

\bibitem[{{Quirk} \& {Patel}(2020)}]{quirk20}
{Quirk}, A. C.~N., \& {Patel}, E. 2020, \mnras, 497, 2870,
  \dodoi{10.1093/mnras/staa2152}

\bibitem[{{Rich} {et~al.}(1996){Rich}, {Mighell}, \& {Neill}}]{rich96}
{Rich}, R.~M., {Mighell}, K.~J., \& {Neill}, J.~D. 1996, in Astronomical
  Society of the Pacific Conference Series, Vol.~92, Formation of the Galactic
  Halo...Inside and Out, ed. H.~L. {Morrison} \& A.~{Sarajedini}, 544

\bibitem[{{Richardson} {et~al.}(2009){Richardson}, {Ferguson}, {Mackey},
  {Irwin}, {Chapman}, {Huxor}, {Ibata}, {Lewis}, \& {Tanvir}}]{richardson09}
{Richardson}, J.~C., {Ferguson}, A.~M.~N., {Mackey}, A.~D., {et~al.} 2009,
  \mnras, 396, 1842, \dodoi{10.1111/j.1365-2966.2009.14788.x}

\bibitem[{{Riess} {et~al.}(2012){Riess}, {Fliri}, \& {Valls-Gabaud}}]{riess12}
{Riess}, A.~G., {Fliri}, J., \& {Valls-Gabaud}, D. 2012, \apj, 745, 156,
  \dodoi{10.1088/0004-637X/745/2/156}

\bibitem[{{Samland} \& {Gerhard}(2003)}]{samland03}
{Samland}, M., \& {Gerhard}, O.~E. 2003, \aap, 399, 961,
  \dodoi{10.1051/0004-6361:20021842}

\bibitem[{{Sbordone}(2005)}]{sbordone05}
{Sbordone}, L. 2005, Memorie della Societa Astronomica Italiana Supplementi, 8,
  61

\bibitem[{{Sbordone} {et~al.}(2004){Sbordone}, {Bonifacio}, {Castelli}, \&
  {Kurucz}}]{sbordone04}
{Sbordone}, L., {Bonifacio}, P., {Castelli}, F., \& {Kurucz}, R.~L. 2004,
  Memorie della Societa Astronomica Italiana Supplementi, 5, 93

\bibitem[{{Searle} \& {Zinn}(1978)}]{Searle78}
{Searle}, L., \& {Zinn}, R. 1978, \apj, 225, 357, \dodoi{10.1086/156499}

\bibitem[{{Simon} \& {Geha}(2007)}]{simon07}
{Simon}, J.~D., \& {Geha}, M. 2007, \apj, 670, 313, \dodoi{10.1086/521816}

\bibitem[{{Sneden}(1973)}]{sneden73}
{Sneden}, C.~A. 1973, PhD thesis, THE UNIVERSITY OF TEXAS AT AUSTIN.

\bibitem[{{Sohn} {et~al.}(2007){Sohn}, {Majewski}, {Mu{\~n}oz}, {Kunkel},
  {Johnston}, {Ostheimer}, {Guhathakurta}, {Patterson}, {Siegel}, \&
  {Cooper}}]{sohn07}
{Sohn}, S.~T., {Majewski}, S.~R., {Mu{\~n}oz}, R.~R., {et~al.} 2007, \apj, 663,
  960, \dodoi{10.1086/518302}

\bibitem[{{Stanek} {et~al.}(1997){Stanek}, {Udalski}, {Szyma{\'N}ski},
  {Ka{\L}u{\.Z}ny}, {Kubiak}, {Mateo}, \& {Krzemi{\'N}ski}}]{stanek97}
{Stanek}, K.~Z., {Udalski}, A., {Szyma{\'N}ski}, M., {et~al.} 1997, \apj, 477,
  163, \dodoi{10.1086/303702}

\bibitem[{{Tanaka} {et~al.}(2010){Tanaka}, {Chiba}, {Komiyama}, {Guhathakurta},
  {Kalirai}, \& {Iye}}]{tanaka10}
{Tanaka}, M., {Chiba}, M., {Komiyama}, Y., {et~al.} 2010, \apj, 708, 1168,
  \dodoi{10.1088/0004-637X/708/2/1168}

\bibitem[{{Tissera} {et~al.}(2017){Tissera}, {Machado}, {Vilchez}, {Pedrosa},
  {Sanchez-Blazquez}, \& {Varela}}]{tissera17}
{Tissera}, P.~B., {Machado}, R. E.~G., {Vilchez}, J.~M., {et~al.} 2017, \aap,
  604, A118, \dodoi{10.1051/0004-6361/201628915}

\bibitem[{{Tissera} {et~al.}(2012){Tissera}, {White}, \&
  {Scannapieco}}]{tissera12}
{Tissera}, P.~B., {White}, S. D.~M., \& {Scannapieco}, C. 2012, \mnras, 420,
  255, \dodoi{10.1111/j.1365-2966.2011.20028.x}

\bibitem[{{Tollerud} {et~al.}(2012){Tollerud}, {Beaton}, {Geha}, {Bullock},
  {Guhathakurta}, {Kalirai}, {Majewski}, {Kirby}, {Gilbert}, {Yniguez},
  {Patterson}, {Ostheimer}, {Cooke}, {Dorman}, {Choudhury}, \&
  {Cooper}}]{tollerud12}
{Tollerud}, E.~J., {Beaton}, R.~L., {Geha}, M.~C., {et~al.} 2012, \apj, 752,
  45, \dodoi{10.1088/0004-637X/752/1/45}

\bibitem[{{Tolstoy} {et~al.}(2009){Tolstoy}, {Hill}, \& {Tosi}}]{Tolstoy09}
{Tolstoy}, E., {Hill}, V., \& {Tosi}, M. 2009, \araa, 47, 371,
  \dodoi{10.1146/annurev-astro-082708-101650}

\bibitem[{{van der Walt} {et~al.}(2011){van der Walt}, {Colbert}, \&
  {Varoquaux}}]{numpy}
{van der Walt}, S., {Colbert}, S.~C., \& {Varoquaux}, G. 2011, Computing in
  Science and Engineering, 13, 22, \dodoi{10.1109/MCSE.2011.37}

\bibitem[{{VandenBerg} {et~al.}(2006){VandenBerg}, {Bergbusch}, \&
  {Dowler}}]{vandenberg06}
{VandenBerg}, D.~A., {Bergbusch}, P.~A., \& {Dowler}, P.~D. 2006, \apjs, 162,
  375, \dodoi{10.1086/498451}

\bibitem[{{Vargas} {et~al.}(2014){Vargas}, {Gilbert}, {Geha}, {Tollerud},
  {Kirby}, \& {Guhathakurta}}]{vargas14}
{Vargas}, L.~C., {Gilbert}, K.~M., {Geha}, M., {et~al.} 2014, \apjl, 797, L2,
  \dodoi{10.1088/2041-8205/797/1/L2}

\bibitem[{{Virtanen} {et~al.}(2019){Virtanen}, {Gommers}, {Oliphant},
  {Haberland}, {Reddy}, {Cournapeau}, {Burovski}, {Peterson}, {Weckesser},
  {Bright}, {van der Walt}, {Brett}, {Wilson}, {Jarrod Millman}, {Mayorov},
  {Nelson}, {Jones}, {Kern}, {Larson}, {Carey}, {Polat}, {Feng}, {Moore}, {Vand
  erPlas}, {Laxalde}, {Perktold}, {Cimrman}, {Henriksen}, {Quintero}, {Harris},
  {Archibald}, {Ribeiro}, {Pedregosa}, {van Mulbregt}, \&
  {Contributors}}]{scipy}
{Virtanen}, P., {Gommers}, R., {Oliphant}, T.~E., {et~al.} 2019, arXiv
  e-prints, arXiv:1907.10121.
\newblock \doarXiv{1907.10121}

\bibitem[{{Wojno} {et~al.}(2020){Wojno}, {Gilbert}, {Kirby}, {Escala},
  {Beaton}, {Tollerud}, {Majewski}, \& {Guhathakurta}}]{wojno20}
{Wojno}, J., {Gilbert}, K.~M., {Kirby}, E.~N., {et~al.} 2020, \apj, 895, 78,
  \dodoi{10.3847/1538-4357/ab8ccb}

\bibitem[{{Yang} {et~al.}(2013){Yang}, {Kirby}, {Guhathakurta}, {Peng}, \&
  {Cheng}}]{yang13}
{Yang}, L., {Kirby}, E.~N., {Guhathakurta}, P., {Peng}, E.~W., \& {Cheng}, L.
  2013, \apj, 768, 4, \dodoi{10.1088/0004-637X/768/1/4}

\bibitem[{{Zolotov} {et~al.}(2009){Zolotov}, {Willman}, {Brooks}, {Governato},
  {Brook}, {Hogg}, {Quinn}, \& {Stinson}}]{Zolotov09}
{Zolotov}, A., {Willman}, B., {Brooks}, A.~M., {et~al.} 2009, \apj, 702, 1058,
  \dodoi{10.1088/0004-637X/702/2/1058}

\bibitem[{{Zucker} {et~al.}(2004){Zucker}, {Kniazev}, {Bell},
  {Mart{\'\i}nez-Delgado}, {Grebel}, {Rix}, {Rockosi}, {Holtzman}, {Walterbos},
  {Ivezi{\'c}}, {Brinkmann}, {Brewington}, {Harvanek}, {Kleinman},
  {Krzesinski}, {Lamb}, {Long}, {Newman}, {Nitta}, \& {Snedden}}]{zucker04}
{Zucker}, D.~B., {Kniazev}, A.~Y., {Bell}, E.~F., {et~al.} 2004, \apjl, 612,
  L117, \dodoi{10.1086/424706}

\end{thebibliography}
\bibliographystyle{aasjournal}

\end{document}